\documentclass[a4paper,11pt]{article}
\pdfoutput=1

\usepackage{aas_macros}
\usepackage{rotating}
\usepackage{pst-all}	
\usepackage{color}  
\usepackage{comment}
\usepackage{braket}
\usepackage{subfigure}  % For subfigures
\usepackage{url}

\usepackage{jcappub} % for details on the use of the package, please
\usepackage[T1]{fontenc} % if needed

\usepackage{tikz}
\usetikzlibrary{matrix, fit, positioning,arrows.meta,arrows,decorations.pathreplacing,shapes.geometric,backgrounds}

\usepackage{bm}

\newcommand{\vk}{\vec{k}}
\newcommand{\vx}{\vec{x}}
\newcommand{\vq}{\vec{q}}
\newcommand{\vPsi}{\vec{\Psi}}
\newcommand{\vPsiLR}{\vec{\Psi}_{\rm LR}}
\newcommand{\vPsiSR}{\vec{\Psi}_{\rm SR}}
\newcommand{\vA}{\vec{A}}

\newcommand{\vomega}{\vec{\omega}}

\newcommand{\bk}{\mathbf{k}}

\newcommand{\avg}[1]{\left\langle #1\right\rangle}

\title{\boldmath Ridged Lagrangian Perturbation Theory (RLPT)}

\author[1,2]{Francisco-Shu Kitaura}
\author[3,4,5,6,1,2]{and Francesco Sinigaglia}

\affiliation[1]{Instituto de Astrof\'{\i}sica de Canarias, Vía Láctea s/n, E-38205, La Laguna, Tenerife, Spain}  
\affiliation[2]{Departamento de Astrof\'{\i}sica, Universidad de La Laguna, E-38206, La Laguna, Tenerife, Spain}
\affiliation[3]{Institute for Fundamental Physics of the Universe, Via Beirut 2, I-34151 Trieste, Italy}
\affiliation[4]{SISSA - International School for Advanced Studies, Via Bonomea 265, 34136 Trieste, Italy}
\affiliation[5]{INAF - Osservatorio Astronomico di Trieste, Via G. B. Tiepolo 11, I-34131 Trieste, Italy}
\affiliation[6]{INFN – National Institute for Nuclear Physics, Via Valerio 2, I-34127 Trieste, Italy}

\emailAdd{fkitaura@iac.es}
\emailAdd{fsinigag@sissa.it}

\abstract{
Galaxy surveys demand fast large-scale structure forward models that preserve large-scale phases while providing realistic nonlinear morphology at fixed force resolution.
Single-step Lagrangian Perturbation Theory (LPT) solvers are efficient, but they typically yield overly diffuse filaments and knots and underpredict small-scale clustering.

We introduce Ridged Lagrangian Perturbation Theory (RLPT), a modular two-step scheme: a standard long-range LPT/ALPT transport is followed by a single post-processing Eulerian ridging update that reconstructs a short-range, curl-free displacement from the realised density field through a smooth scale separation and a Poisson inversion.
This explicit completion layer is inexpensive, preserves the large-scale solution, and provides a small set of transparent parameters to tune the short-range response.

We test RLPT against particle-mesh and $N$-body references and find that one additional ridging step systematically improves both nonlinear power and field-level agreement relative to 2LPT/ALPT baselines.
Finally, we demonstrate that ridging can be repurposed as a deterministic subgrid relocation model: even when the underlying dark-matter field is only "good enough" on the mesh, ridging enables controlled tuning of tracer clustering beyond the nominal resolution, which is particularly relevant for mock-galaxy production and observational systematics sensitive to close pairs.
}

\begin{document}
\maketitle
\flushbottom

% =========================================================
\section{Introduction}
\label{sec:intro}

The cosmic web is the large-scale structure (LSS) pattern emerging from gravitational instability in an expanding universe.
Its morphology --- voids, sheets, filaments, and knots --- encodes cosmological information inherited from the initial conditions and processed by nonlinear dynamics \citep[e.g.][]{Bernardeau_2002,2018MNRAS.473.1195L}.
Ongoing and upcoming surveys such as DESI, Euclid, Roman and J-PAS map the three-dimensional distribution of galaxies over unprecedented volumes and redshift ranges \citep[e.g.][]{DESI,Euclid,Wang_2022,jpas}.

Since the observable Universe is unique, robust analyses require large ensembles of realistic mock catalogues to estimate covariance matrices, validate pipelines, and control observational systematics.
High-fidelity $N$-body simulations remain the standard for nonlinear gravity \citep[e.g.][]{Angulo_2022}, but their cost is prohibitive when mass-producing mocks across large volumes and parameter grids (different initial conditions, cosmological parameters, primordial non-Gaussianities, gravity models, neutrino physics, etc.).
This motivates approximate gravity solvers and hybrid pipelines that trade detailed trajectories for speed while preserving the relevant statistical and morphological content of LSS, including accelerated particle mesh (hereafter PM)  approaches such as COLA \citep[][]{Tassev_2013}, FastPM \citep{2013JCAP...06..036T,2016MNRAS.463.2273F}, or GLAM \citep[][]{Klypin_2018} and schemes that add a short-range correction step to recover missing small-scale forces \citep[e.g.][]{2018JCAP...11..009D,2020JCAP...04..002F,Chuang_2019}.

A number of successful mock-generation frameworks therefore rely on fast approximations (e.g.\ 2LPT {\color{black} and its regularised Augmented Lagrangian
Perturbation Theory (ALPT) variant,} or accelerated PM approaches) coupled to bias-mapping or learning-based methods \citep[e.g.][]{1995A&A...296..575B,2014MNRAS.439L..21K,2015MNRAS.450.1836K,2015MNRAS.446.2621C,2019MNRAS.483L..58B,2020MNRAS.491.2565B,2019PNAS..11613825H}.
Related work has explored iterative, Eulerian extensions of ALPT (eALPT) in which the displacement is applied in several Eulerian mapping steps, effectively introducing a small-scale regularisation and generating viscosity- and vorticity-like contributions through the composition of maps \citep[e.g.][]{Kitaura_2024}. While this multi-step approach can reach excellent field-level agreement with only a few iterations, its performance degrades when the force/mesh resolution becomes too coarse (typically for cell sizes $\gtrsim 2.5,h^{-1},\mathrm{Mpc}$), and it is not designed as a strictly {post-processing} correction that can be applied on top of a single-step ALPT realisation at an arbitrary target redshift.

A practical limitation persists across many fast solvers at fixed force resolution (e.g.\ PM meshes of a few $h^{-1}$Mpc):
even if large-scale phases are captured accurately, the nonlinear density field often lacks sufficient small-scale sharpening.
Filaments become too thick, knots too shallow, and the power spectrum typically exhibits a deficit toward high wavenumbers.
This is especially relevant because mock-population, field-level biasing, and subgrid placement schemes are sensitive to the morphology of the underlying matter field \citep[][]{Forero_2024}, even if non-local bias schemes are able to compensate for inaccuracies in the gravity solver \cite[][]{Coloma_2024}.

%\paragraph{Core idea.}
{Motivated by the need to improve the small-scale clustering when working with coarse meshes, we explore in this work {Ridged} Lagrangian Perturbation Theory (hereafter RLPT)}. RLPT is inspired by the scale-separation philosophy of ALPT and by the Eulerian multi-mapping viewpoint of eALPT, but it implements a  simple operator split tailored to the fixed-resolution regime: a fast long-range LPT-based transport provides accurate large-scale phases, and a single additional Eulerian step supplies the missing short-range response. Concretely, the ridging operator is applied as a post-processing correction to the realised density field and costs essentially one additional Poisson solve (one forward and one inverse FFT on the mesh), while remaining straightforward to generalise to a small number of iterative applications if additional refinement is required.

{In addition, in the spirit of modeling small scales beyond the mesh, we explore in this work for the first time the usage of smooth particle hydrodynamics (hereafter SPH) kernels in LPT}. These kernels have been used in SPH for decades to address the challenge of modelling short-range dynamics without relying on fine meshes by using compact-support kernels and local gradient estimates in particle neighbourhoods \citep[see,e.g.,][]{annurev:/content/journals/10.1146/annurev.aa.30.090192.002551,annurev:/content/journals/10.1146/annurev-astro-081309-130914,10.1111/j.1365-2966.2012.21439.x,10.1145/2461912.2461984}. This perspective motivates ``smooth particle ridging'' operators ({hereafter} SPRLPT) that act below the nominal cell size, enabling controlled short-range updates beyond the mesh resolution. Such SPH-inspired schemes provide a natural route to FFT-free (or hybrid) subgrid corrections for fast large-scale structure solvers, and they appear particularly promising for tracer-level applications. We illustrate this potential in a proof-of-concept study.

The paper is organised as follows.
Section~\ref{sec:background} reviews the perturbative context and motivates an explicit short-range completion at fixed force resolution.
Section~\ref{sec:method} introduces RLPT as a two-step operator split: long-range $n$LPT/A-($n$)LPT transport followed by a post-processing Eulerian ridging update, and it also describes an optional vortical extension and a smooth-particle (SPH-inspired) ridging variant.
Section~\ref{sec:results} presents three numerical studies: (i) two particle-mesh benchmarks against \textsc{FastPM} in $L=100$ and $200\,h^{-1}\mathrm{Mpc}$ boxes, including a vorticity test at high $k$; (ii) an \textsc{Abacus} $N$-body comparison on a coarse $360^3$ mesh highlighting the role of tetrahedral mass assignment; and (iii) a proof-of-concept subgrid application to halo positioning and tracer small-scale control.
We conclude in Section~\ref{sec:discussion} with implications for mock production and field-level modelling. {Throughout the paper, we consistently adopt both for LPT and for the \textsc{FastPM} N-body calculations the same {Planck2018} cosmological parameters \citep{Planck2018} used to run the base \textsc{Abacus} simulation considered herein.}

% =========================================================
\section{Background and motivation}
\label{sec:background}

Fast PM gravity solvers have become a workhorse for large-scale structure applications, offering a favourable accuracy–cost trade-off compared to full $N$-body runs \citep[][]{Tassev_2013,2013JCAP...06..036T,2016MNRAS.463.2273F,Klypin_2018}.
Nevertheless, even in their accelerated variants, PM approaches typically remain substantially more expensive than single-step LPT-based solvers: they require repeated force evaluations (time stepping), 3D FFTs at each step, and non-negligible memory for mesh fields and particle state, so that both runtime and memory footprint can exceed LPT baselines by orders of magnitude for comparable volumes and resolutions \citep[see Table~1 in][]{Blot_2019}.
This becomes especially true if one aims at simulating the cosmic volume and target density of the full DESI survey \citep[see][]{KitauraSinigaglia_DESI_2026}, or if one attempts Bayesian {field-level} inference from galaxy clustering, which requires many forward-model evaluations \citep[see, e.g.,][]{2025arXiv250603969R}.

Ideally, one would like simple, physically motivated analytic models that can be calibrated with only a handful of parameters, yet still reach percent-level accuracy relative to full $N$-body calculations on the range of scales most relevant for cosmological analyses (e.g.\ up to quasi-nonlinear wavenumbers, $k\sim 0.2$--$0.5\,h\,\mathrm{Mpc}^{-1}$, depending on redshift and tracer). 
Such models should preserve the correct large-scale phases while providing a controlled and interpretable description of unresolved short-range dynamics, enabling both rapid mock production and efficient forward modelling in field-level inference.

With this goal in mind, we now summarise the perturbative transport frameworks that provide the natural long-range backbone for such models, and we highlight the specific fixed-resolution failure mode that motivates our Eulerian completion.

LPT describes gravitational evolution through a displacement field expanded in perturbative orders of the initial conditions \citep{Bernardeau_2002}.
At leading order, the Zel'dovich approximation captures large-scale flows but breaks down after shell crossing.
Higher-order LPT (e.g.\ 2LPT) improves tidal evolution but still fails to represent post-shell-crossing dynamics \citep[e.g.][]{1995A&A...296..575B,1995MNRAS.276..115C}.
ALPT \citep{Kitaura_2013} partially regularises this regime by combining a long-range LPT displacement with a short-range spherical-collapse displacement controlled by a smoothing scale.

In survey applications, the dominant limitation is often the force/mesh resolution rather than particle number:
large-scale transport is already ``good enough'', while the field lacks nonlinear sharpening below a few mesh cells.
RLPT makes this separation explicit by keeping a robust long-range transport model (Step~1) and restoring missing small-scale growth through a controlled Eulerian operator acting on the post-transport density field (Step~2).

% =========================================================
\section{Method}
\label{sec:method}

RLPT is a two-step forward model for gravity at fixed spatial/force resolution.
Starting from Lagrangian particle coordinates $\vq$ and initial conditions, we apply:
(i) a long-range LPT-based transport operator to obtain particle positions $\vx_{(1)}$, and
(ii) a post-processing Eulerian ridging operator that reconstructs a purely short-range {potential} displacement from the realised overdensity field and applies it once to obtain $\vx_{(2)}$.

\begin{sidewaysfigure}
\hspace{-1.5cm}
\vspace{-0.2cm}
    \begin{tabular}{cccc}
\subfigure{\includegraphics[width=0.28\textwidth]{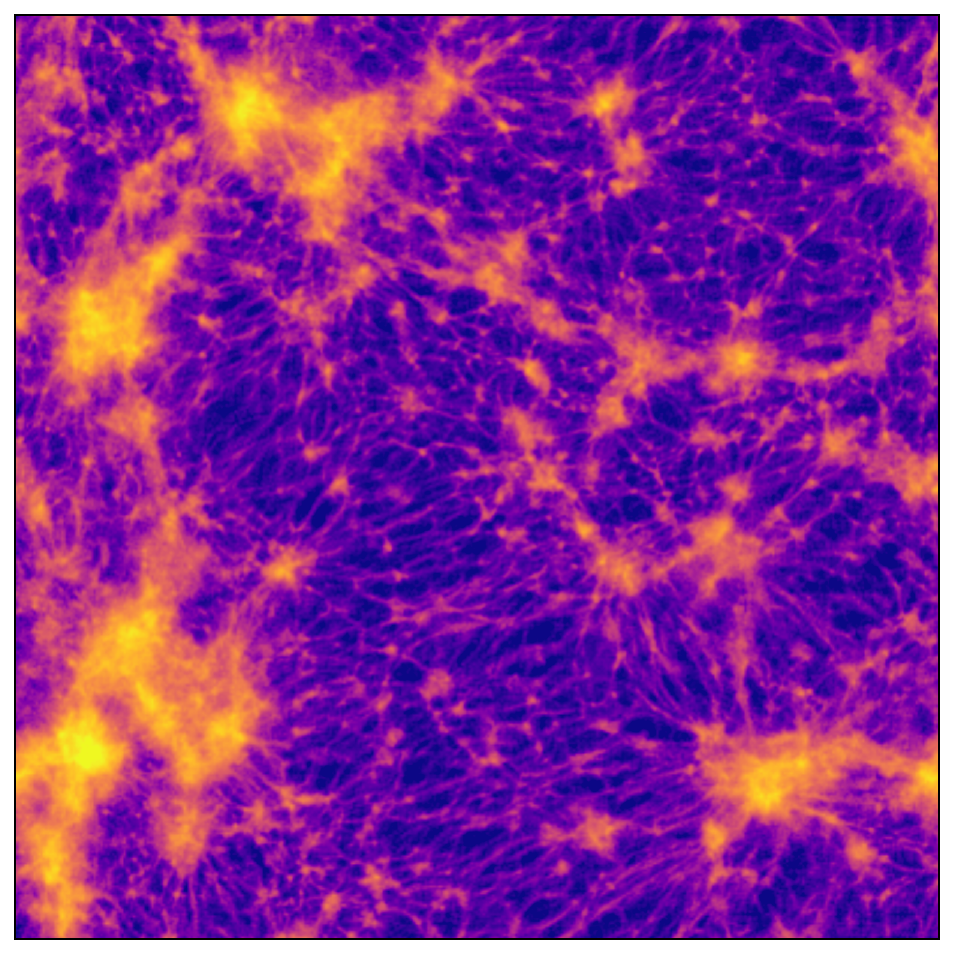}} \put(-178,167){\fcolorbox{white}{white}{2LPT}} 
\put(-178,154){\fcolorbox{white}{white}{z=0}} 
\hspace{-0.4cm} 
& 
\subfigure{\includegraphics[width=0.28\textwidth]{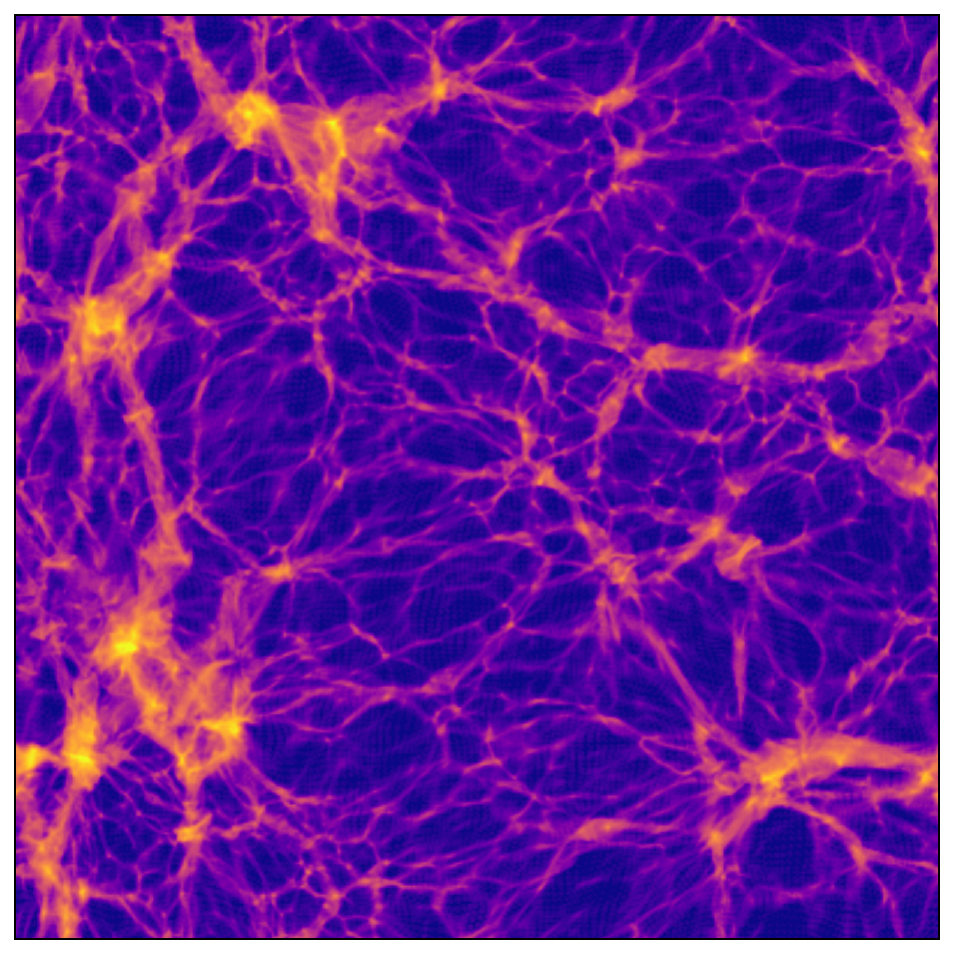}}  \put(-178,167){\fcolorbox{white}{white}{ALPT}} 
\put(-178,154){\fcolorbox{white}{white}{z=0}} 
\hspace{-0.4cm} &
\subfigure{\includegraphics[width=0.28\textwidth]{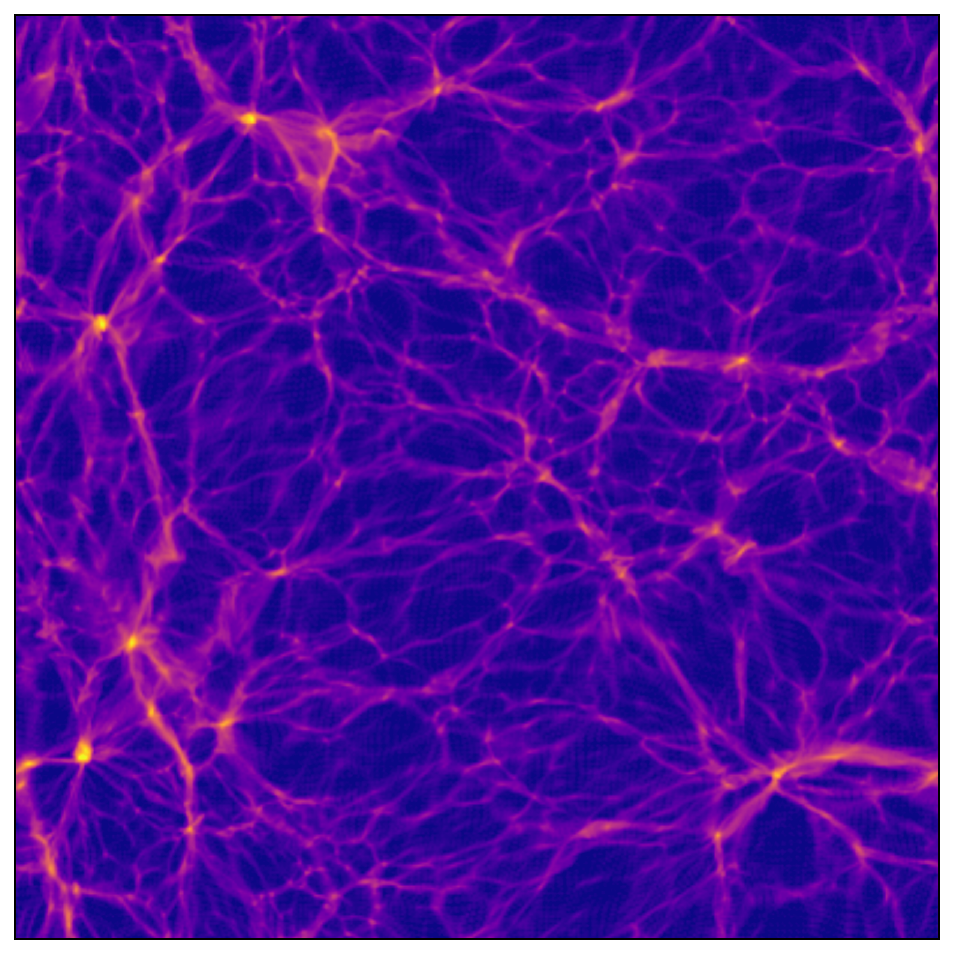}} \put(-178,167){\fcolorbox{white}{white}{rALPTwv}} 
\put(-178,154){\fcolorbox{white}{white}{z=0}} 
\hspace{-0.4cm} & \subfigure{\includegraphics[width=0.28\textwidth]{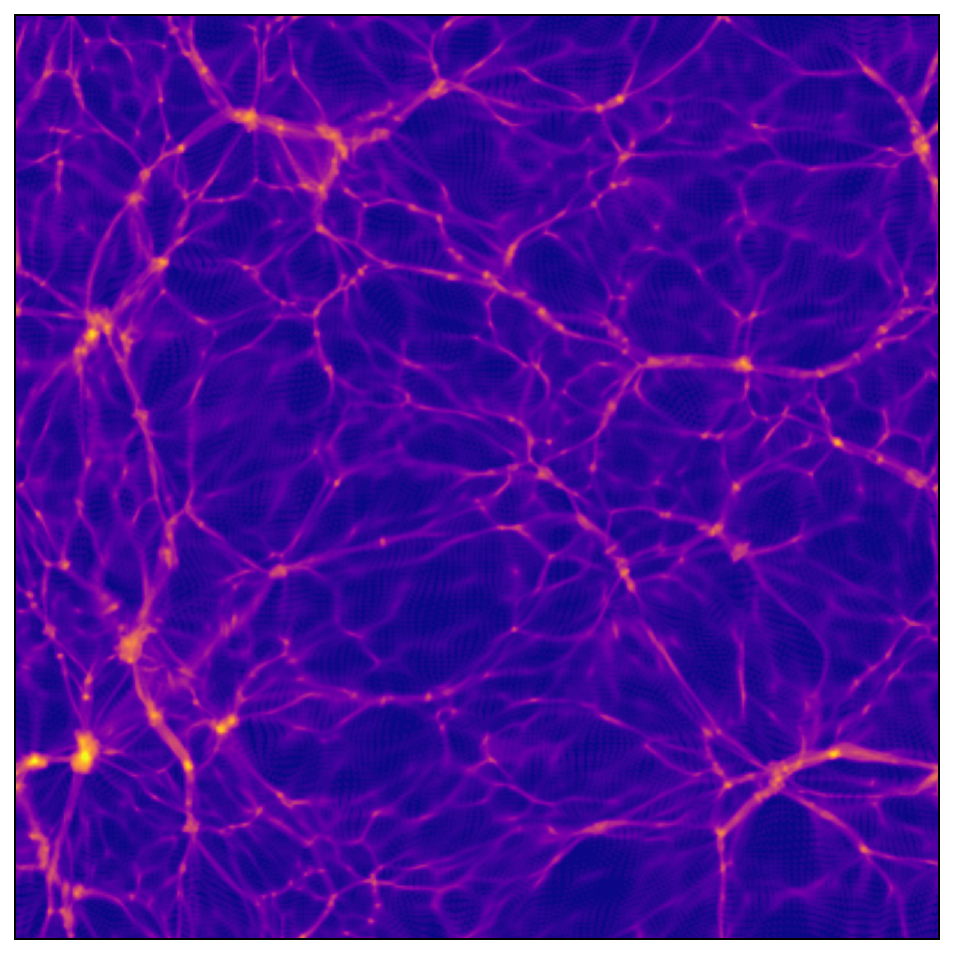}} 
\put(-178,167){\fcolorbox{white}{white}{fastPM}} 
\put(-178,154){\fcolorbox{white}{white}{z=0}}  
\vspace{-0.0cm}
\\
\subfigure{\includegraphics[width=0.28\textwidth]{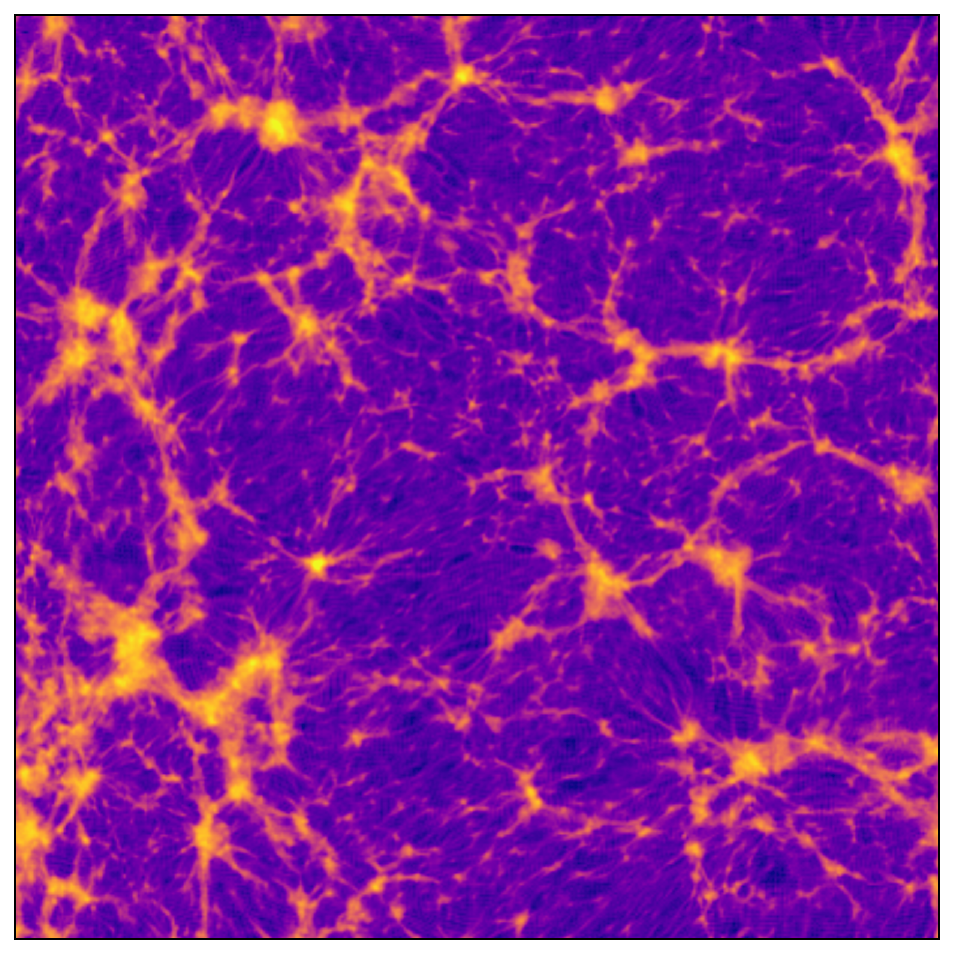}} \put(-178,167){\fcolorbox{white}{white}{2LPT}} \put(-178,154){\fcolorbox{white}{white}{z=1}} 
\hspace{-0.4cm} & \hspace{-0.1cm}
\subfigure{\includegraphics[width=0.28\textwidth]{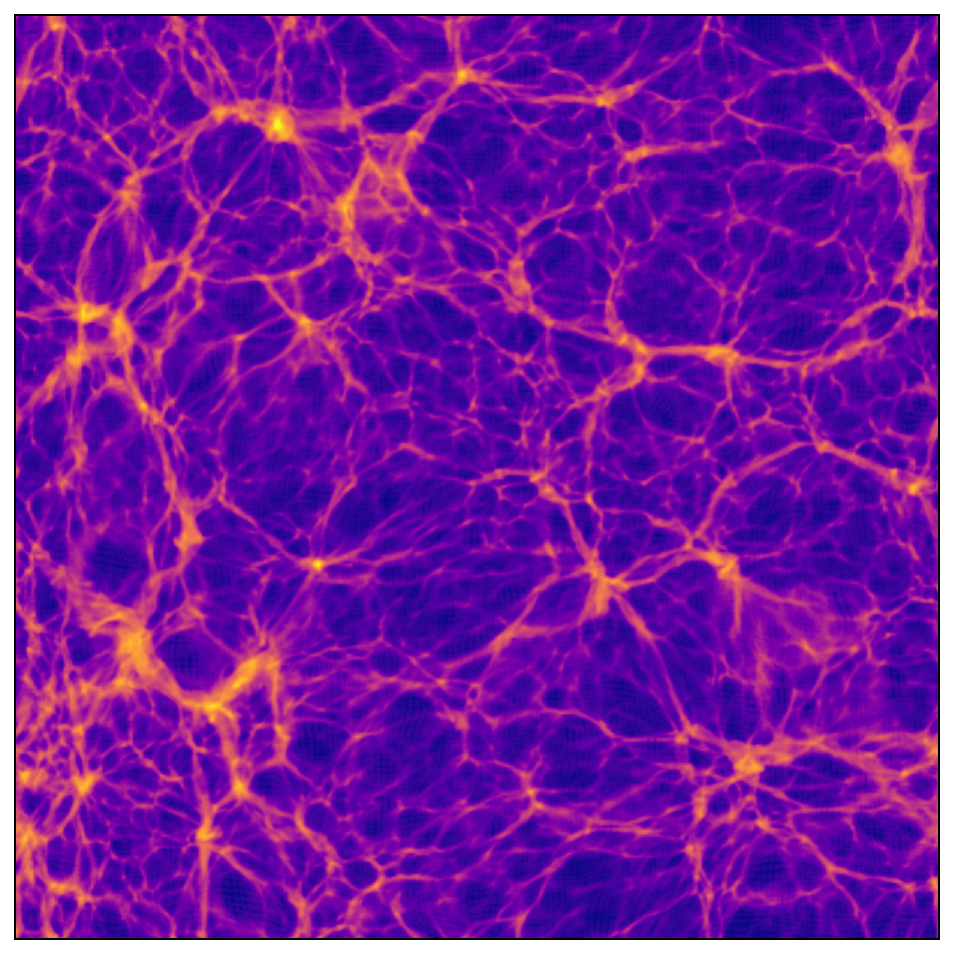}}  \put(-178,167){\fcolorbox{white}{white}{ALPT}} \put(-178,154){\fcolorbox{white}{white}{z=1}} \hspace{-0.4cm} & \subfigure{\includegraphics[width=0.28\textwidth]{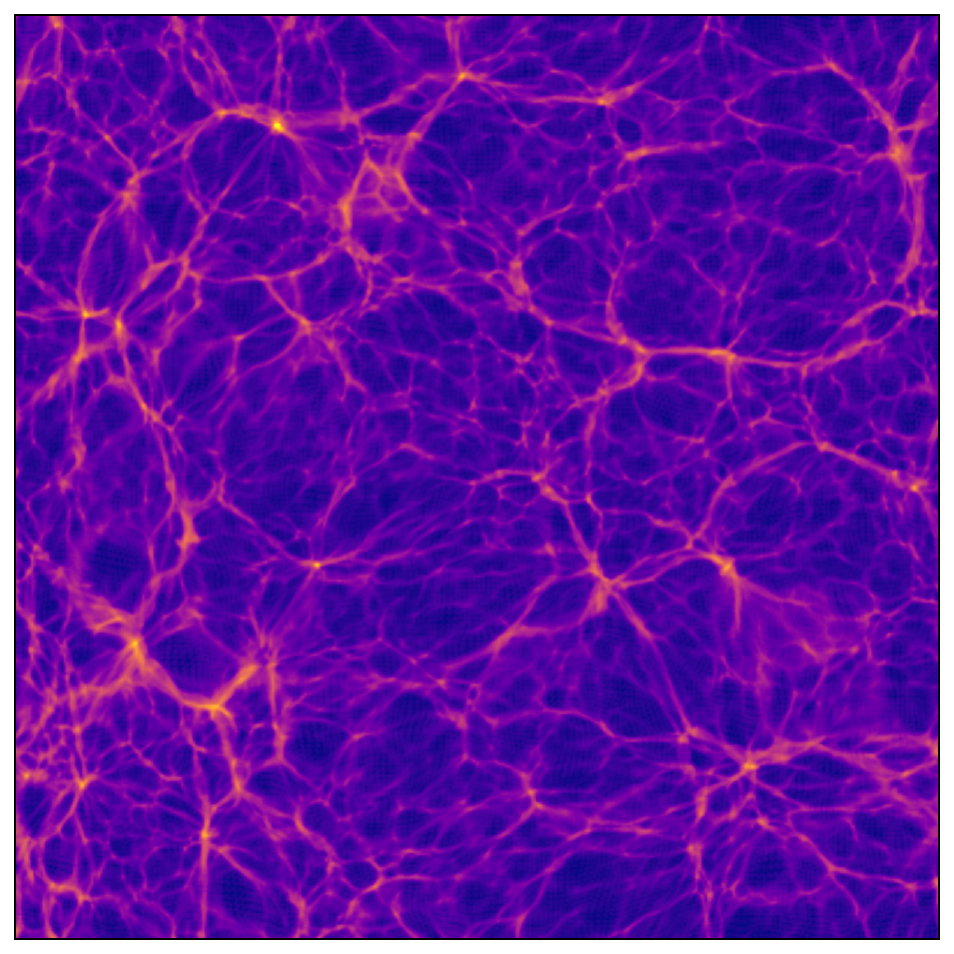}}  \put(-178,167){\fcolorbox{white}{white}{rALPT}}  \put(-178,154){\fcolorbox{white}{white}{z=1}} 
\hspace{-0.4cm}
&
\subfigure{\includegraphics[width=0.28\textwidth]{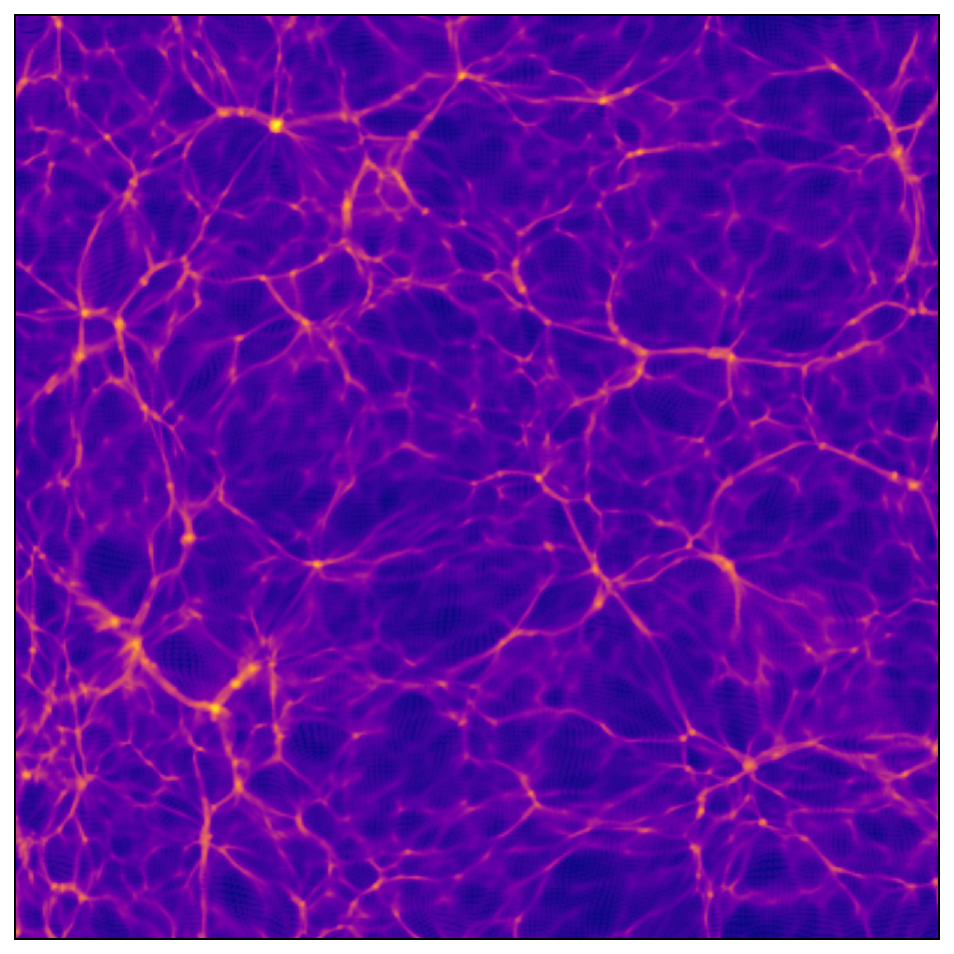}} 
\put(-178,167){\fcolorbox{white}{white}{fastPM}}  \put(-178,154){\fcolorbox{white}{white}{z=1}} \end{tabular}
\put(-734,174){\rotatebox[]{90}{\large$\longrightarrow$}}    
\put(-739,90){\rotatebox[]{90}{\large${\rm L=}100\,h^{-1}\,{\rm Mpc}$}}
\put(-734,16){\rotatebox[]{90}{\large$\longleftarrow$}}
\put(-706,190){\rotatebox[]{0}{\large$\longrightarrow$}}    
\put(-685,190){\rotatebox[]{0}{\large${\rm dL}=0.25\,h^{-1}\,{\rm Mpc}$}}
\put(-588,190){\rotatebox[]{0}{\large$\longleftarrow$}}
\caption{Matter overdensity field. {\bf Upper panels:} $z=0$; {\bf lower panels:} $z=1$. From left to right: {\bf 2LPT}; {\bf ALPT}; {\bf ALPT with ridging applied} (two-step gravity solver; the upper panel includes the vorticity model); {\bf FastPM} solution (50 time steps with force resolution equal to the mesh resolution). The simulation box has side length $100,h^{-1}\mathrm{Mpc}$ and contains $400^3$ particles, corresponding to a cell size of $0.25,h^{-1}\mathrm{Mpc}$.}
    \label{fig:dmV100}
\end{sidewaysfigure}

\begin{figure}[]
\vspace{-1cm}
    \centering
    \begin{tabular}{c}
    \includegraphics[scale=.5]{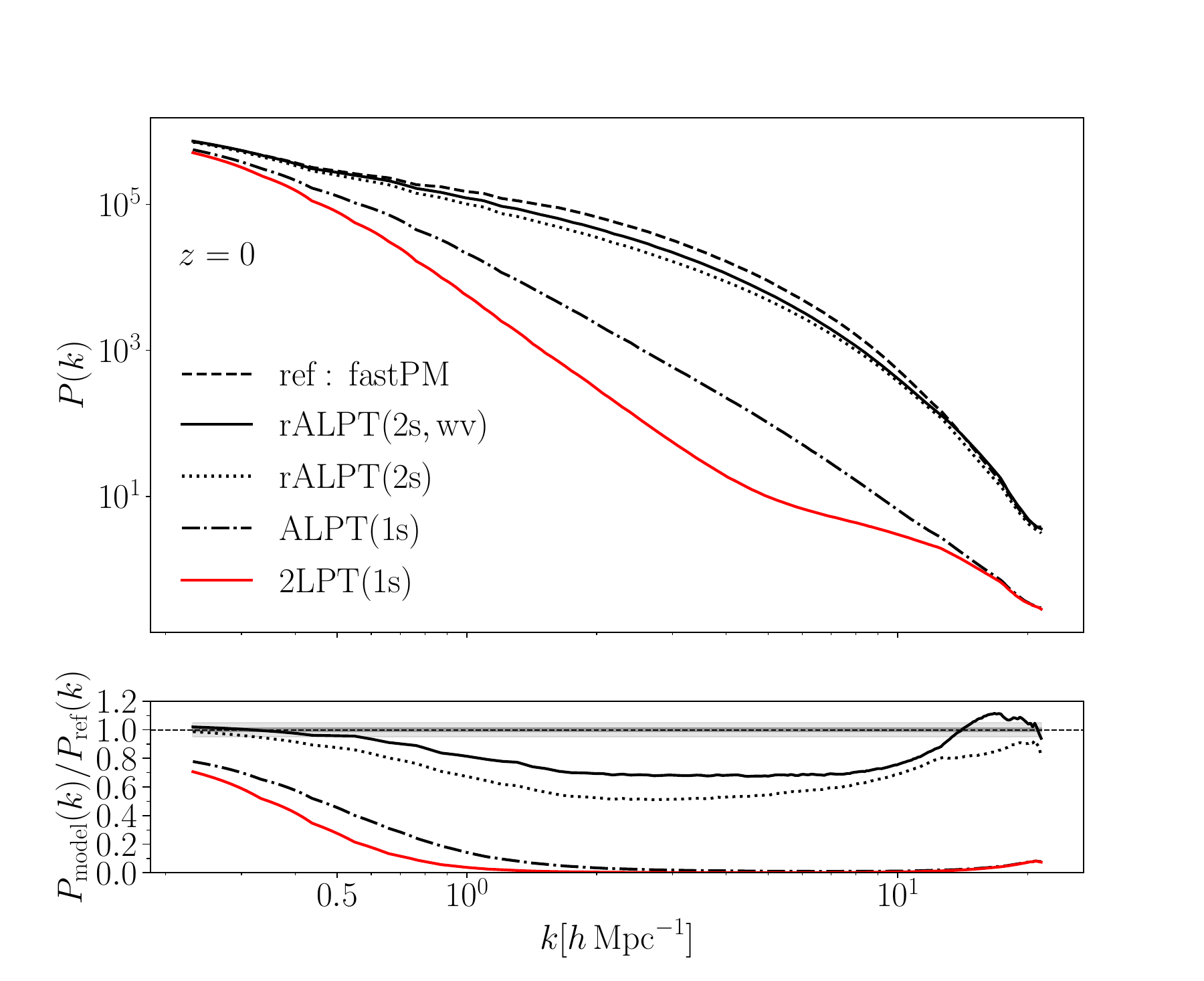}
     \put(-180,280){$L=100\,h^{-1}\mathrm{Mpc}$, $N=400^3$} 
     \vspace{-1.cm}
    \\
\includegraphics[scale=.5]{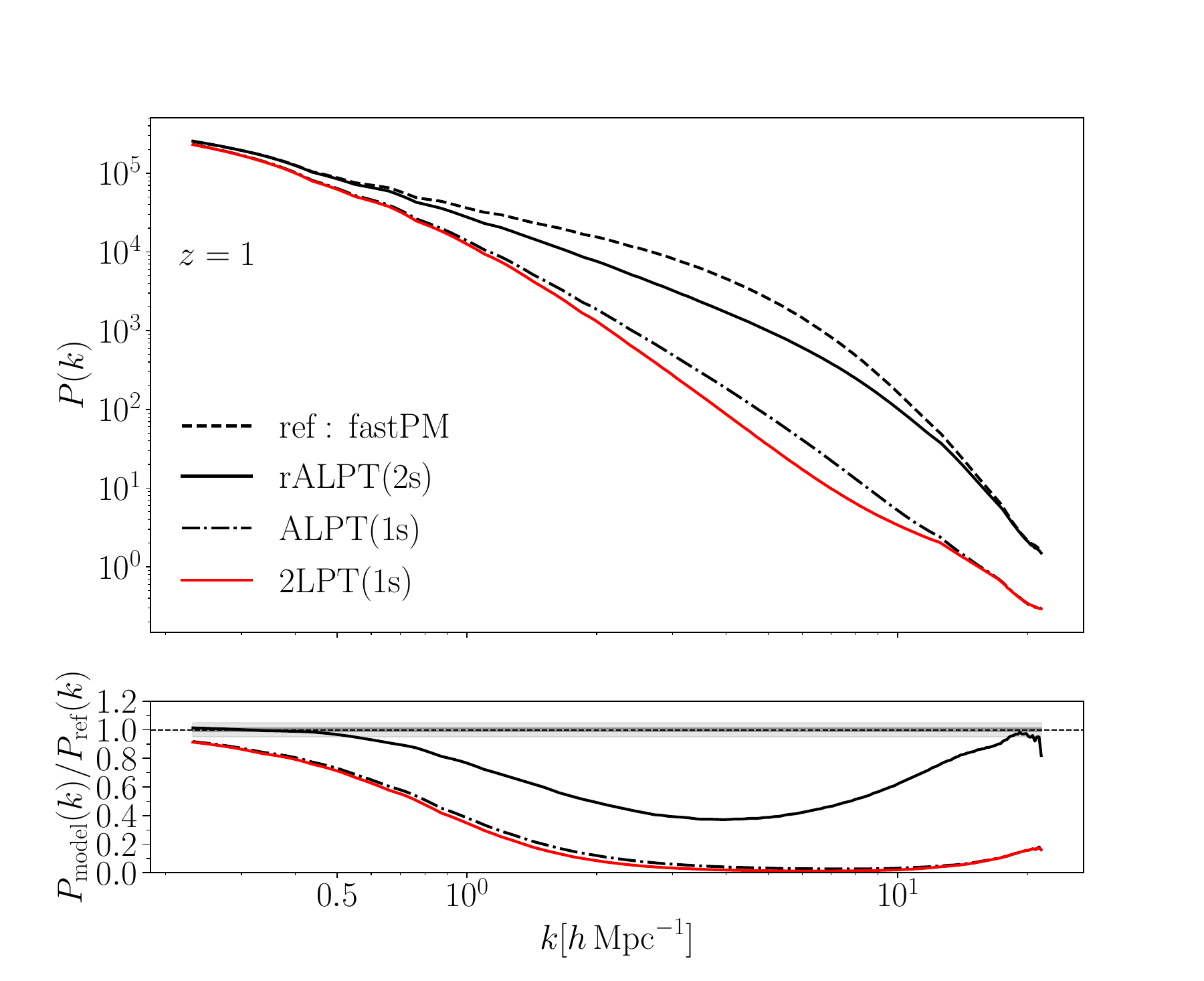}
     \put(-180,280){$L=100\,h^{-1}\mathrm{Mpc}$, $N=400^3$} 
    \end{tabular}
    \vspace{-.6cm}
\caption{Power spectra of the different LPT-based approximations compared with the \textsc{FastPM} particle-mesh solution with a volume of $L=100\,h^{-1}\,{\rm Mpc}$ side and $400^3$ particles. {\bf Upper panel:} $z=0$; {\bf lower panel:} $z=1$.}
    \label{fig:pkV100}
\end{figure}

\begin{figure}[]
\vspace{-1cm}
    \centering
    \begin{tabular}{c}
    \includegraphics[scale=.6]{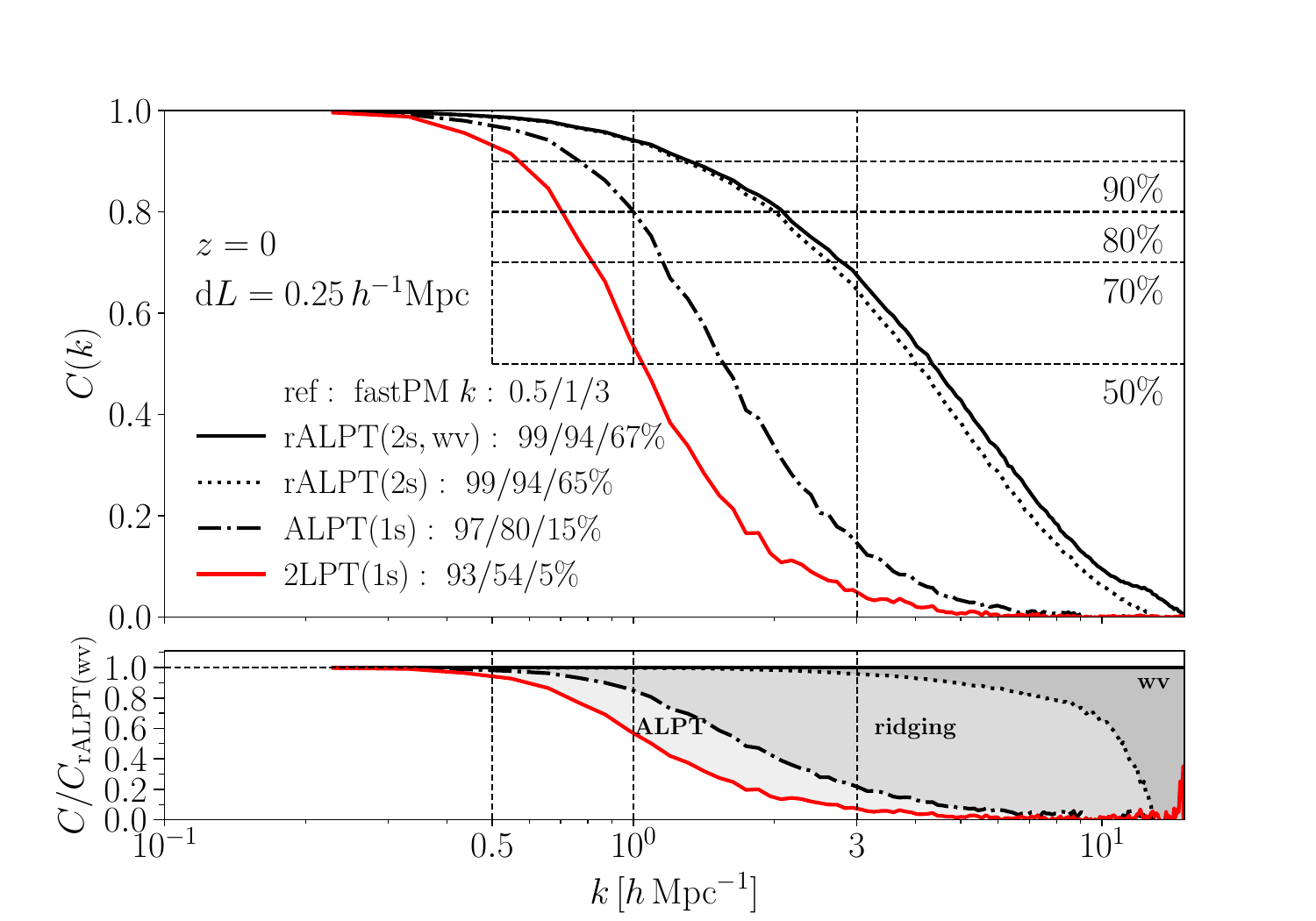}\vspace{-0.7cm}\\
\includegraphics[scale=.6]{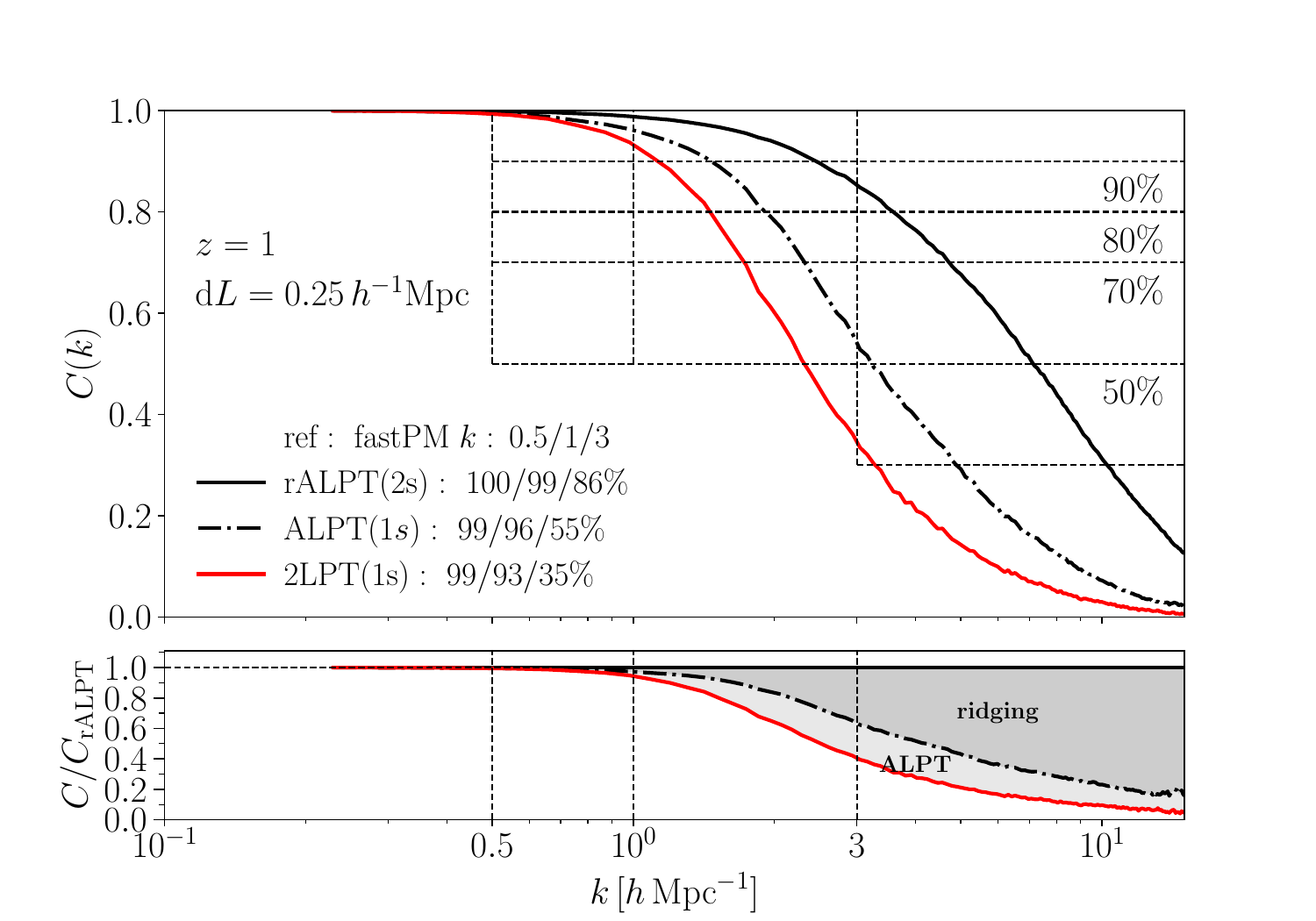}
    \end{tabular}
\caption{Cross power spectra (defined as $C(k)=\avg{\hat{\delta}_{\rm A}(\bk)\,\hat{\delta}^{\,*}_{\rm B}(\bk)}/\sqrt{P_{\rm A}(k)P_{\rm B}(k)}.$ for two fields $\mathrm{A}$ and $\mathrm{B}$) of the different LPT-based approximations compared with the \textsc{FastPM} particle-mesh solution with a volume of $L=100\,h^{-1}\,{\rm Mpc}$ side and $400^3$ particles. {\bf Upper panel:} $z=0$; {\bf lower panel:} $z=1$.}
    \label{fig:ckV100}
\end{figure}

\begin{sidewaysfigure}
\hspace{-1.5cm}
\vspace{-0.2cm}
    \begin{tabular}{cccc}
\subfigure{\includegraphics[width=0.28\textwidth]{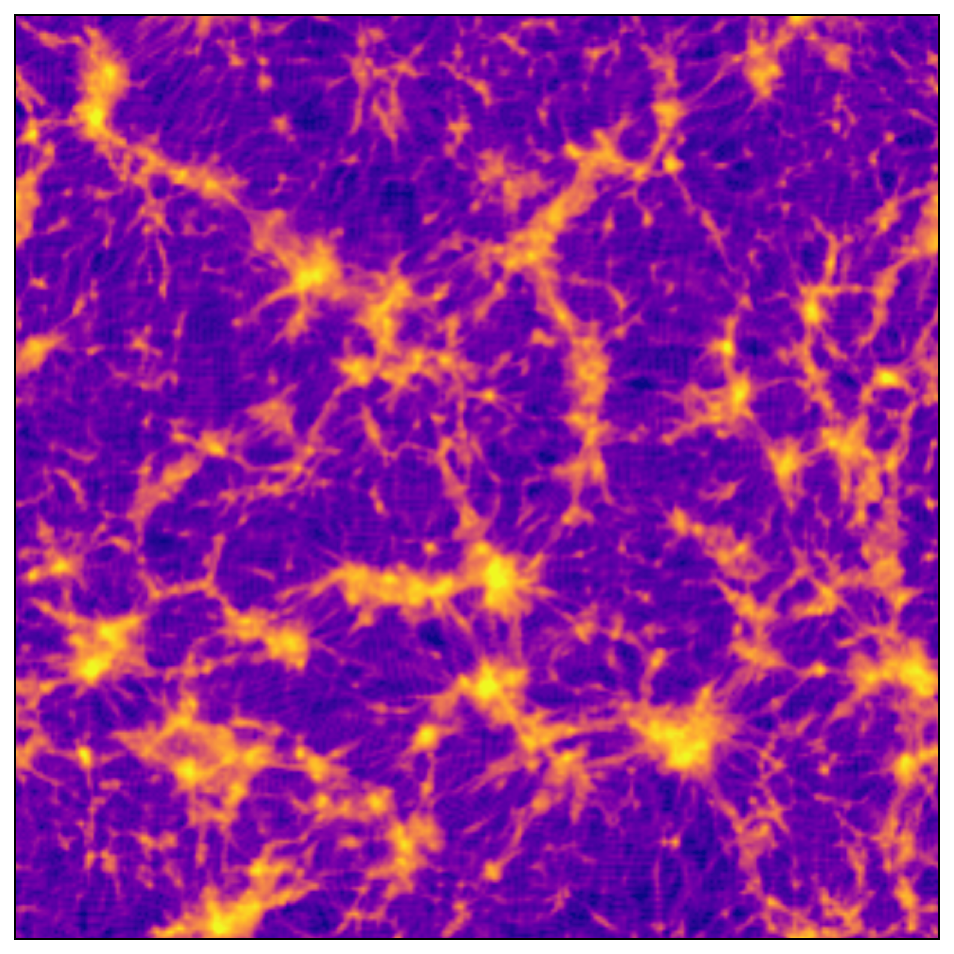}} \put(-178,167){\fcolorbox{white}{white}{2LPT}} 
\put(-178,154){\fcolorbox{white}{white}{z=0}} 
\hspace{-0.4cm} 
& 
\subfigure{\includegraphics[width=0.28\textwidth]{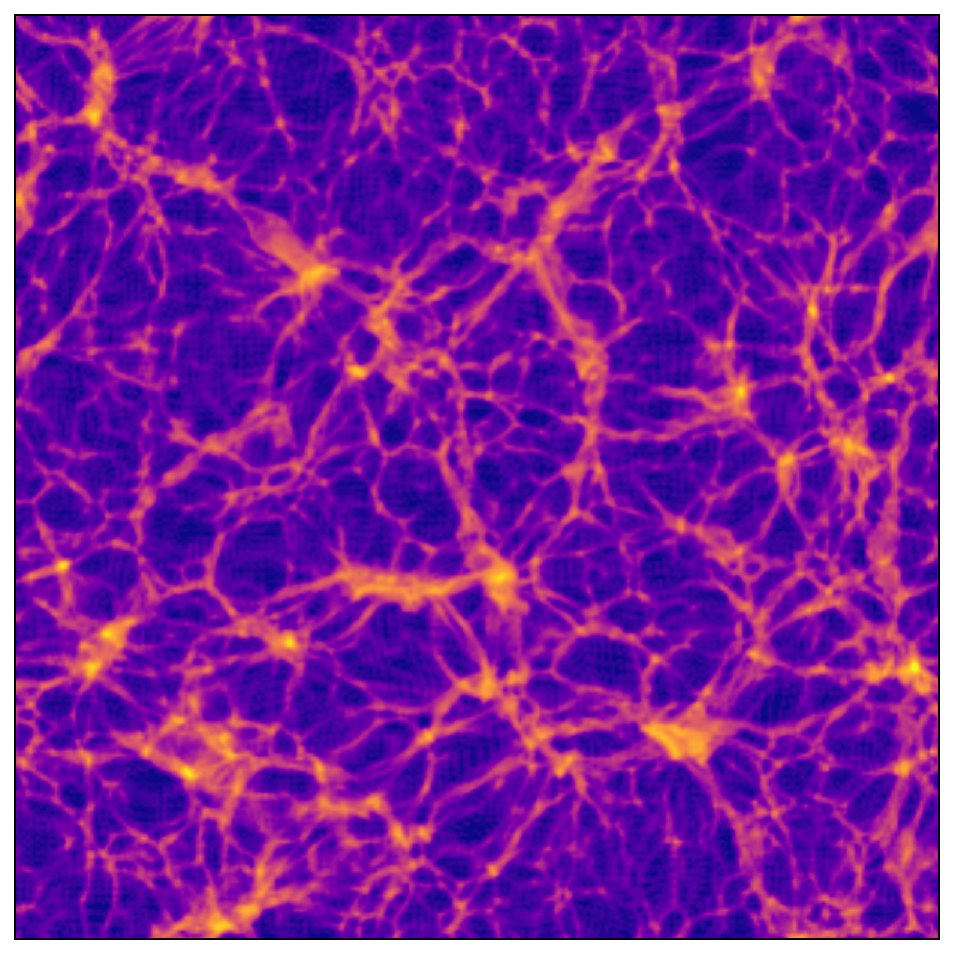}}  \put(-178,167){\fcolorbox{white}{white}{ALPT}} 
\put(-178,154){\fcolorbox{white}{white}{z=0}} 
\hspace{-0.4cm} &
\subfigure{\includegraphics[width=0.28\textwidth]{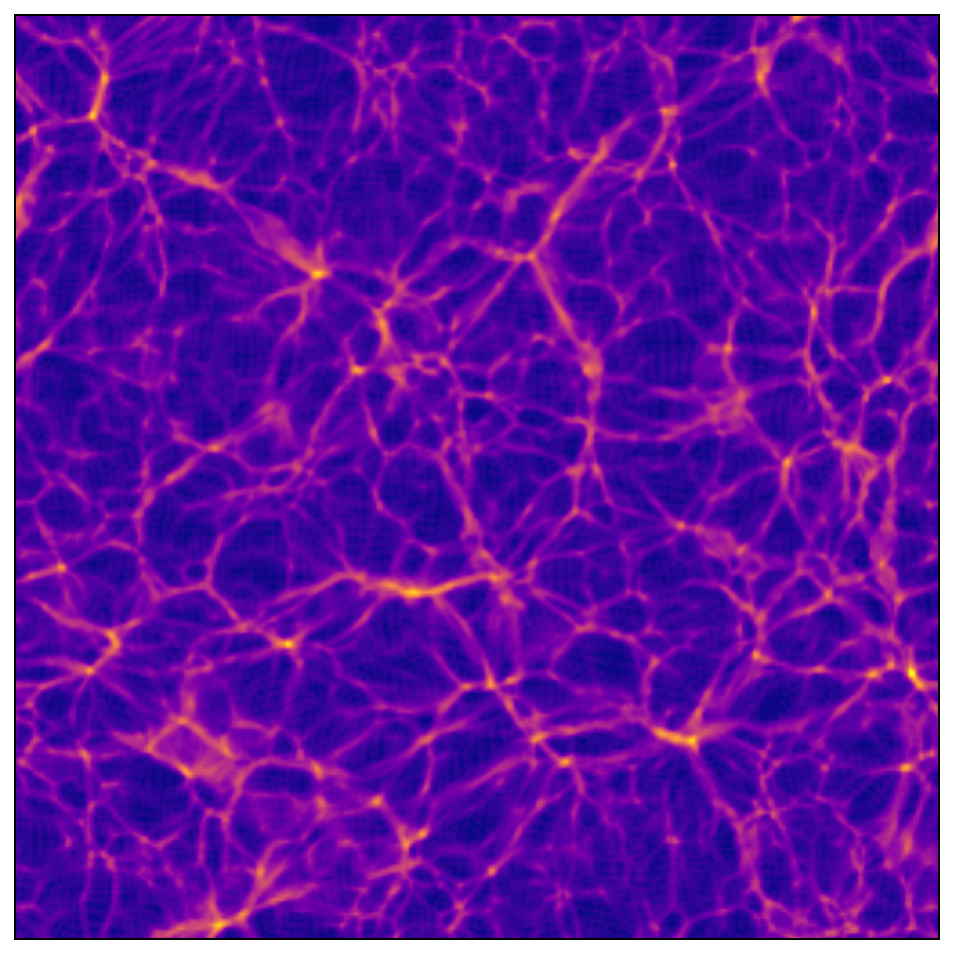}} \put(-178,167){\fcolorbox{white}{white}{rALPT}} 
\put(-178,154){\fcolorbox{white}{white}{z=0}} 
\hspace{-0.4cm} & \subfigure{\includegraphics[width=0.28\textwidth]{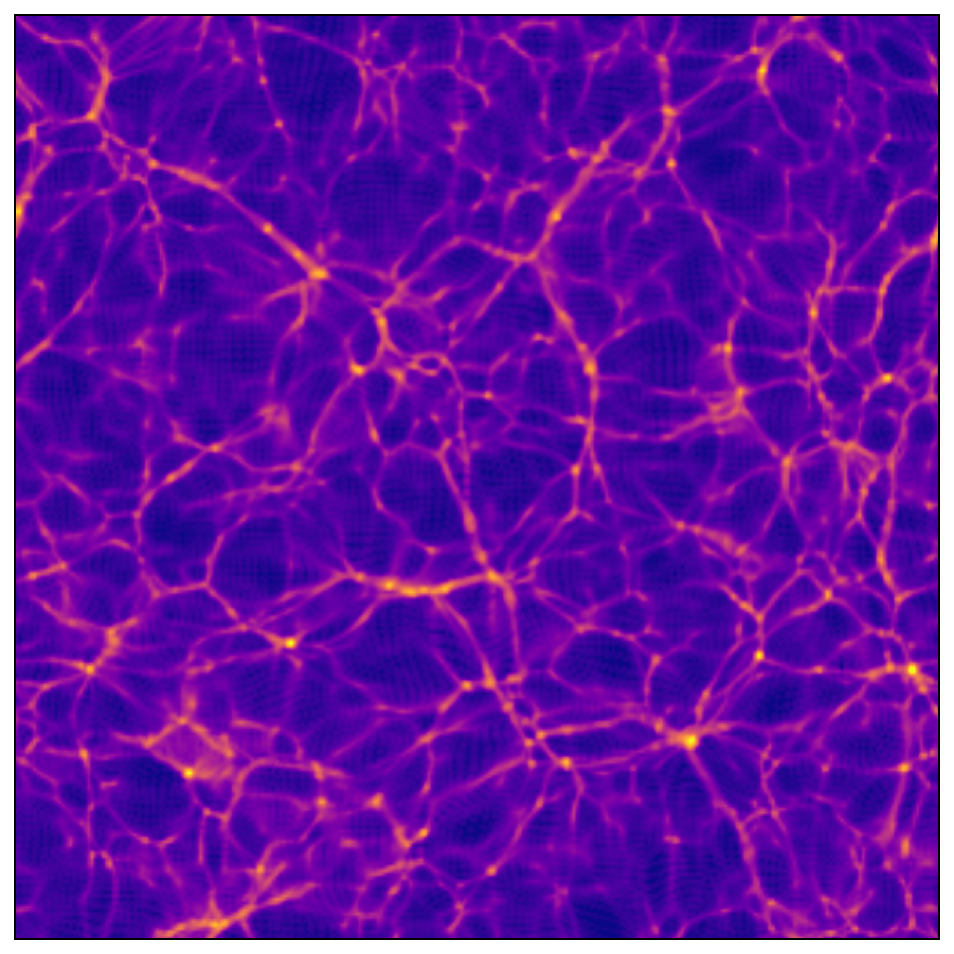}} 
\put(-178,167){\fcolorbox{white}{white}{fastPM}} 
\put(-178,154){\fcolorbox{white}{white}{z=0}}  
\vspace{-0.0cm}
\\
\subfigure{\includegraphics[width=0.28\textwidth]{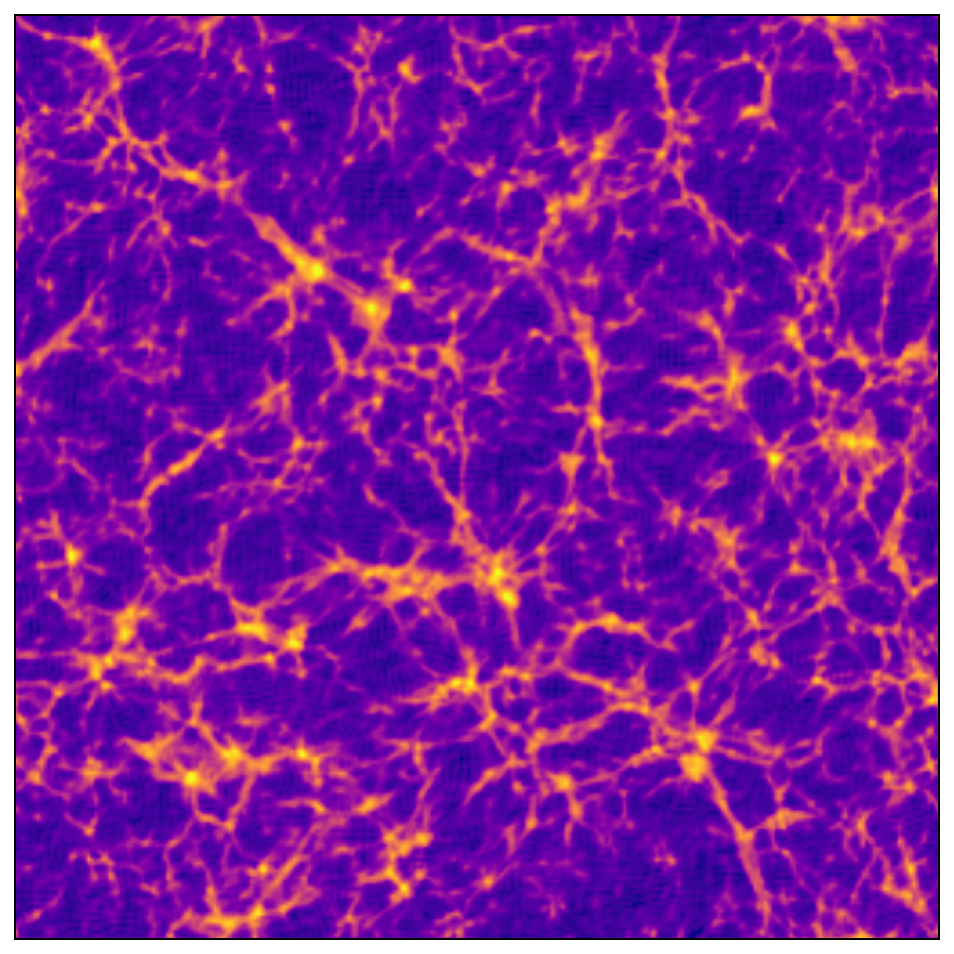}} \put(-178,167){\fcolorbox{white}{white}{2LPT}} \put(-178,154){\fcolorbox{white}{white}{z=1}} 
\hspace{-0.4cm} & \hspace{-0.1cm}
\subfigure{\includegraphics[width=0.28\textwidth]{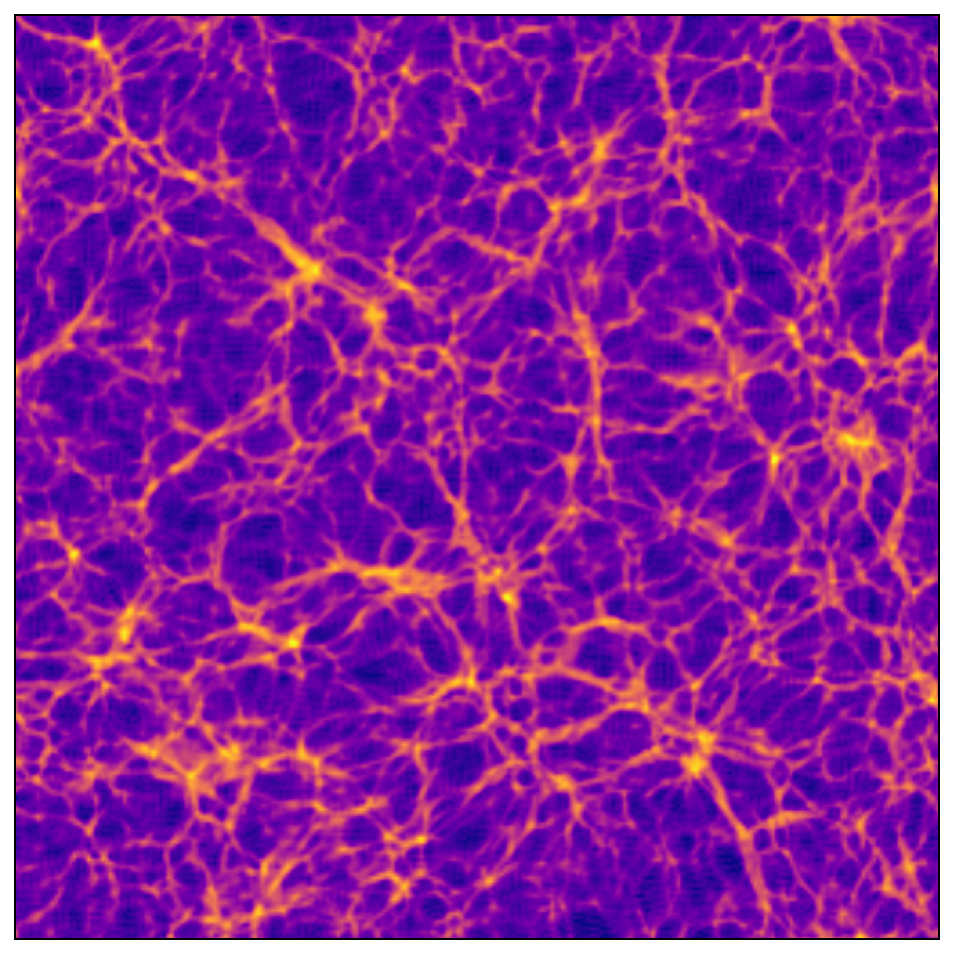}}  \put(-178,167){\fcolorbox{white}{white}{ALPT}} \put(-178,154){\fcolorbox{white}{white}{z=1}} \hspace{-0.4cm} & \subfigure{\includegraphics[width=0.28\textwidth]{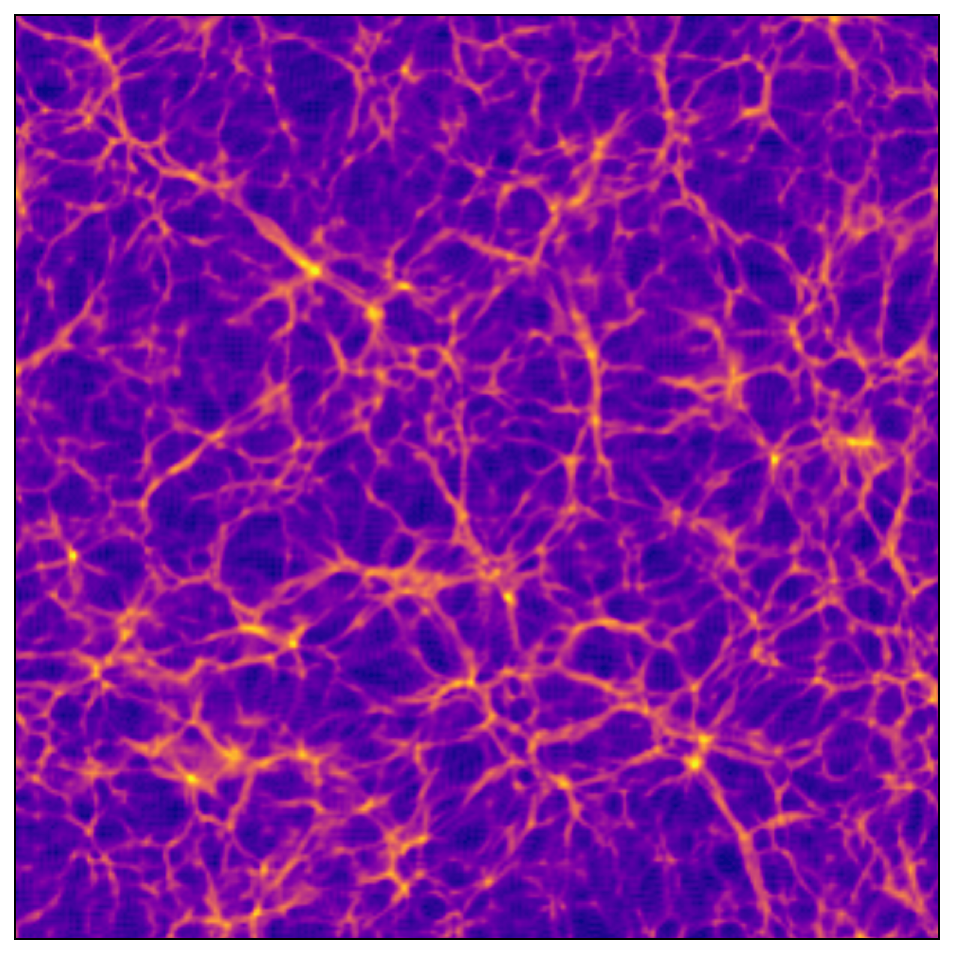}}  \put(-178,167){\fcolorbox{white}{white}{rALPT}}  \put(-178,154){\fcolorbox{white}{white}{z=1}} 
\hspace{-0.4cm}
&
\subfigure{\includegraphics[width=0.28\textwidth]{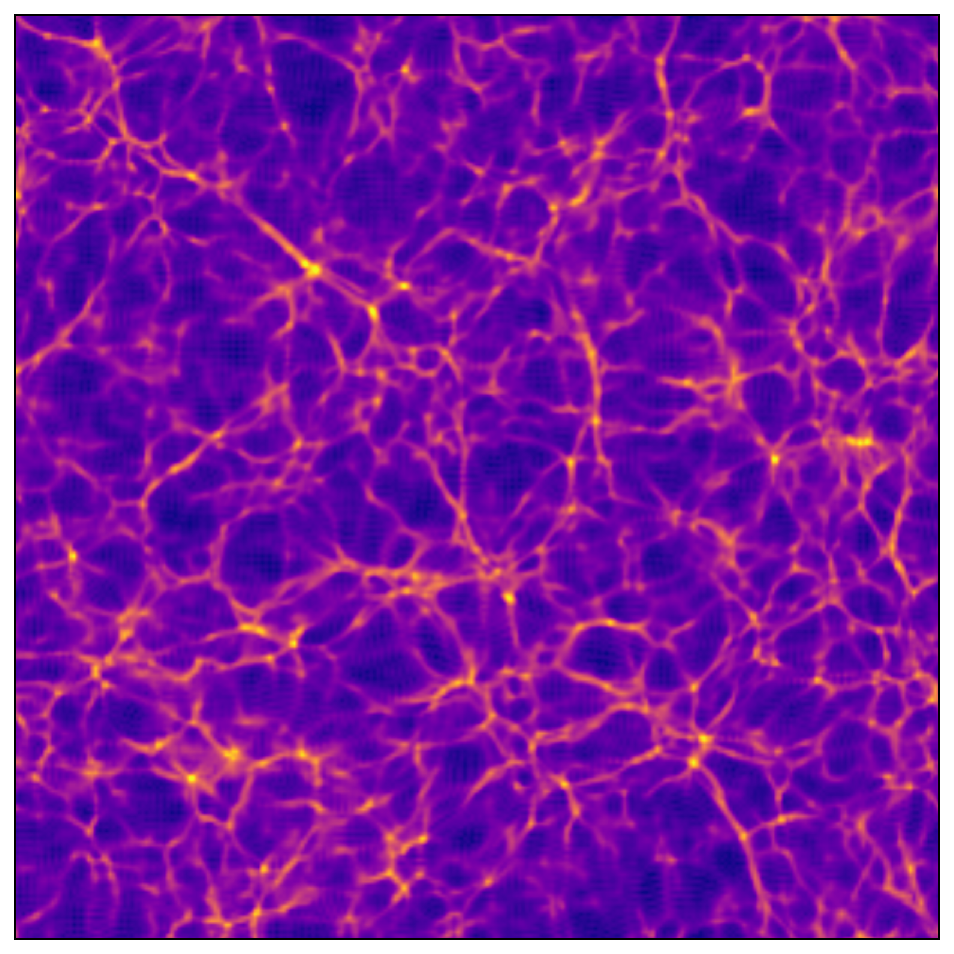}} 
\put(-178,167){\fcolorbox{white}{white}{fastPM}}  \put(-178,154){\fcolorbox{white}{white}{z=1}} \end{tabular}
\put(-734,174){\rotatebox[]{90}{\large$\longrightarrow$}}    
\put(-739,90){\rotatebox[]{90}{\large${\rm L=}200\,h^{-1}\,{\rm Mpc}$}}
\put(-734,16){\rotatebox[]{90}{\large$\longleftarrow$}}
\put(-706,190){\rotatebox[]{0}{\large$\longrightarrow$}}    
\put(-685,190){\rotatebox[]{0}{\large${\rm dL}=0.78\,h^{-1}\,{\rm Mpc}$}}
\put(-588,190){\rotatebox[]{0}{\large$\longleftarrow$}}
\caption{Matter overdensity field. {\bf Upper panels:} $z=0$; {\bf lower panels:} $z=1$. From left to right: {\bf 2LPT}; {\bf ALPT}; {\bf ALPT with ridging applied} (two-step gravity solver; the upper panel includes the vorticity model); {\bf FastPM} solution (50 time steps with double force resolution to the mesh resolution). The simulation box has side length $200,h^{-1}\mathrm{Mpc}$ and contains $256^3$ particles, corresponding to a cell size of $0.78,h^{-1}\mathrm{Mpc}$.}
    \label{fig:dmV200}
\end{sidewaysfigure}

\begin{figure}[]
\vspace{-1cm}
    \centering
    \begin{tabular}{c}
    \includegraphics[scale=.5]{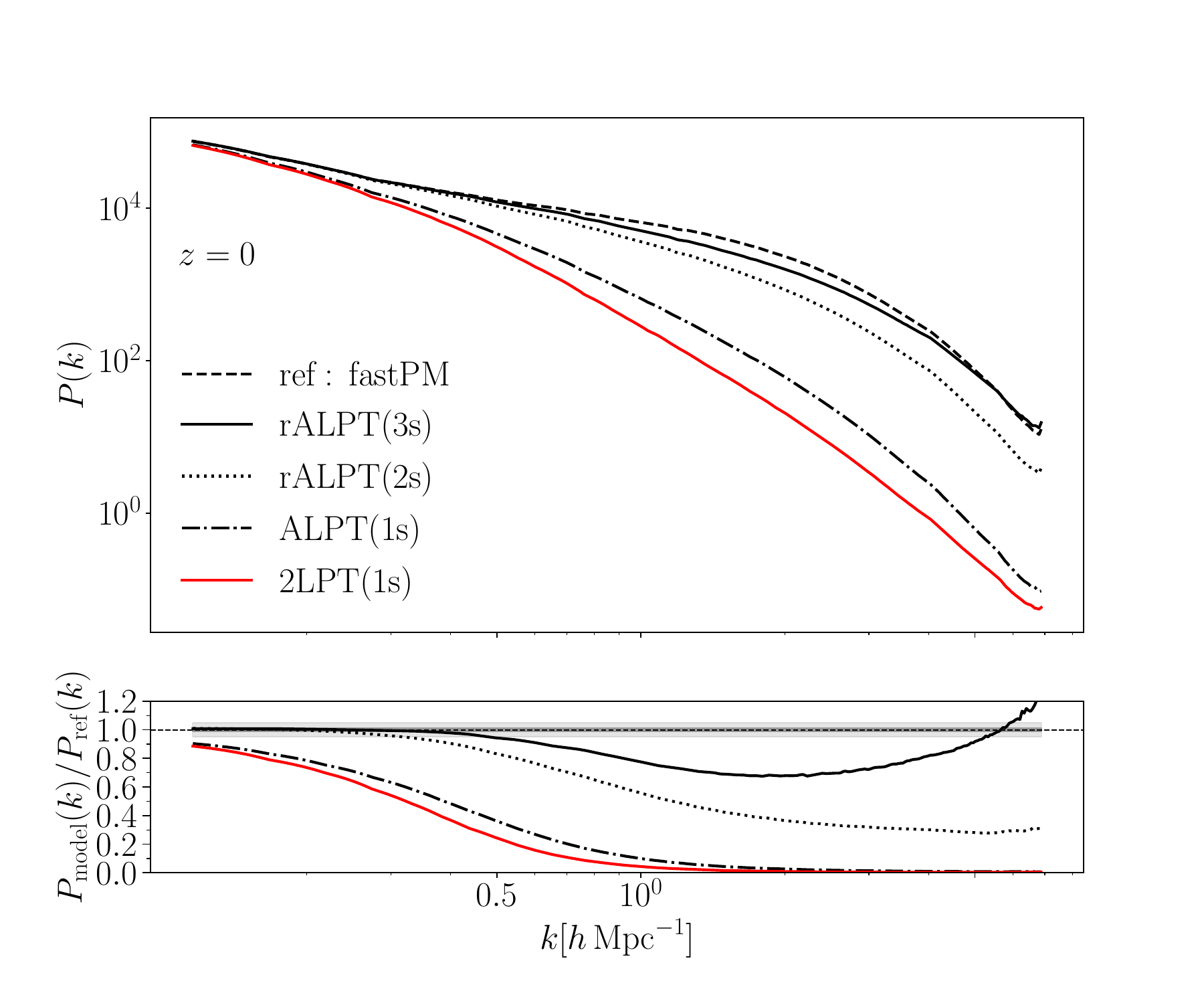}
     \put(-180,280){$L=200\,h^{-1}\mathrm{Mpc}$, $N=256^3$} 
     \vspace{-1.cm}
    \\
\includegraphics[scale=.5]{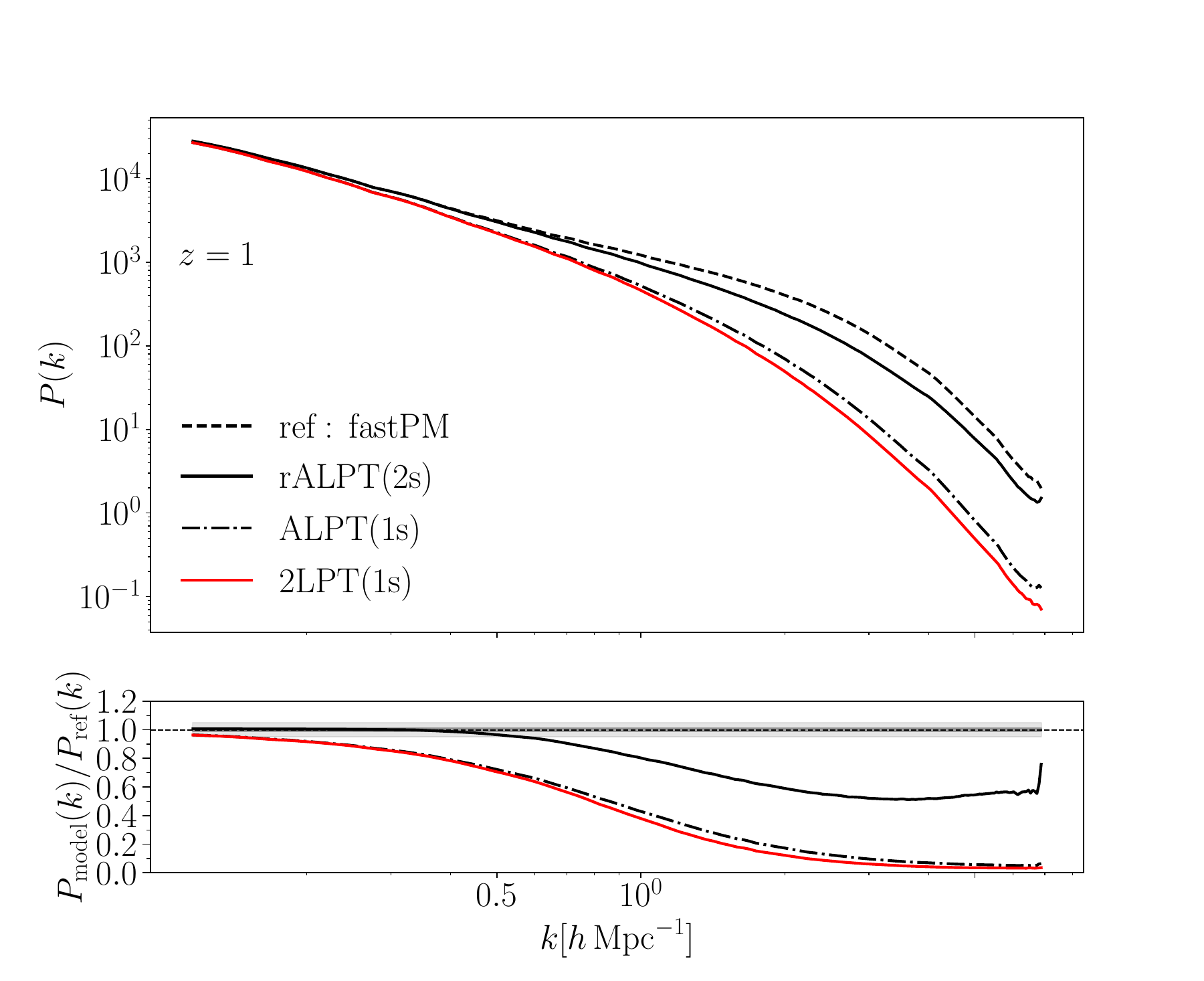}
     \put(-180,280){$L=200\,h^{-1}\mathrm{Mpc}$, $N=256^3$} 
    \end{tabular}
    \vspace{-.6cm}
\caption{Power spectra of the different LPT-based approximations compared with the \textsc{FastPM}particle-mesh solution with a volume of $L=200\,h^{-1}\,{\rm Mpc}$ side and $256^3$ particles. {\bf Upper panel:} $z=0$; {\bf lower panel:} $z=1$.}
    \label{fig:pkV200}
\end{figure}

\begin{figure}[]
\vspace{-1cm}
    \centering
    \begin{tabular}{c}
    \includegraphics[scale=.6]{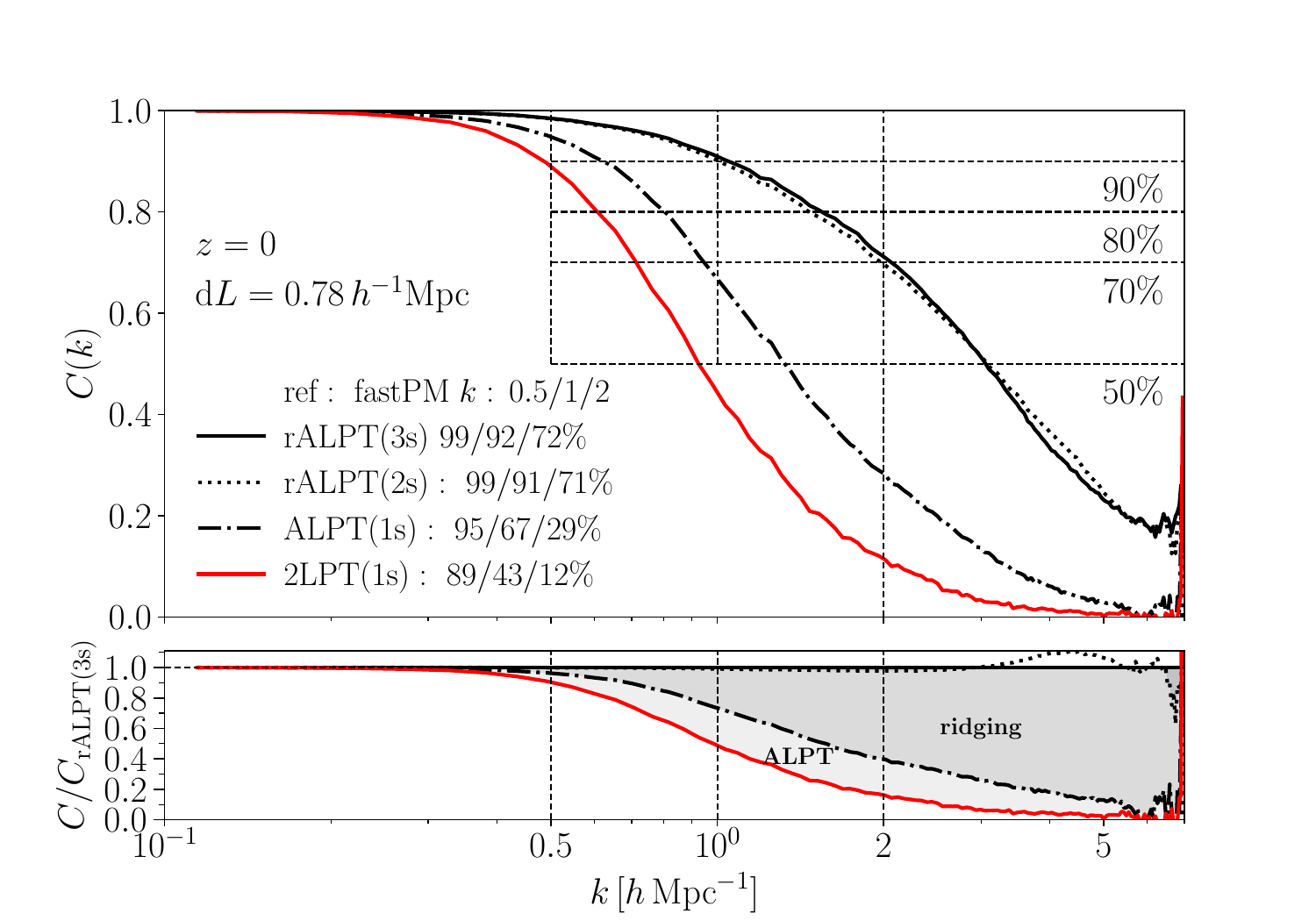}\vspace{-0.7cm}\\
\includegraphics[scale=.6]{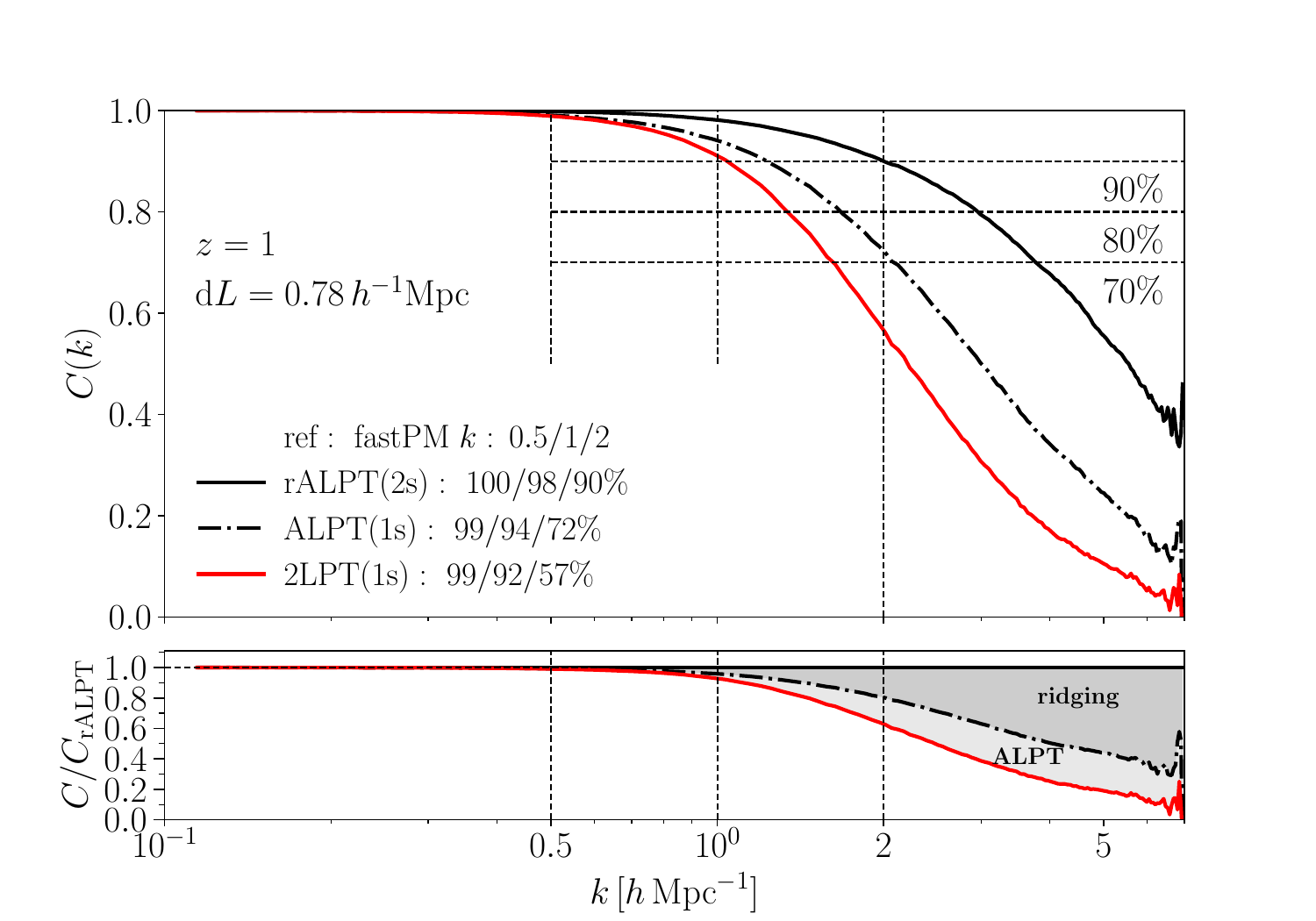}
    \end{tabular}
\caption{Cross power spectra of the different LPT-based approximations compared with the \textsc{FastPM}particle-mesh solution with a volume of $L=200\,h^{-1}\,{\rm Mpc}$ side and $256^3$ particles. {\bf Upper panel:} $z=0$; {\bf lower panel:} $z=1$.}
    \label{fig:ckV200}
\end{figure}

\begin{sidewaysfigure}
\hspace{.0cm}
\vspace{-0.2cm}
    \begin{tabular}{ccc}
\subfigure{\includegraphics[width=0.32\textwidth]{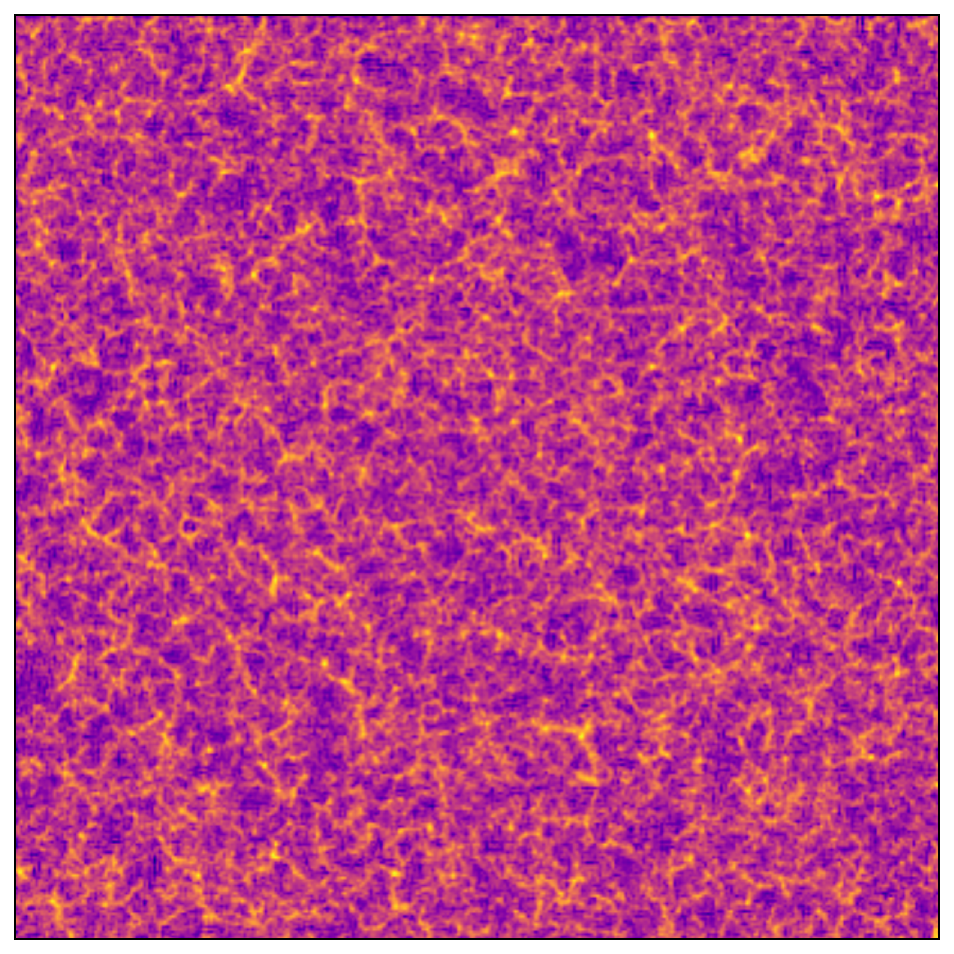}} \put(-203,193){\fcolorbox{white}{white}{2LPT}} 
\put(-203,180){\fcolorbox{white}{white}{z=0.2}} 
\hspace{-0.4cm} 
& 
\subfigure{\includegraphics[width=0.32\textwidth]{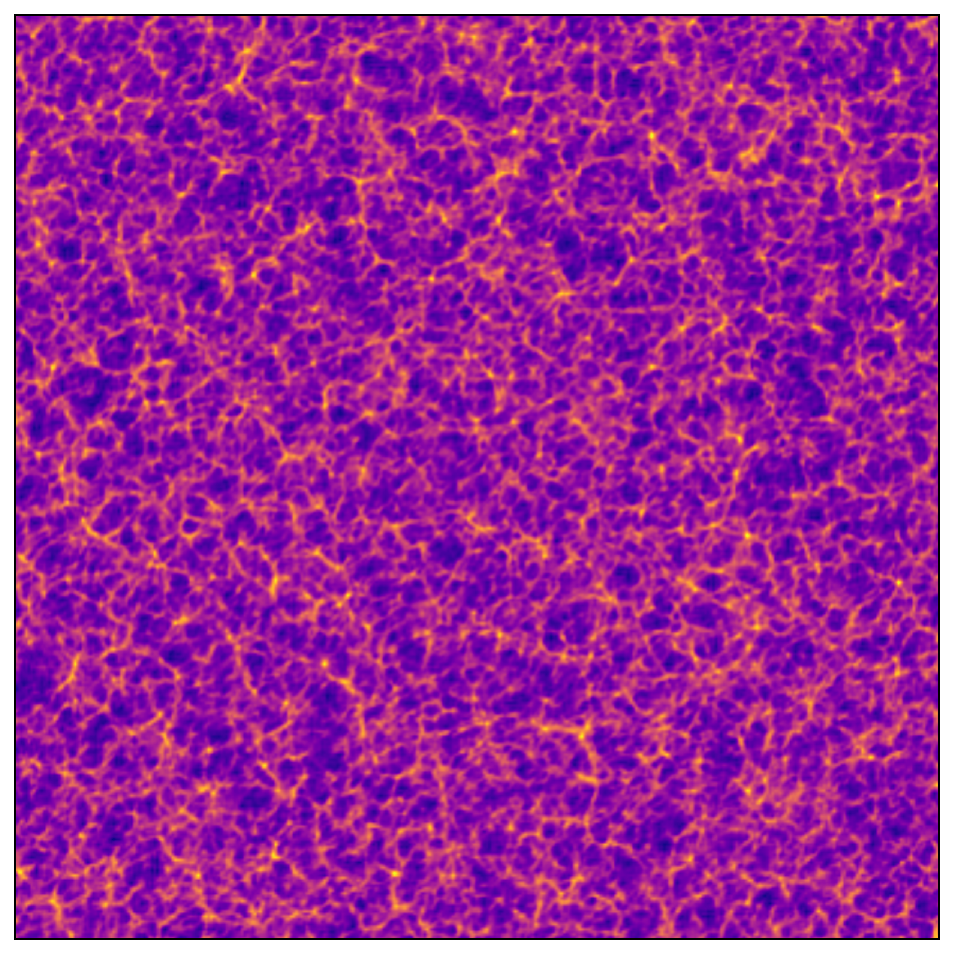}}  
\put(-203,193){\fcolorbox{white}{white}{rALPT}} 
\put(-203,180){\fcolorbox{white}{white}{z=0.2}} 
\hspace{-0.4cm}  & \subfigure{\includegraphics[width=0.32\textwidth]{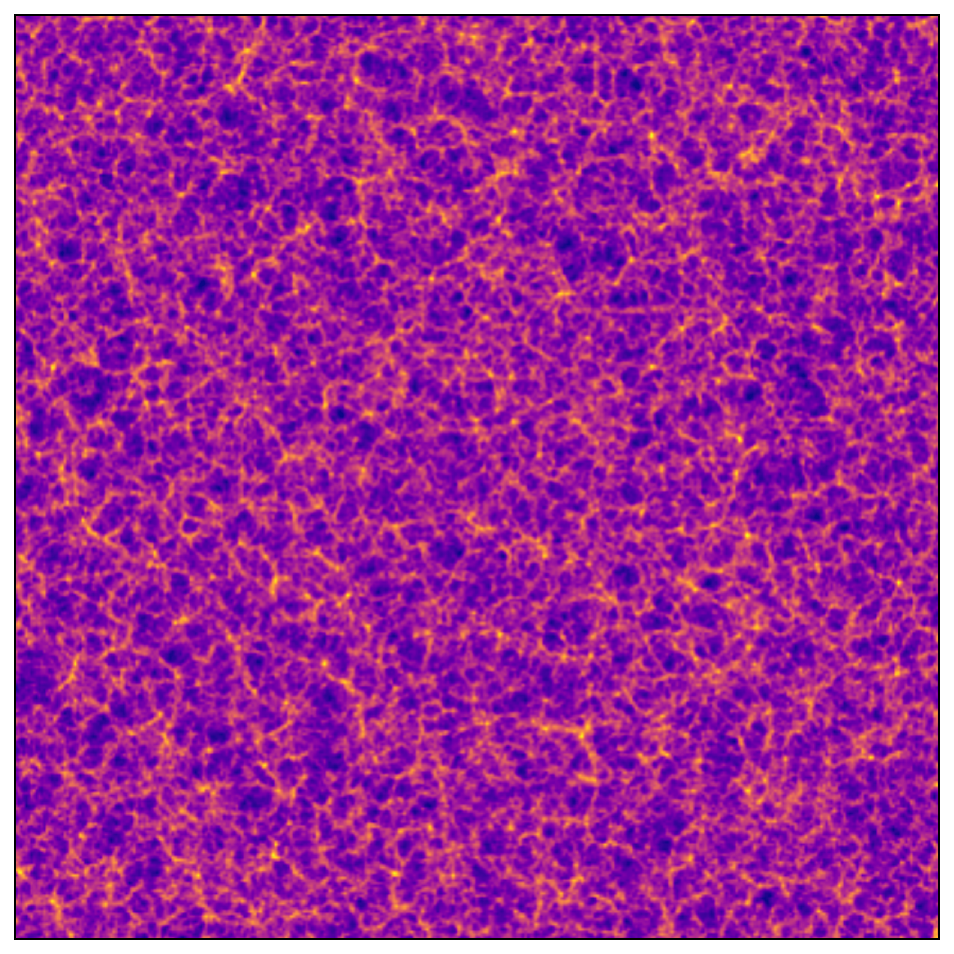}} 
\put(-203,193){\fcolorbox{white}{white}{ABACUS}} 
\put(-203,180){\fcolorbox{white}{white}{z=0.2}}  
\vspace{-0.0cm}
\\
\subfigure{\includegraphics[width=0.32\textwidth]{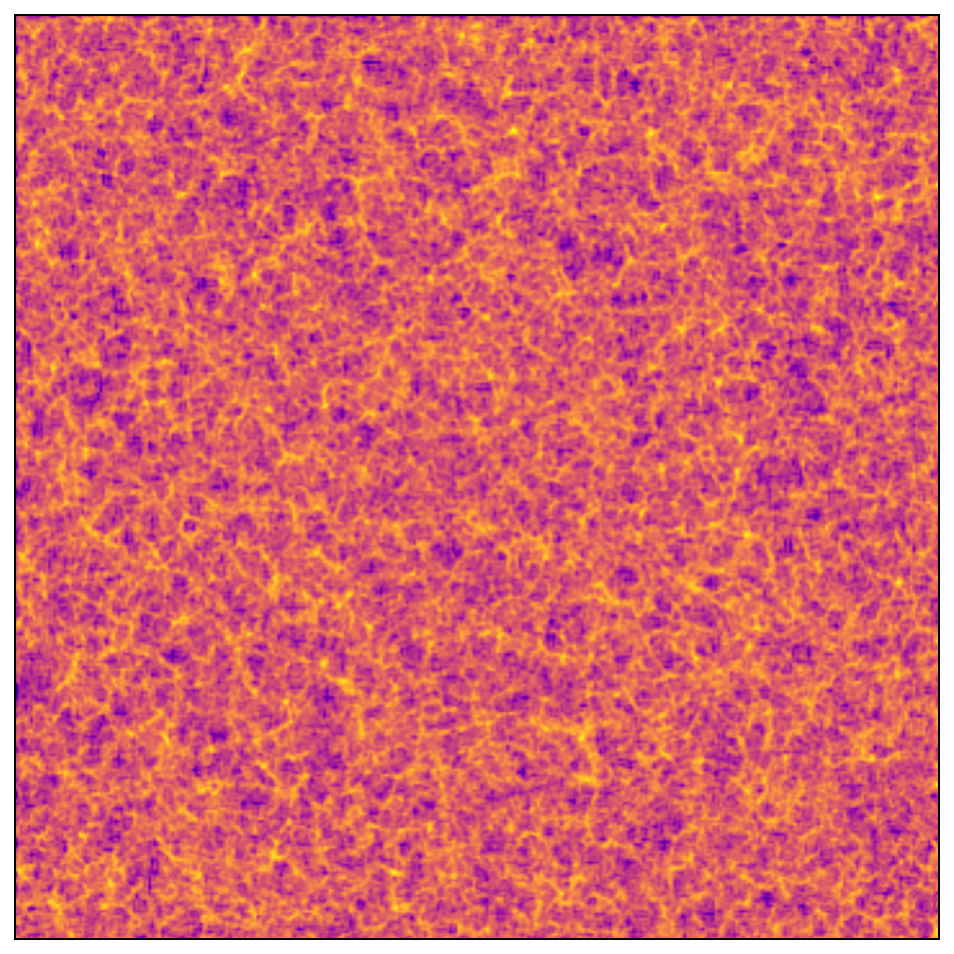}} \put(-203,193){\fcolorbox{white}{white}{2LPT}} \put(-203,180){\fcolorbox{white}{white}{z=1.1}} 
\hspace{-0.4cm} & \hspace{-0.1cm}
\subfigure{\includegraphics[width=0.32\textwidth]{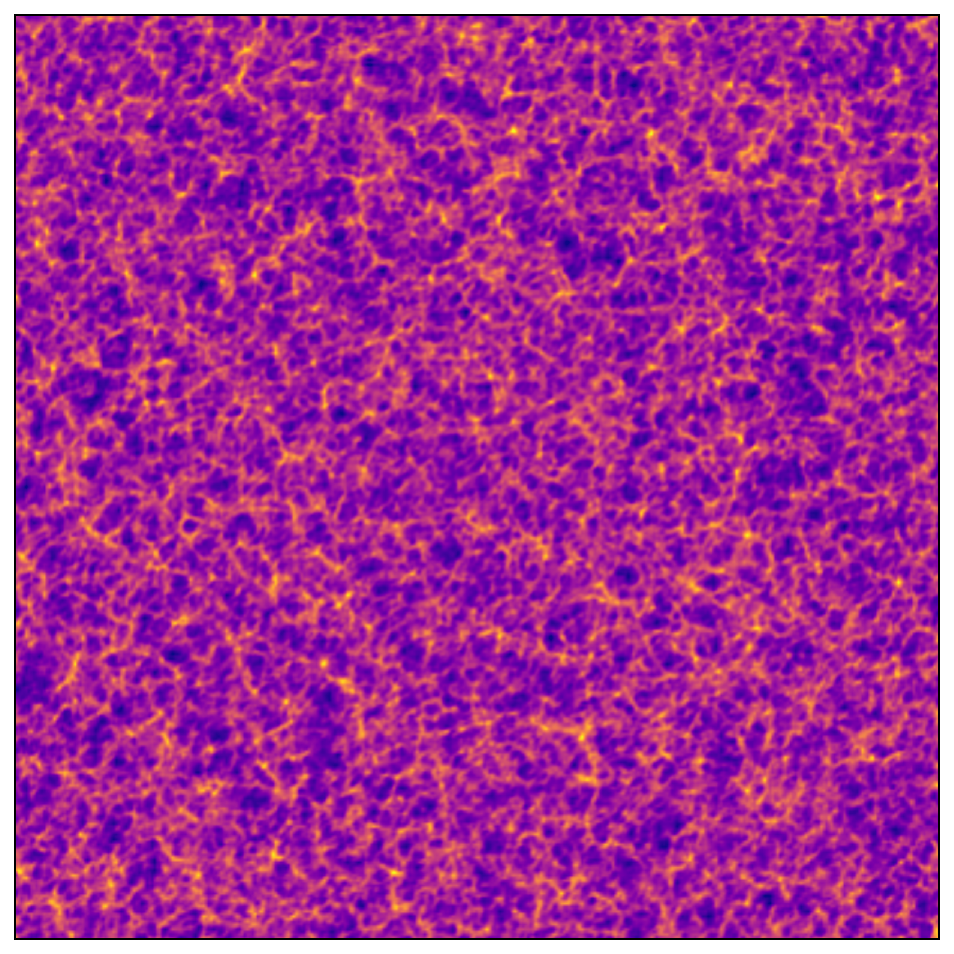}}  \put(-203,193){\fcolorbox{white}{white}{r2LPT}} \put(-203,180){\fcolorbox{white}{white}{z=1.1}} \hspace{-0.4cm} 
&
\subfigure{\includegraphics[width=0.32\textwidth]{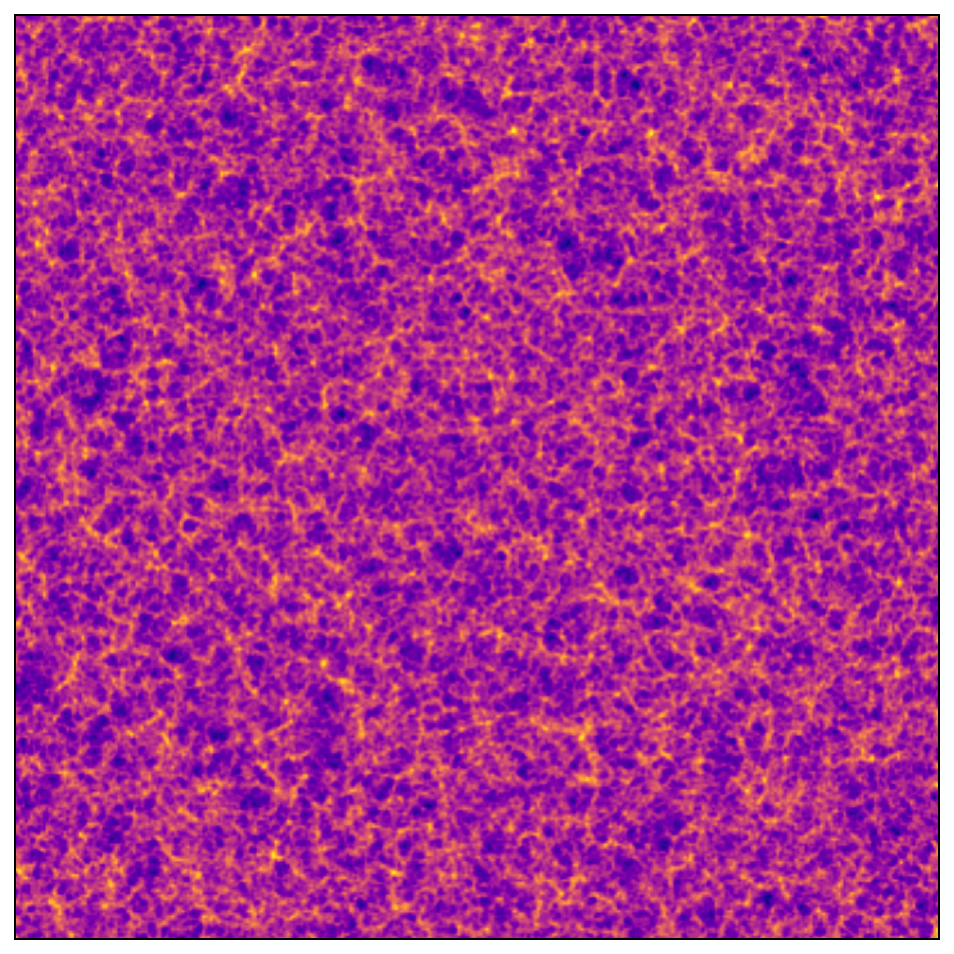}} 
\put(-203,193){\fcolorbox{white}{white}{ABACUS}}  \put(-203,180){\fcolorbox{white}{white}{z=1.1}} \end{tabular}
\put(-629,197){\rotatebox[]{90}{\large$\longrightarrow$}}    
\put(-634,110){\rotatebox[]{90}{\large${\rm L=}2000\,h^{-1}\,{\rm Mpc}$}}
\put(-629,19){\rotatebox[]{90}{\large$\longleftarrow$}}
\put(-586,213){\rotatebox[]{0}{\large$\longrightarrow$}}    
\put(-565,213){\rotatebox[]{0}{\large${\rm dL}=5.55\,h^{-1}\,{\rm Mpc}$}}
\put(-468,213){\rotatebox[]{0}{\large$\longleftarrow$}}
\caption{Matter over-density field: {\bf upper panels} at $z=0.2$; {\bf lower panels} at $z=1.1$, {\bf left}:  2LPT without tetra-hedra tesselation; {\bf center}: ridged LPT (2 steps-gravity solver z=0.2:  smooth particle ridging, z=1.1: ridging); {\bf right}: full $N$-body simulation. The \textsc{Abacus}  simulation used $6912^3$ particles and has been down-sampled to obtain the dark matter field on a mesh with $360^3$ cells with side length 2000 $h^{-1}$Mpc and a corresponding cell side resolution of 5.55  $h^{-1}$Mpc. The sp/r/n-LPT calculations used $360^3$ particles.}
    \label{fig:dmV2000}
\end{sidewaysfigure}

\begin{figure}[]
\vspace{-1cm}
    \centering
    \begin{tabular}{c}
    \includegraphics[scale=.5]{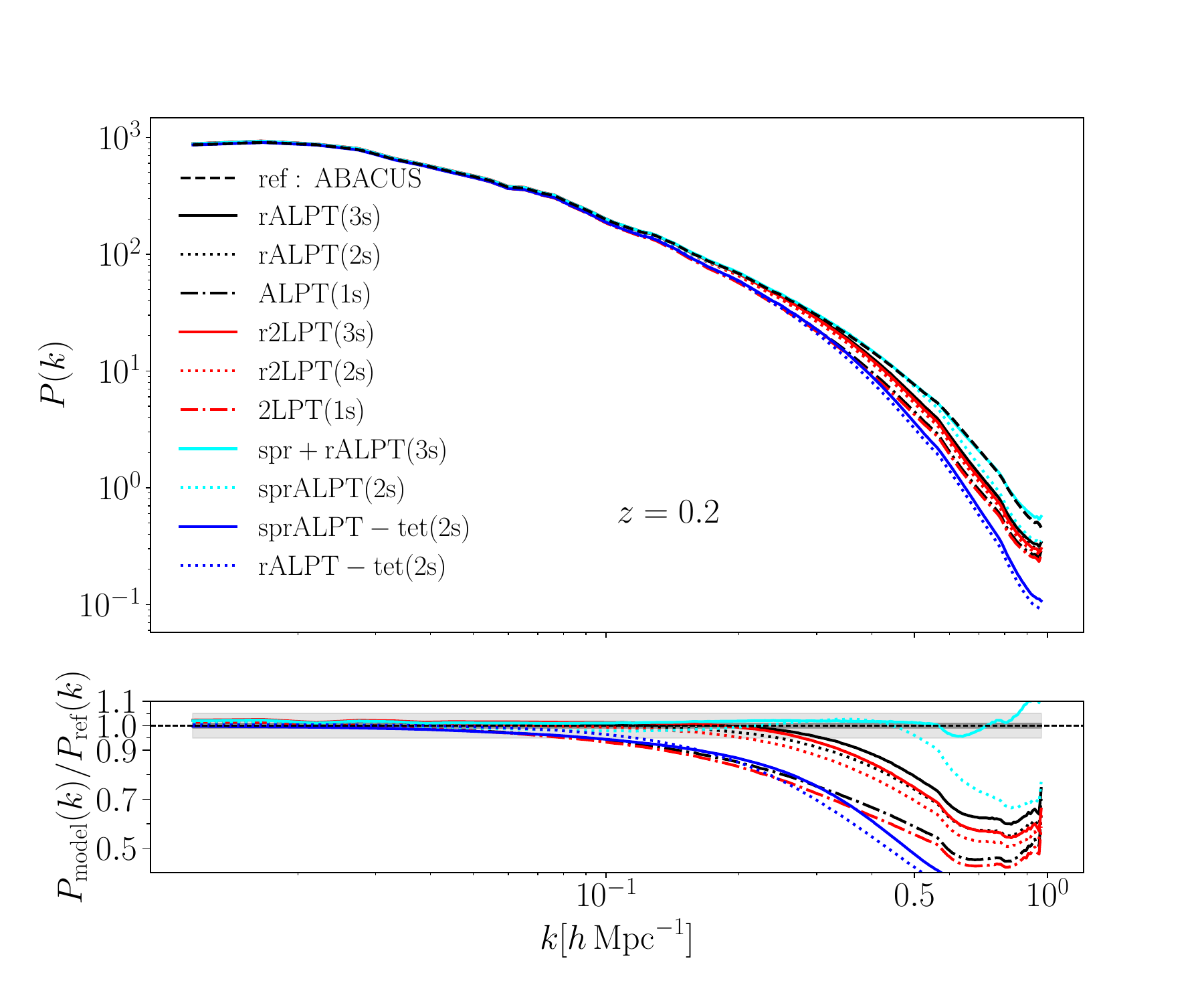}
     \put(-190,280){$L=2000\,h^{-1}\mathrm{Mpc}$, $N=360^3$} 
     \vspace{-1.cm}
    \\
\includegraphics[scale=.5]{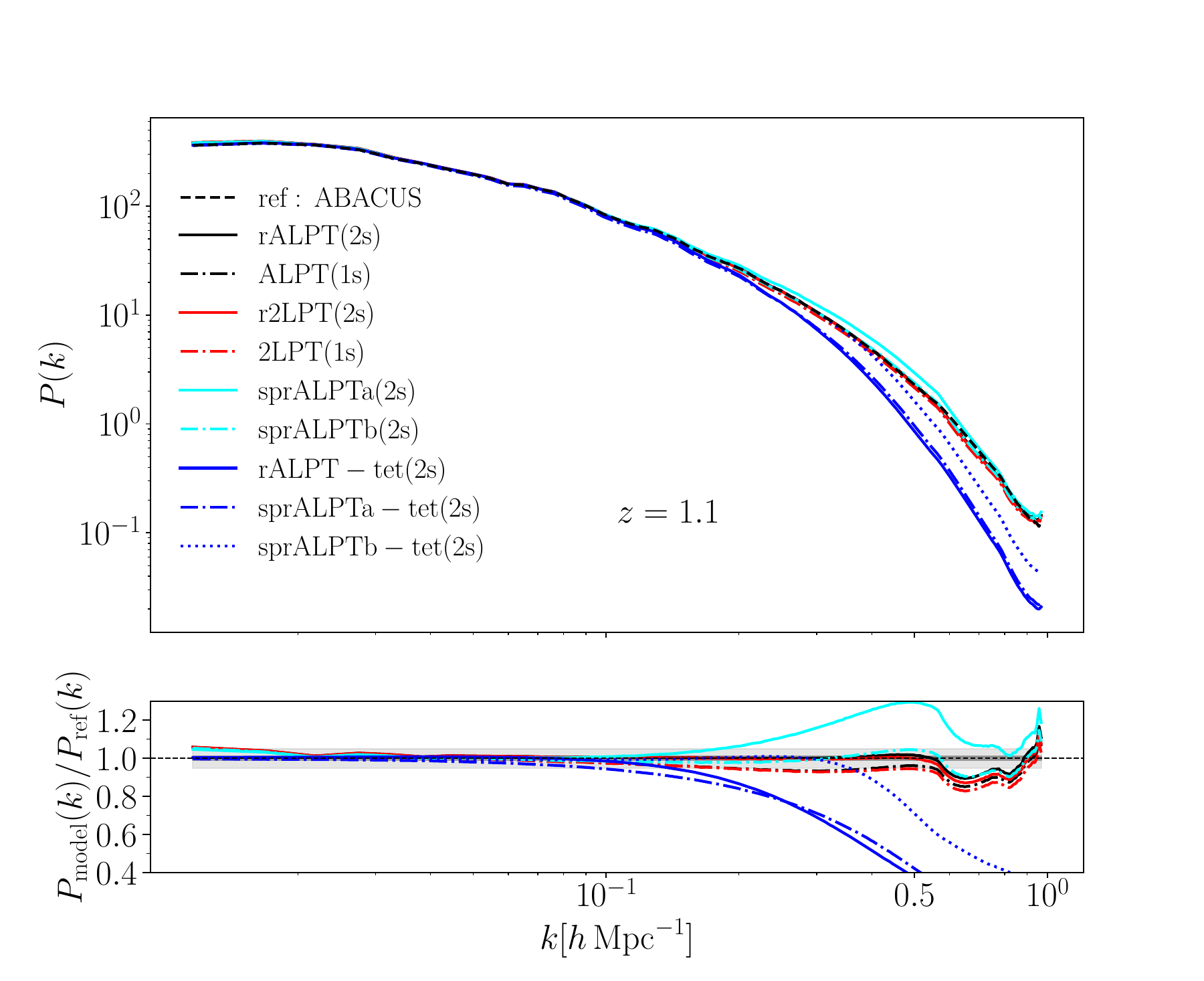}
     \put(-190,280){$L=2000\,h^{-1}\mathrm{Mpc}$, $N=360^3$} 
    \end{tabular}
    \vspace{-.6cm}
  \caption{Power spectra of the different LPT-based approximations compared with the \textsc{Abacus}  full $N$-body simulation with a volume of $L=2000\,h^{-1}\,{\rm Mpc}$ side and a mesh of $360^3$ cells. {\bf Upper panel:} $z=0.2$; {\bf lower panel:} $z=1.1$.}
    \label{fig:pkABACUS}
\end{figure}

\begin{figure}[]
\vspace{-1cm}
    \centering
    \begin{tabular}{c}
    \includegraphics[scale=.6]{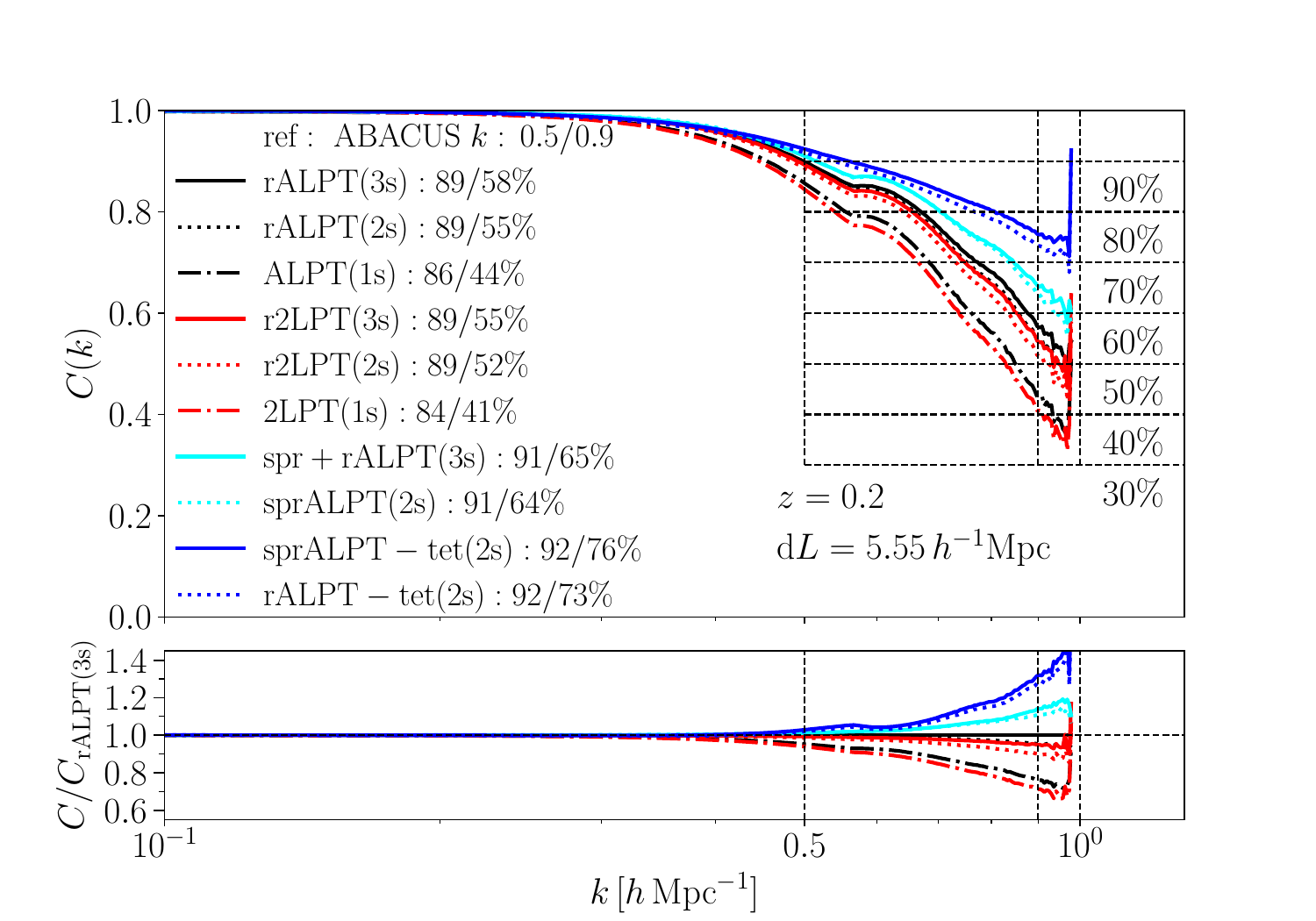}\vspace{-0.7cm}\\
\includegraphics[scale=.6]{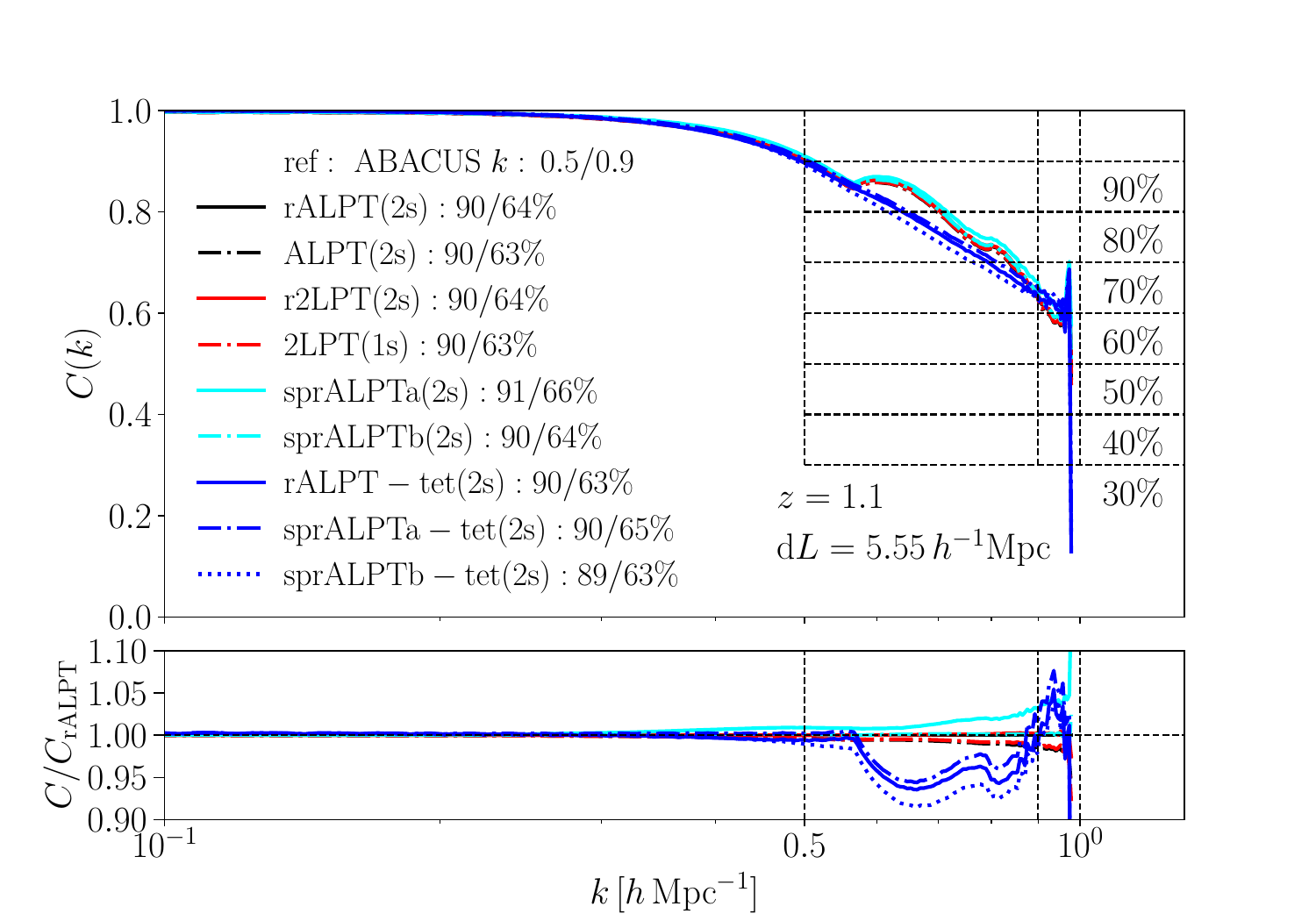}
    \end{tabular}
  \caption{Cross power spectra of the different LPT-based approximations compared with the { \textsc{Abacus} } full $N$-body simulation with a volume of $L=2000\,h^{-1}\,{\rm Mpc}$ side and a mesh of $360^3$ cells. {\bf Upper panel:} $z=0.2$; {\bf lower panel:} $z=1.1$.}
    \label{fig:ckABACUS}
\end{figure}

\begin{figure}[ht!]
\vspace{-1cm}
    \centering
    \begin{tabular}{c}
    \includegraphics[width=0.9\textwidth]{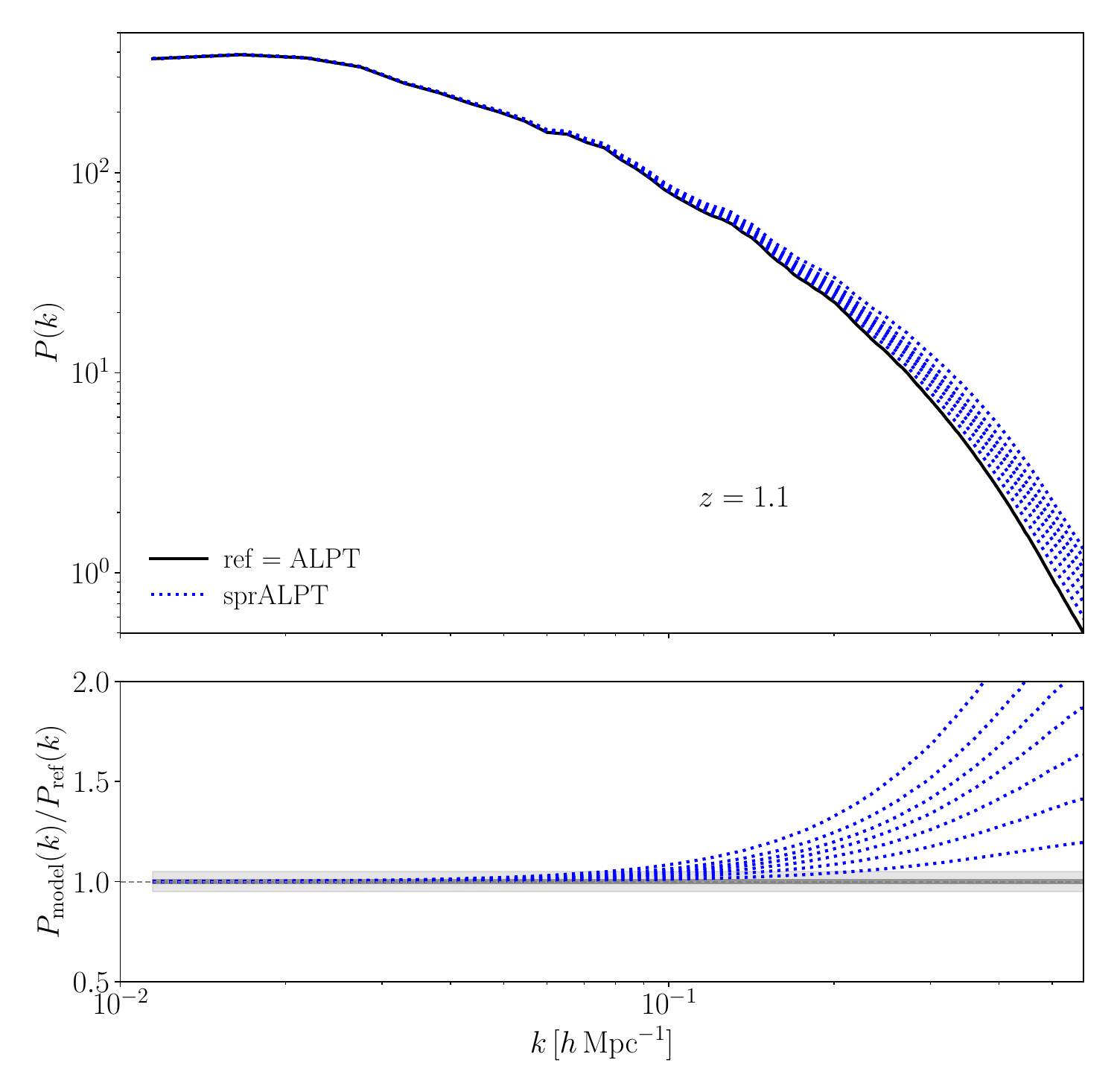}
    \end{tabular}
    \caption{Power spectra  corresponding to smooth particle ridging applied to ALPT at $z=1.1$ sharing the same initial conditions as the \textsc{Abacus}  simulation running over a set of short range growth factors.}
    \label{fig:pkspralptruns}
\end{figure}

\begin{figure}
    \vspace{-1cm}    
    \centering
 \begin{tabular}{cc}
\subfigure{\includegraphics[width=0.5\textwidth]{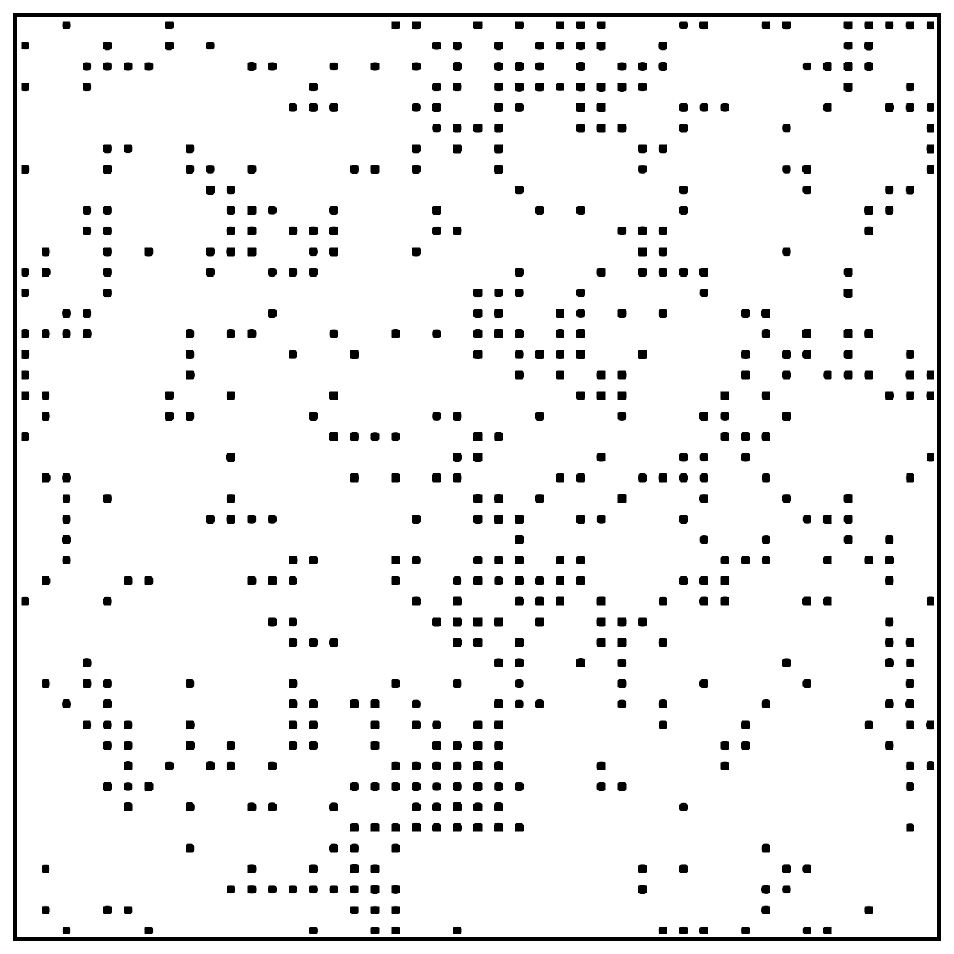}}\put(-218,207){\fcolorbox{white}{white}{${\rm Nh-ABACUS}$}}  
\hspace{-0.4cm} & \subfigure{\includegraphics[width=0.5\textwidth]{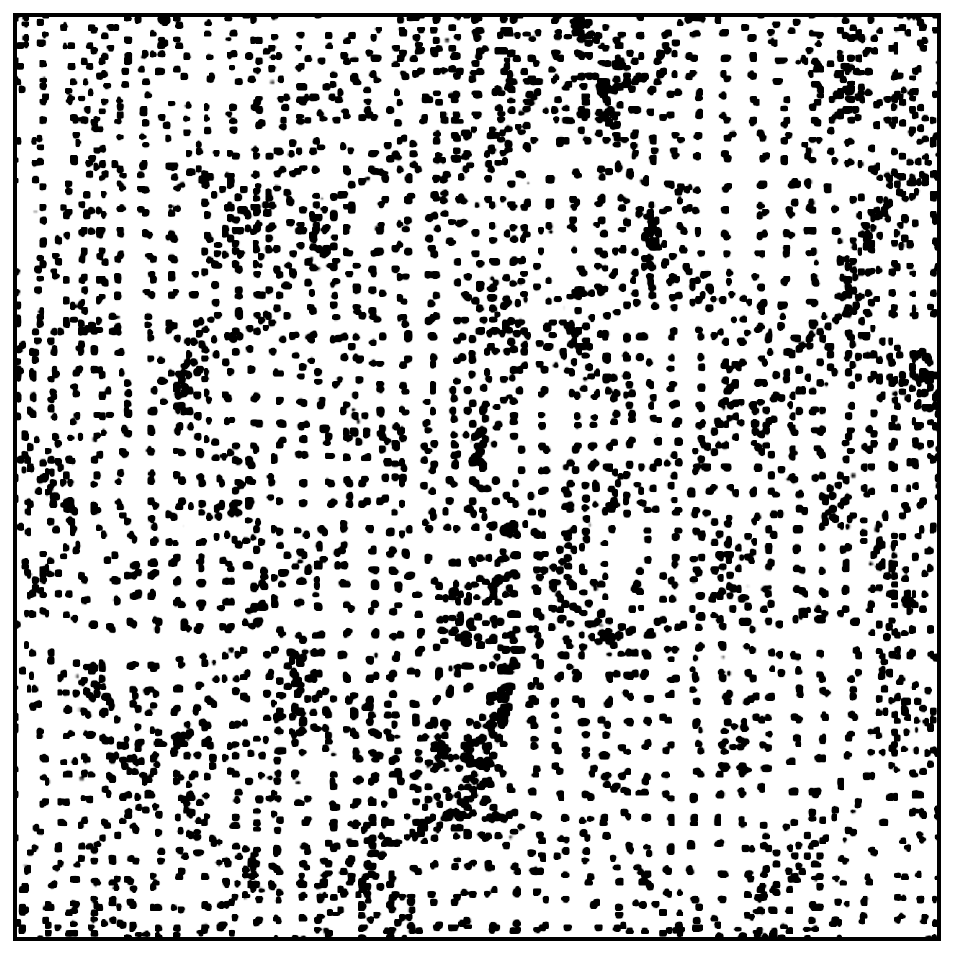}}
\put(-218,207){\fcolorbox{white}{white}{${\rm DM-ALPT}$}}
\end{tabular}
\put(-455,91){\rotatebox[]{90}{\large$\longrightarrow$}}    \put(-460,-0){\rotatebox[]{90}{\large${\rm L}=250\,[\subset2000]\,h^{-1}\,{\rm Mpc}$}}
\put(-455,-90){\rotatebox[]{90}{\large$\longleftarrow$}}
\put(-426,113){\rotatebox[]{0}{\large$\longrightarrow$}}    \put(-400,113){\rotatebox[]{0}{\large${\rm dL}=0.69\,[\subset5.55]\,h^{-1}\,{\rm Mpc}$}}
\put(-260,113){\rotatebox[]{0}{\large$\longleftarrow$}}
\\
\begin{tabular}{cc}
\subfigure{\includegraphics[width=0.5\textwidth]{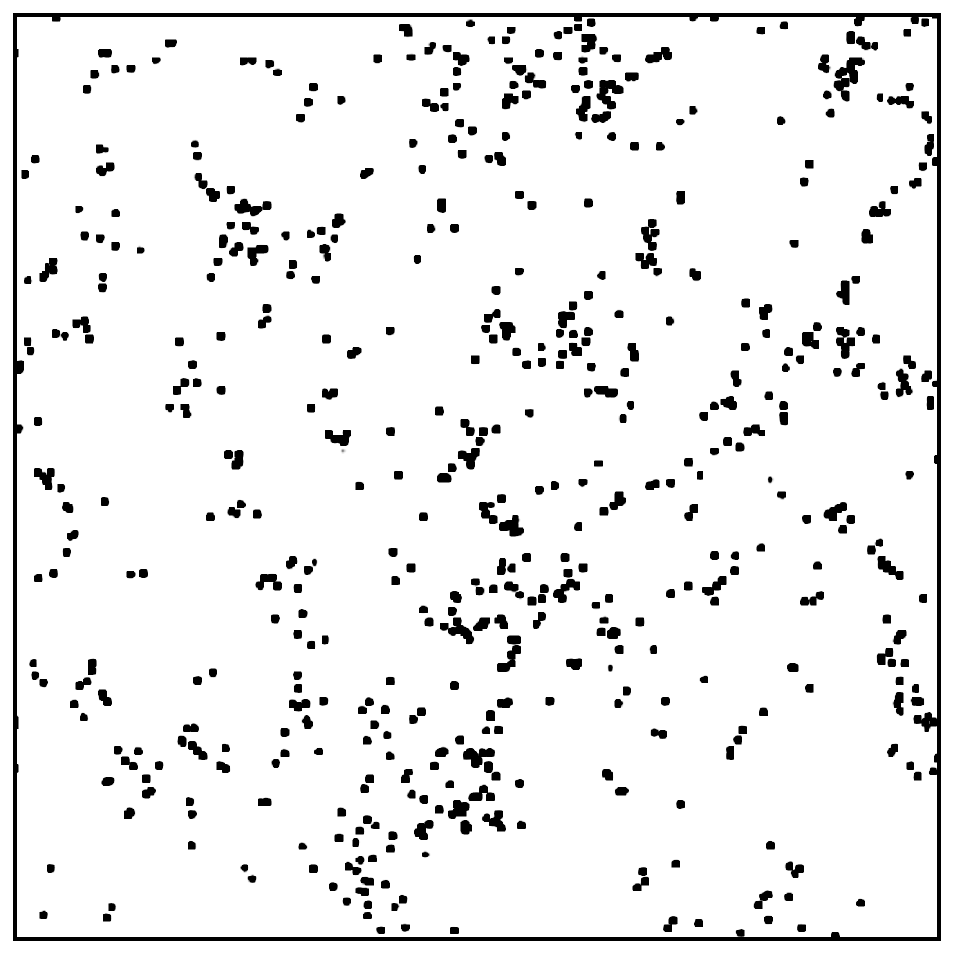}} \put(-218,207){\fcolorbox{white}{white}{${\rm h-ABACUS}$}} 
\hspace{-0.4cm} & 
\subfigure{\includegraphics[width=0.5\textwidth]{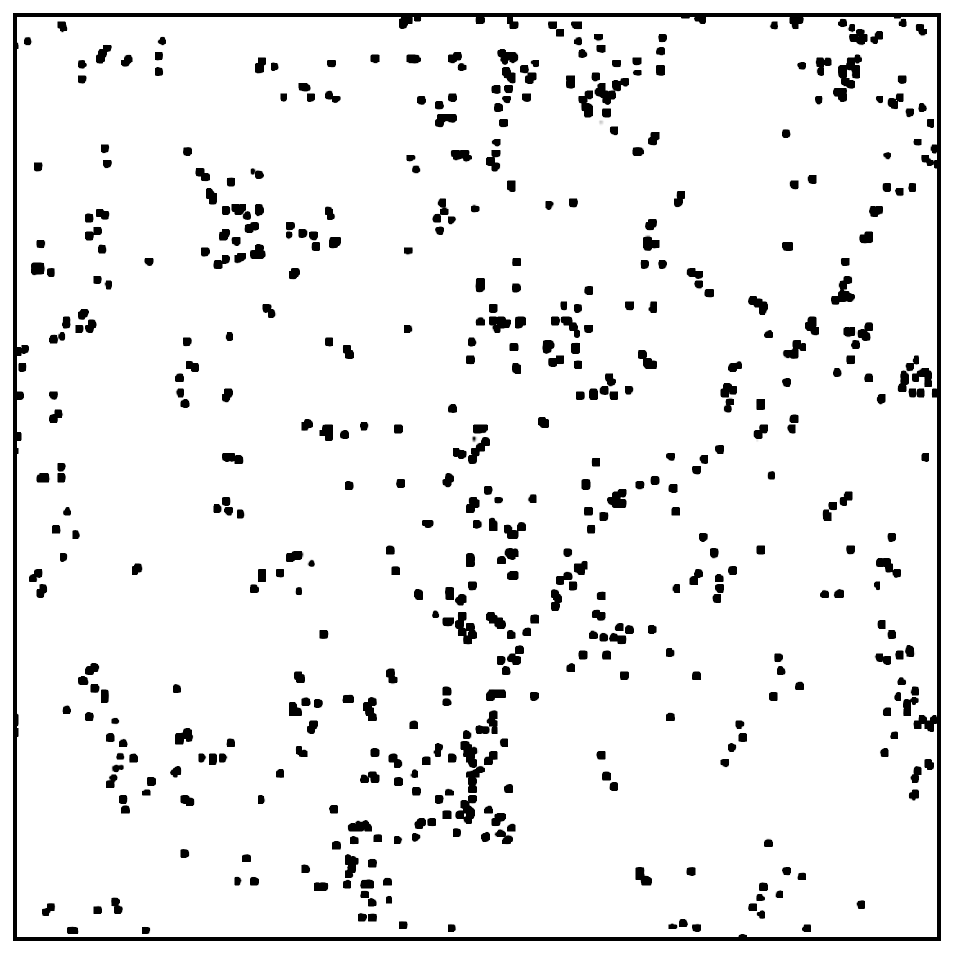}}\put(-218,207){\fcolorbox{white}{white}{${\rm Nh-DM-sprALPT}$}}
%\, (-k^{-2}, -k^{0}, -k^2, -k^4)$}}
\end{tabular}
\caption{Halo distribution reconstruction. Slice of a $250\,h^{-1}\mathrm{Mpc}$ subvolume extracted from the \textsc{Abacus}  simulation, represented on a $360^3$ mesh with cell size $d\mathrm{L}=0.69,h^{-1}\mathrm{Mpc}$. The {\bf upper left panel} shows the \textsc{Abacus} halo number counts at low-resolution cell centers  ($d\mathrm{L}=5.55,h^{-1}\mathrm{Mpc})$; the {\bf upper right panel} shows the ALPT dark matter particle counts in the  high resolution mesh. {\bf Lower left panel}: halo number counts from the \textsc{Abacus}  simulation on the high resolution mesh. {\bf Lower right panel}: reconstruction obtained from halo counts measured on a low-resolution mesh ($d\mathrm{L}=5.55,h^{-1}\mathrm{Mpc}$), where halos are assigned to ALPT particle positions within each cell (cloning particles when required), followed by smooth particle ridging and a small positional scatter.}
\label{fig:halos}
\end{figure}

\begin{figure}[ht!]
\vspace{-1cm}
    \centering
    \begin{tabular}{c}
    \includegraphics[scale=0.5]{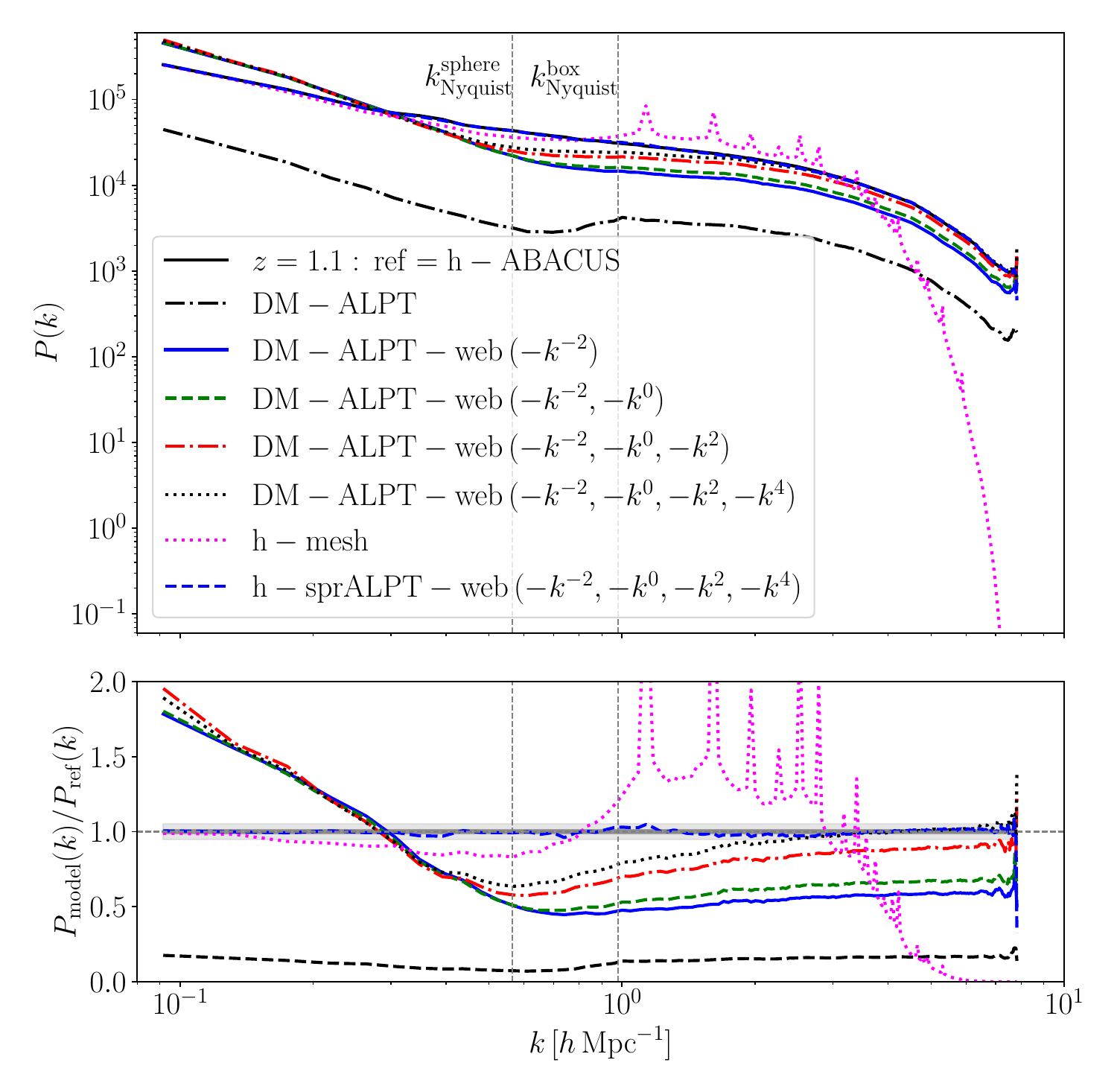}
    \end{tabular}
\caption{Power spectra of the different halo and dark matter fields. The solid line shows the halo distribution from the \textsc{Abacus}  simulation on a mesh with resolution $d\mathrm{L}=0.69,h^{-1}\mathrm{Mpc}$. The dash–dotted line corresponds to the ALPT dark matter particle distribution from the low-resolution mesh gridded onto the high-resolution mesh. The subsequent DM–ALPT curves represent the same dark matter field constrained to match the \textsc{Abacus}  halo number densities in progressively finer cosmic-web hierarchies. The magenta dotted line shows the halo field obtained by placing halos at the centers of the low-resolution cells, while the dashed blue line corresponds to the assignment of ALPT particles (with cloning when necessary) followed by smooth particle ridging.}
    \label{fig:pkhalosprlpt}
\end{figure}

\subsection{Step 1: long-range Lagrangian transport}
\label{sec:step1}

We transport particles as
\begin{equation}
\vx_{(1)}=\vq+\vPsiLR(\vq,z).
\label{eq:x1_generic}
\end{equation}
The long-range displacement $\vPsiLR$ can be computed with $n$-th order LPT (``$n$LPT'') using recursion relations and fast solvers \citep{2021JCAP...04..033S}.
If desired, one may regularise the small-scale behaviour through the ALPT construction \citep{Kitaura_2013} (A-($n$)LPT), by blending the long-range $n$LPT displacement with a spherical-collapse displacement using a smoothing kernel.
In the numerical studies below we use the common 2LPT and ALPT (A-2LPT) baselines.

After the long-range transport, we construct the realised Eulerian overdensity field $\delta_{(1)}(\vx)$ from the particle distribution at $\vx_{(1)}$ by depositing mass onto a regular mesh.
{By default} we use a tetrahedra tessellation (``tet'') density estimator: the initial particle lattice is decomposed into tetrahedra whose vertices are advected by the displacement; the mass of each deformed tetrahedron is then distributed to the grid by integrating its volume overlap with mesh cells.
This yields a continuous, piecewise-linear mass distribution that reduces shot noise and better preserves thin caustics and anisotropic features at fixed resolution.
Unless stated otherwise, the chosen mass-assignment scheme is {CIC with tetrahedral tessellation} (CIC+tet).
When we explicitly use standard point-particle cloud-in-cell deposition, we will refer to it as {plain CIC}.
We denote by $\delta_{(1)}(\vk)$ the Fourier transform of the gridded field (including the corresponding assignment window).

\subsection{Step 2: Eulerian ridging (short-range potential completion)}
\label{sec:step2}

%From particles at $\vx_{(1)}$ we construct a mesh overdensity $\delta_{(1)}(\vx)$ (CIC/TSC or tetrahedra).
The ridging step then builds a purely short-range \emph{potential} displacement from this realised Eulerian field.
Our implementation follows the Eulerian-remapping logic of eALPT \citep[e.g.][]{Kitaura_2024} in that we first apply a local spherical-collapse (SC) regularisation in configuration space, and only \emph{then} enforce an explicit scale separation prior to the displacement reconstruction.

\subsubsection{Spherical-collapse remapping in configuration space}

The ridging step constructs an incremental displacement from the already
transported particle configuration, motivating the use of its realised
Eulerian density to describe the structures on which the correction acts.
We adopt the spherical-collapse (SC) functional form as an effective
nonlinear closure. It recovers the linear response at small amplitudes
and, with the standard saturation prescription, regularises collapse
at high densities. Its motivation comes from the SC approximation of Bernardeau 94 
\citep[][]{Bernardeau_1994}, its numerical validation for the relation between
the initial linear density and the final Lagrangian displacement
divergence \citep[][]{neyrinck_2013}, and, more directly, its successful use
with updated density fields in iterative Eulerian remapping with eALPT
\citep[][]{Kitaura_2024}. Alternative mappings, such as the empirical
half-exponential relation considered by Neyrinck 13 \citep[][]{neyrinck_2013}, are
possible; SC provides a simple collapse-based prescription with an
established numerical precedent.

The original SC approximation describes the evolution of a finite region
in terms of its mean enclosed density. Its pointwise application to the
realised Eulerian density is an additional modelling assumption and does
not reproduce the exact evolution of a general spherical density profile,
for which each shell responds to the enclosed mean density rather than
the local density. Here, its purpose is to regularise the additional
displacement after the initial LPT transport: saturation of the effective
divergence source at $\psi=-3$ limits the compressive response in dense
regions, helping to mitigate excessive contraction and artificial shell
crossing in subsequent updates. This prescription neither guarantees
the absence of shell crossing nor describes post-shell-crossing dynamics
exactly. Its practical accuracy is assessed through our power-spectrum
and field-level correlation benchmarks.

%The scale-separation filter controls the transition to the long-range solution, whereas the nonlinear mapping determines the density-dependent short-range response.

Specifically, we apply the SC functional form
\citep[e.g.][]{Mohayaee_2006,2013MNRAS.435L..78K}
to the realised Eulerian density $\delta_{(1)}(\mathbf{x})$
to construct an effective nonlinear source for the short-range completion:
\begin{equation}
\psi(\vx)\equiv \nabla\cdot\vPsi(\vx)
=
3\left[\sqrt{1-\frac{2}{3}D_{\rm SR}(z)\,\delta_{(1)}(\vx)}-1\right],
\label{eq:sc_map}
\end{equation} 
where $D_{\rm SR}(z)$ is a calibratable short-range nonlinear-response
parameter. Its placement inside the mapping controls both the nonlinear
response and the effective saturation threshold. This differs from an
overall rescaling of the reconstructed displacement, as can be seen by
expanding for $\left|D_{\rm SR}\delta_{(1)}\right|\ll 1$: 
\begin{align}
\psi(\bm{x})
&=
3\left[
\sqrt{1-\frac{2}{3}D_{\rm SR}\delta_{(1)}(\bm{x})}
-1
\right]
\nonumber\\
&=
-D_{\rm SR}\,\delta_{(1)}
-\frac{1}{6}D_{\rm SR}^{2}\,\delta_{(1)}^{2}
-\frac{1}{18}D_{\rm SR}^{3}\,\delta_{(1)}^{3}
-\frac{5}{216}D_{\rm SR}^{4}\,\delta_{(1)}^{4}
+\mathcal{O}\!\left(\delta_{(1)}^{5}\right).
\label{eq:dsr_expansion}
\end{align}
By contrast, applying the same parameter as an overall multiplicative
factor after the spherical-collapse-inspired mapping would give
\begin{align}
D_{\rm SR}\,
3\left[
\sqrt{1-\frac{2}{3}\delta_{(1)}}
-1
\right]
&=
-D_{\rm SR}\,\delta_{(1)}
-\frac{1}{6}D_{\rm SR}\,\delta_{(1)}^{2}
-\frac{1}{18}D_{\rm SR}\,\delta_{(1)}^{3}
\nonumber\\
&\quad
-\frac{5}{216}D_{\rm SR}\,\delta_{(1)}^{4}
+\mathcal{O}\!\left(\delta_{(1)}^{5}\right).
\end{align}
Thus, the two prescriptions agree at linear order but differ from
second order onwards. In the adopted formulation, $D_{\rm SR}$ controls
the nonlinear hierarchy of the local closure itself, rather than merely
rescaling the final displacement.

When the argument of the square root becomes negative we saturate to the collapsed limit $\psi=-3$ (equivalently $\nabla\cdot\vPsi\to -3$), matching the standard prescription used in ALPT-like regularisations.

\subsubsection{Explicit short-range filtering}

To suppress the long-wavelength contribution to the additional
displacement and limit its impact on the large-scale density field,
we apply an explicit Eulerian scale separation prior to the
Poisson inversion.
In Fourier space we adopt an error-function low-pass filter,
\begin{equation}
K(k)=\frac{1}{2}\left[1-\mathrm{erf}\!\left(\frac{k-k_{\rm t}}{\sqrt{2}\,\Delta k}\right)\right],
\label{eq:erf_filter}
\end{equation}
where $k_{\rm t}$ sets the transition scale and $\Delta k$ controls the width of the transition.
We define the short-range component of the SC source as
\begin{equation}
\psi_{\rm SR}(\vk)=\left[1-K(k)\right]\psi(\vk)\,.
\label{eq:psi_sr}
\end{equation}

\subsubsection{Short-range potential displacement}
We then reconstruct a purely short-range {potential} displacement by Poisson inversion of the SC-remapped source,
\begin{equation}
\vPsiSR(\vk)= -i\frac{\vk}{k^2}\,\psi_{\rm SR}(\vk)\,, 
%\mathcal{T}_{\rm N}(k),
\qquad \vk\neq \mathbf{0},
\label{eq:psiSR_core}
\end{equation}
with $\vPsiSR(\mathbf{0})=\mathbf{0}$.
%Here $\mathcal{T}_{\rm N}(k)$ denotes an optional high-$k$ taper near Nyquist introduced for numerical stability; however, in our numerical experiments this did not improve the results and we therefore set $\mathcal{T}_{\rm N}(k)=1$ throughout. 

In the absence of vorticity, the second-step update is purely potential and can be written as
\begin{equation}
\vx_{(2)}=\vx_{(1)}+
\vPsi_{\rm SR}^{\parallel}\!\left(\vx_{(1)}\right) \quad ,
\label{eq:x2_core}
\end{equation}
so that the short-range displacement is curl-free.

\subsection{Optional extension: vortical ridging}
\label{sec:vortical_method}

The baseline ridging displacement $\vPsiSR$ is curl-free by construction (a potential flow) at the Eulerian stage.
However, adding an Eulerian update after the first mapping means that the {total} displacement from $\vq$ to $\vx_{(2)}$ need not remain globally irrotational, consistent with vorticity generation in iterative Eulerian LPT approaches \citep[e.g.][]{Kitaura_2024}.
The possibility of generating non-zero vorticity from mode coupling beyond the leading irrotational approximations has a long history in the LPT literature \citep[e.g.][]{1995MNRAS.276..115C}.
As an alternative to expensive $N$-body simulations, effective field theory has emerged as a powerful framework in large-scale structure, including treatments of the stress tensor for the prediction of summary statistics \citep[see, e.g.,][]{Carrasco_2012,Baumann_2012,Pajer_2013,2014JCAP...05..022P,2014JCAP...03..006M,2015JCAP...10..039A,2015PhRvD..92l3007B,2016JCAP...03..017B,2016JCAP...05..027F}. See also other perturbative approaches to modelling the curl \citep[][]{2011JCAP...04..032M,2012JCAP...01..019P,Rampf_2012,2017PhRvD..95f3527C}, as well as non-perturbative ones \citep[][]{2005A&A...438..443B}.

As a clarifying remark, EFT of LSS is a systematically improvable framework for predicting {correlators} (summary statistics) by renormalising the impact of unresolved short modes, whereas RLPT is a {field-level} closure: it introduces an explicit, scale-separated operator acting directly on a realised density field (and can be calibrated to statistics a posteriori). In the remainder of this section we focus on the field-level motivation for adding a solenoidal displacement component in RLPT, independently of whether EFT is used as a target description for calibration.

If desired, RLPT can be extended by adding a calibrated solenoidal correction, as detailed below.

\subsubsection{Helmholtz decomposition and update rule}

More generally, any displacement field can be decomposed through the Helmholtz theorem as
\begin{equation}
\vPsi(\vx)=\nabla\phi(\vx)+\nabla\times \vA(\vx),
\qquad \nabla\cdot \vA = 0,
\end{equation}
where the first term is irrotational and the second is divergence-free, carrying all the vorticity.
Motivated by this decomposition, we generalise Eq.~\eqref{eq:x2_core} to
\begin{equation}
\vx_{(2)}=\vx_{(1)}+
\vPsi_{\rm SR}^{\parallel}\!\left(\vx_{(1)}\right)
+
D_{\rm vort}(z)\,\vPsi_{\rm SR}^{\perp}\!\left(\vx_{(1)}\right),
\label{eq:x2_vort}
\end{equation}
where $\vPsi_{\rm SR}^{\parallel}\equiv\nabla\phi_{\rm SR}$ is the curl-free short-range displacement already present in the model, while
$\vPsi_{\rm SR}^{\perp}\equiv\nabla\times\vA_{\rm SR}$ is a purely solenoidal short-range contribution.
The new growth factor $D_{\rm vort}(z)$ controls the amplitude of the vortical correction.

\subsubsection{A minimal, calibratable vorticity-source model}
A practical route is to define $\vPsi_{\rm SR}^{\perp}$ from a short-range vorticity field $\vomega_{\rm SR}$.
In Fourier space, for a divergence-free displacement one may write
\begin{equation}
\vPsi_{\rm SR}^{\perp}(\vk)
=
i\,\frac{\vk\times \vomega_{\rm SR}(\vk)}{k^2}\,,
%\mathcal{T}_{\rm N}(k),
\qquad \vk\neq \mathbf{0},
\label{eq:psi_perp_from_omega}
\end{equation}
which guarantees $\vk\cdot\vPsi_{\rm SR}^{\perp}=0$.
It remains to specify $\vomega_{\rm SR}$ in a way that is (i) local/fast to compute and (ii) parametrised with few degrees of freedom.

A minimal ansatz is that vorticity is sourced by nonlinear mode-coupling of short-range density and tidal fields.
One convenient choice is to build a pseudo-source from the post-transport fields:
\begin{equation}
\vomega_{\rm SR}(\vx)
=
\nabla\times\left[
\left(\nabla\phi_{\rm SR}\right)\,\mu_\omega(\vx)
\right],
\qquad
\mu_\omega(\vx)=  \mu_1\,\delta_{\rm LR}(\vx) + \mu_2\,s^2(\vx),
\label{eq:omega_model}
\end{equation}
where $s^2\equiv s_{ij}s_{ij}$ is the tidal shear invariant computed from the (smoothed) potential/tidal tensor used elsewhere in the method.
This model is purely local in real space once $\phi_{\rm SR}$ and the auxiliary fields are available, and introduces only a handful of parameters $(\mu_1,\mu_2)$.
One may also impose short-range support explicitly by applying the same $(1-K)$ filter to $\mu_\omega$ or $\vomega_{\rm SR}$ in Fourier space.

For the $L=100\,h^{-1}\mathrm{Mpc}$ benchmark (Section~\ref{sec:pm_100}) we adopt an even simpler vorticity-source proxy directly in configuration space,
\begin{equation}
\vomega_{\rm SR}(\vx) \equiv \nabla\delta_{(1)}(\vx)\times \vPsi_{\rm SR}^{\parallel}(\vx),
\label{eq:omega_source}
\end{equation}
which is inexpensive to evaluate once $\delta_{(1)}$ and $\vPsi_{\rm SR}^{\parallel}$ are available on the mesh.
This choice is intended as a proof-of-concept and provides a stable solenoidal direction aligned with short-scale density gradients and ridging flows.
The amplitude $D_{\rm vort}$ can then be calibrated to maximise high-$k$ cross-correlation while preserving the large-scale agreement.

Vortical ridging will be investigated in further detail in future work, including: (i) the redshift and resolution dependence of $D_{\rm vort}$, (ii) calibration targets based on solenoidal displacement or velocity-vorticity spectra, and (iii) more general source models.
In particular, variations of the model naturally admit more complex expressions for $\vomega_{\rm SR}$ inspired by perturbative expansions, including non-local tidal operators and higher-derivative terms (beyond $\nabla\delta_{(1)}$), which may better capture the structure of vorticity generation in the mildly nonlinear regime \citep[e.g.][]{1995MNRAS.276..115C}.

% ---------------------------------------------------------
\subsection{Smooth-particle ridging (SPRLPT): a SPH-inspired short-range completion}
\label{sec:spr}

{\color{black}
SPRLPT is a local particle-space short-range closure inspired by Smooth Particle
Hydrodynamics (SPH), rather than an SPH discretisation of the mesh-based RLPT
operator of Section~\ref{sec:step2}. Both approaches are designed to provide a
controlled short-range completion after the long-range transport, but they use
different constructions: mesh-based RLPT derives a longitudinal displacement
from an effective scalar source through a Poisson inversion, whereas SPRLPT
estimates short-range displacements directly from particle neighbourhoods using
compact-support kernels and kernel-density gradients. There is therefore no exact
limit in which SPRLPT reduces to the SC-inspired mapping of Section~\ref{sec:step2}.
For smooth, approximately isotropic overdensities, however, both approaches can
produce qualitatively similar inward displacements toward density ridges.

SPRLPT is particularly useful when one wishes to operate below the nominal mesh
scale or directly on tracer positions. In our numerical experiments it is used
either as an alternative to the mesh-based ridging step or, where explicitly
indicated, as an additional subgrid relocation applied after mesh-based ridging.
These cases are identified explicitly in the figure legends.

SPH-inspired kernels provide a natural way to estimate local densities and
gradients from particle neighbourhoods without introducing an additional fine
mesh. We therefore refer to this particle-space variant as
 SPRLPT.
}

\subsubsection{Scheme}

Starting from the Step~1 positions $\{\vx_{i,(1)}\}$, SPRLPT constructs a short-range displacement
$\vPsi^{\rm spr}_{\rm SR}(\vx)$ directly from local particle neighbourhoods:
(i) estimate a kernel density $\rho_i$ for each particle,
(ii) compute a pressure-like, ridge-contracting increment from density gradients,
(iii) add a viscosity-like regularisation that damps noisy neighbour-to-neighbour fluctuations,
and (iv) update positions with the same short-range growth factor $D_{\rm SR}$ used elsewhere in RLPT.
Neighbour queries are accelerated with a KD-tree (we use \texttt{nanoflann}\footnote{\url{https://jlblancoc.github.io/nanoflann/}}).

For each particle $i$, we define a neighbourhood $\mathcal{N}(i)$ given by all particles within a smoothing radius $h$. 
In practice we cap the number of returned neighbours to $N_{\rm ngb}$ %(called \texttt{nmaxfriends}) 
for stability and performance.

\subsubsection{Kernel density estimate}
We estimate a local density with a compact-support kernel $W(r,h)$,
\begin{equation}
\rho_i \equiv \sum_{j\in\mathcal{N}(i)} m_j\, W(r_{ij},h),
\qquad r_{ij}\equiv |\vx_{j,(1)}-\vx_{i,(1)}|\,,
\label{eq:spr_density}
\end{equation}
{\color{black} where \(m_j=m_p\) is the particle mass. All particles in the applications considered here have equal mass; we retain the index \(j\) only to keep the expression in its standard SPH form and to allow the straightforward generalisation to unequal weights/masses.}
In our default implementation we use a 3D Wendland-type kernel \cite{Wendland1995} with support $q\equiv r/h<2$,
\begin{equation}
W_{\rm W}(r,h)=
\begin{cases}
\displaystyle
\frac{21}{16\pi h^3}
\left(1-\frac{1}{2}q\right)^4(2q+1),
& 0\le q<2,\\[6pt]
0,&q\ge2,
\end{cases}
\qquad q\equiv\frac{r}{h}.
\label{eq:spr_wendland}
\end{equation}

Its radial derivative is
\begin{equation}
\frac{\partial W_{\rm W}}{\partial r}(r,h)=
\begin{cases}
\displaystyle
-\frac{105}{16\pi h^4}\,
q\left(1-\frac{1}{2}q\right)^3,
&0\le q<2,\\[6pt]
0,&q\ge2.
\end{cases}
\label{eq:spr_wendland_grad}
\end{equation}

{\color{black} The factor $q$ ensures that
$\partial W_{\rm W}/\partial r\rightarrow0$ as $r\rightarrow0$, so that
the kernel gradient is regular at coincident particle positions.
The prefactor $21/(16\pi h^3)$ normalises the kernel to unity in three
dimensions for support $0\le q<2$.}

Wendland kernels are a robust default because they have compact support and smooth derivatives and are well known in SPH for their stability against particle pairing and noise amplification in gradient estimates.
We have also explored sigmoid-like smooth weightings with very similar outcomes; unless stated otherwise we adopt the Wendland form above.

\subsubsection{Pressure-like driving (ridge contraction)}

We define an effective pressure-like scalar that increases with overdensity,
\begin{equation}
P_i \equiv K_{\rm spr}\,(\rho_i-\rho_0),
\label{eq:spr_pressure}
\end{equation}
where $K_{\rm spr}$ 
%(called \texttt{stiffness}) 
sets the strength of the contraction and $\rho_0$ is a reference density (set to unity after normalisation).
We compute a displacement-like pressure increment as
\begin{equation}
\vPsi^{\rm press}_{{\rm SR},i}
=
\sum_{j\in\mathcal{N}(i)}
m_j\,
\frac{P_i+P_j}{2\,\rho_j}\,
\nabla_i W(r_{ij},h),
\label{eq:spr_press}
\end{equation}
with
\begin{equation}
\nabla_i W(r_{ij},h)
=
-\frac{\partial W(r_{ij},h)}{\partial r_{ij}}\,
\hat{\mathbf r}_{ij},
\qquad
\hat{\mathbf r}_{ij}
\equiv
\frac{\vx_{j,(1)}-\vx_{i,(1)}}{r_{ij}}.
\label{eq:spr_kernel_gradient}
\end{equation}
This term drives short-range contraction along local density ridges and provides the core ``ridging'' effect in particle space.

\subsubsection{Viscosity-like regularisation}
To suppress noisy relative updates at finite neighbour counts, we add a viscosity-like damping term (purely numerical, analogous in spirit to artificial viscosity in SPH).
Let $\vPsi^{\rm base}_{{\rm SR},i}$ denote a baseline displacement for particle $i$ (set to zero in a single-step run, or to the previous iterate when the method is iterated).
We apply
\begin{equation}
\vPsi^{\rm visc}_{{\rm SR},i}
=
-\eta_{\rm spr}
\sum_{j\in\mathcal{N}(i)}
\frac{W(r_{ij},h)}{\rho_j\,(r_{ij}^2+\epsilon)}
\left(\vPsi_{{\rm SR},i}-\vPsi^{\rm base}_{{\rm SR},i}\right),
\label{eq:spr_visc}
\end{equation}
where $\eta_{\rm spr}$ controls the damping strength and $\epsilon$ regularises the $r\to 0$ limit.
In practice we compute a pressure-driven increment $\Delta\vPsi^{\rm press}_{{\rm SR},i}$ from Eq.~\eqref{eq:spr_press}, and set $\vPsi_{{\rm SR},i}=\vPsi^{\rm base}_{{\rm SR},i}+\Delta\vPsi^{\rm press}_{{\rm SR},i}$ within the same pass.

\subsubsection{SPRLPT short-range displacement and update}

The final particle-space short-range displacement is
\begin{equation}
\vPsi^{\rm spr}_{{\rm SR},i}
\equiv
\vPsi^{\rm press}_{{\rm SR},i}+\vPsi^{\rm visc}_{{\rm SR},i}.
\label{eq:spr_total}
\end{equation}
We then apply the same RLPT short-range update rule as in Eq.~\eqref{eq:x2_core}, but with $\vPsi_{\rm SR}\rightarrow\vPsi^{\rm spr}_{\rm SR}$:
\begin{equation}
\vx_{i,(2)}=\vx_{i,(1)}+D_{\rm SR}\,\vPsi^{\rm spr}_{{\rm SR},i},
\label{eq:spr_update}
\end{equation}
followed by periodic wrapping into the simulation box.

SPRLPT is particularly useful when one wants to inject controlled short-range response {beyond the nominal mesh resolution}, e.g., for deterministic subgrid repositioning of {dark matter particles and/or }tracers within coarse cells in mock-production pipelines.
Because it is local and kernel-based, it naturally supports ``beyond-the-mesh'' adjustments while keeping the large-scale transport (Step~1) fixed.

\subsection{Relation of SPRLPT to conservative SPH formulations}
\label{sec:spr_sph_conservation}

{\color{black}
We stress that SPRLPT is an SPH-inspired \emph{displacement closure},
rather than a time integration of the hydrodynamical Euler equations.
The quantities denoted by $P_i$ and by
$\vPsi^{\rm press}_{{\rm SR},i}$ are therefore effective constructs used
to generate a controlled contraction of particles toward density ridges;
they should not be interpreted literally as a thermodynamic pressure
and the force resulting from it.

Consequently, the SPRLPT update used in this work is not required to obey
pairwise Newtonian action--reaction symmetry. In particular, the
pressure-like prescription
\begin{equation}
\vPsi^{\rm press}_{{\rm SR},i}
=
\sum_{j\in\mathcal N(i)}
m_j
\frac{P_i+P_j}{2\rho_j}
\nabla_i W_{ij}
\end{equation}
is deliberately regarded here as a local displacement estimator rather
than as a conservative SPH force law. Its purpose is to determine the
direction and amplitude of a short-range ridge-contraction update.
Because no physical acceleration is integrated in time, exact pairwise
momentum conservation is not a defining requirement of the present
algorithm.

Nevertheless, if one wished to construct a strictly conservative
SPH-like variant of SPRLPT, the pressure-like contribution could instead
be written using the standard symmetric pair weighting
\begin{equation}
\vPsi^{{\rm press,cons}}_{{\rm SR},i}
=
A_{\rm spr}
\sum_{j\in\mathcal N(i)}
m_j
\left(
\frac{P_i}{\rho_i^2}
+
\frac{P_j}{\rho_j^2}
\right)
\nabla_i W_{ij},
\label{eq:spr_press_conservative}
\end{equation}
where $A_{\rm spr}$ is an effective response amplitude.
Since
\begin{equation}
\nabla_i W_{ij}=-\nabla_j W_{ji},
\end{equation}
the pair coefficient in parentheses is symmetric under
$i\leftrightarrow j$. For equal particle masses this construction gives
equal and opposite pairwise updates and therefore preserves the
action--reaction structure characteristic of conservative SPH
discretisations.

An alternative commonly used symmetric structure is
\begin{equation}
\vPsi^{{\rm press,cons}}_{{\rm SR},i}
=
A_{\rm spr}
\sum_{j\in\mathcal N(i)}
m_j
\frac{P_i+P_j}{\rho_i\rho_j}
\nabla_i W_{ij},
\label{eq:spr_press_conservative_alt}
\end{equation}
up to an overall numerical normalisation that can be absorbed into
$A_{\rm spr}$.
Both forms are closer to a conventional SPH pressure operator than the
minimal displacement estimator adopted in the present proof-of-concept
implementation.

A similar distinction applies to the regularisation term. The
implementation used in this work damps the displacement of particle $i$
relative to a baseline update,
\begin{equation}
\vPsi^{\rm damp}_{{\rm SR},i}
\propto
-
\sum_{j\in\mathcal N(i)}
w_{ij}
\left(
\vPsi_{{\rm SR},i}
-
\vPsi^{\rm base}_{{\rm SR},i}
\right),
\end{equation}
and is therefore more appropriately interpreted as a local numerical
relaxation than as a physical viscosity.

A genuinely pairwise SPH-inspired damping operator could instead be
constructed from relative particle updates,
\begin{equation}
\vPsi^{{\rm visc,cons}}_{{\rm SR},i}
=
-\eta_{\rm spr}
\sum_{j\in\mathcal N(i)}
m_j\,
\mathcal{W}_{ij}
\left(
\vPsi_{{\rm SR},i}
-
\vPsi_{{\rm SR},j}
\right),
\label{eq:spr_visc_pairwise}
\end{equation}
where $\mathcal{W}_{ij}=\mathcal{W}_{ji}$ is a symmetric positive kernel
weight, for example
\begin{equation}
\mathcal{W}_{ij}
=
\frac{W(r_{ij},h)}
{\bar\rho_{ij}\,(r_{ij}^2+\epsilon)},
\qquad
\bar\rho_{ij}
=
\frac{\rho_i+\rho_j}{2}.
\end{equation}
Such a form damps relative particle-to-particle fluctuations while
leaving a coherent common displacement unchanged.

The conservative expressions above define natural extensions of SPRLPT,
but they constitute different short-range closures and are not used for
the numerical results presented in this work. A systematic comparison
between the minimal ridge-displacement prescription adopted here and
fully pair-symmetric SPH-like variants will be explored in future work.
}

% =========================================================

% =========================================================
\section{Numerical studies}
\label{sec:results}

We present three numerical studies designed to isolate the impact of the ridging step.
In all cases we compare single-step baselines (2LPT and ALPT) to RLPT, and we report both the matter power spectrum and field-level cross-correlation with a reference gravity solver.

{\color{black} The baseline RLPT closure is intentionally parametrised by only a small number of free quantities: primarily the transition scale and width of the short-range filter and the amplitude of the short-range response (with additional parameters only when the optional vortical or smooth-particle extensions are enabled).} For each benchmark we adopt fixed ridging parameters chosen to yield stable, representative performance for that resolution and redshift, guided by standard field-level diagnostics (power-spectrum agreement and cross-correlation).
A systematic optimisation across redshift, resolution, and cosmology is left to future work. We emphasise that the qualitative improvements reported here are robust to modest variations around these fiducial choices.
Unless stated otherwise, ridging adds only one post-processing step on top of the chosen baseline.

{\color{black} {\bf Notation used in the figures.}

We use rALPT for RLPT with ALPT as the Step~1 backbone, and r2LPT for RLPT with 2LPT as the Step~1 backbone.
The suffix (Ns) denotes the total number of mapping/update steps. Thus ALPT(1s) is the original one-step ALPT transport, rALPT(2s) consists of one ALPT transport followed by one ridging update, and rALPT(3s) contains two successive ridging updates after the initial ALPT transport. The suffix wv (“with vorticity”) denotes inclusion of the optional solenoidal correction described in Section \ref{sec:vortical_method}. Hence rALPT(2s,wv) is a two-step ALPT+ridging calculation in which the second step contains both the potential and calibrated vortical components.}

% ---------------------------------------------------------
\subsection{Particle-mesh benchmarks at high resolution}
\label{sec:pm_bench}
We benchmark RLPT against PM reference solutions in two high-resolution volumes:
(i) $L=100\,h^{-1}\mathrm{Mpc}$ with $400^3$ particles ($\mathrm{d}L=0.25\,h^{-1}\mathrm{Mpc}$), and
(ii) $L=200\,h^{-1}\mathrm{Mpc}$ with $256^3$ particles ($\mathrm{d}L=0.78\,h^{-1}\mathrm{Mpc}$).
The PM references are obtained with \textsc{FastPM} \citep{2016MNRAS.463.2273F}, {evolving 2LPT initial conditions generated at $z=99$ down to $z=1$ and $z=0$}, using $50$ time steps and identical initial phases, to avoid large-scale cosmic variance effects.
In case (i) the force resolution is set to $\mathrm{d}L$, whereas in case (ii) we adopt a finer force mesh corresponding to a force resolution of $0.39\,h^{-1}\mathrm{Mpc}$ (i.e.\ twice the force resolution of the particle/mesh spacing).

\subsubsection{$L=100\,h^{-1}\mathrm{Mpc}$ with $400^3$ particles}
\label{sec:pm_100}

Figure~\ref{fig:dmV100} illustrates, by direct visual inspection, how cosmic-web morphology improves from 2LPT to ALPT and to rALPT: filaments become thinner and knots more pronounced, bringing the field closer to the \textsc{FastPM} reference at both $z=0$ and $z=1$.

The power spectra in Fig.~\ref{fig:pkV100} support the visual impression, showing a substantial recovery of power at high wavenumbers at both redshifts.
%Including the explicit vortical correction (rALPT-wv) yields an additional, albeit smaller, power increase on small scales.
Overall, the rALPT variants approach percent-level agreement with the PM reference up to $k\simeq 0.3$--$0.4\,h\,\mathrm{Mpc}^{-1}$.

The cross power spectra in Fig.~\ref{fig:ckV100} quantify the field-level information gain achieved by ridging.
At $z=0$, the cross-correlation coefficient $C(k)$ with the PM reference increases from $C\simeq 0.54$ (2LPT) to $C\simeq 0.94$ (rALPT) at $k=0.5\,h\,\mathrm{Mpc}^{-1}$, and from $C\simeq 0.05$ to $C\simeq 0.65$ at $k=1\,h\,\mathrm{Mpc}^{-1}$.
At $z=1$ we find the same behaviour, but with a reduced separation between the different approximations.

{\color{black} This reduced gain is expected. At higher redshift the field is less evolved, shell crossing and strongly nonlinear mode coupling are less developed, and the single-step LPT/ALPT baseline already captures a larger fraction of the PM solution. Ridging therefore has less missing short-range information to restore. The relevant comparison is consequently the residual error with respect to the baseline rather than the absolute proximity of the ridged solution to the reference.}

%~\ref{fig:dmV100}--\ref{fig:ckV100} show that ridging systematically restores missing nonlinear power and increases field-level cross-correlation relative to 2LPT and ALPT at both $z=0$ and $z=1$.
%In this setup, RLPT achieves percent-level agreement in $P(k)$ to at least $k\simeq0.3\,h\,\mathrm{Mpc}^{-1}$ and substantially improves cross-correlation over a broad range.

We use this high-resolution case study to test the optional solenoidal extension (Section~\ref{sec:vortical_method}).
At $z=0$, adding a calibrated vortical correction increases the cross-correlation with the PM reference at very high wavenumbers, beyond the regime where purely potential short-range corrections tend to saturate.
This is consistent with the expectation that solenoidal components become progressively more relevant toward the deeply nonlinear regime and finer scales, and it mirrors vorticity-related improvements reported for iterative Eulerian LPT frameworks \citep[e.g.][]{Kitaura_2024}.

A more systematic exploration of vortical ridging is left for future work, including alternative (nonlocal) source models for $\vomega_{\rm SR}$ motivated by perturbative expansions, and calibration over a wider parameter space and set of resolutions/redshifts.

% --- Your existing figures (kept as-is) ---

\subsubsection{$L=200\,h^{-1}\mathrm{Mpc}$ with $256^3$ particles}
\label{sec:pm_200}

{\color{black} Figure \ref{fig:dmV200} shows the corresponding density slices for this second benchmark. As in the higher-resolution case of Fig.~\ref{fig:dmV100}, ridging sharpens the filamentary network and compact overdense regions relative to the one-step LPT baselines, although the visual differences are smaller at $z=1$.}

Figures~\ref{fig:pkV200}--\ref{fig:ckV200} show the analogous PM benchmark at lower particle density but still high spatial resolution.
RLPT again improves the power spectrum and increases the cross-correlation relative to the single-step 2LPT and ALPT baselines, with percent-level agreement in $P(k)$ at least up to $k\simeq0.4\,h\,\mathrm{Mpc}^{-1}$ in this setup.

This box coincides with (and is therefore directly comparable to) the configuration studied in the eALPT work \citep{Kitaura_2024}.
The key distinction is methodological: eALPT achieves its accuracy through multiple Eulerian remappings (thereby modifying the effective large-scale evolution {including the long-range component of the displacement field} and generating vorticity through step composition), whereas RLPT applies a single, explicitly scale-filtered Eulerian completion as a post-processing correction on top of a chosen one-step baseline.
In this fixed-resolution regime, the latter provides a lightweight alternative that preserves the large-scale phases of the baseline transport while substantially improving the short-scale response at the cost of essentially one additional {solution of the Poisson equation}.
If we compare the two-step eALPT results shown in Fig.~3 of 
%Kitaura et al. 24 
\citep[][]{Kitaura_2024} to our two-step ridged ALPT (rALPT$(2{\rm s})$) in this configuration, we find that both approaches achieve a similar field-level agreement as quantified by the cross power spectrum.
However, rALPT$(2{\rm s})$ extends the range of percent-level accuracy in the matter power spectrum from $k\simeq 0.2$ to $k\simeq 0.4\,h\,{\rm Mpc}^{-1}$ in this benchmark.

% ---------------------------------------------------------
\subsection{\textsc{Abacus}  $N$-body benchmark on a coarse mesh}
\label{sec:abacus}

We next compare RLPT to a high-fidelity $N$-body reference.
As in the study of 
%Forero-Sánchez et al. 2024
\citep{Forero_2024} and {in the forthcoming companion paper presenting a novel spectral hierarchical cosmic web classification \citep{KitauraSinigaglia_2026}}, we rely on {dark matter fields and} halo catalogues from the publicly available \textsc{AbacusSummit} suite \citep{Maksimova_2021}, produced with the \texttt{Abacus} $N$-body code \citep{Garrison_2018,Garrison2021}. \textsc{AbacusSummit} comprises a set of 97 cosmologies around a central ``base'' cosmology consistent with Planck 2018 \citep{Planck2018}, each evolved in a $(2\,h^{-1}\mathrm{Gpc})^3$ volume with $6912^3$ particles, corresponding to a particle mass of $\simeq 2\times10^{9}\,h^{-1}M_\odot$ and designed to meet the resolution requirements of DESI-like tracers \citep{Levi2019}. 

In this section we explicitly show both plain CIC and CIC+tet variants, because the deposition choice is part of the fixed-resolution forward model.

In this work we use one realisation of the base-cosmology data at two snapshots, $z=0.2$ and $z=1.1$, and we make use of both the halo catalogues and the corresponding dark matter density field for our comparisons. For the halo-based tests, we construct a subsample by selecting the most massive haloes until reaching a target number density of $3\times10^{-3}\,(h/\mathrm{Mpc})^{3}$, yielding approximately $2.4\times10^{7}$ haloes in the full box.

We also exploit the matching initial conditions to generate approximate low-resolution realisations with our \texttt{WebON} pipeline \citep{Kitaura_2024}. The \textsc{Abacus} initial density field, originally provided on a $576^3$ mesh, is downsampled with a sharp Fourier-space filter to a lower-resolution grid of $N_{\rm low}=360^3$ cells, corresponding to a cell size $\Delta x\simeq 5.5\,h^{-1}\mathrm{Mpc}$. {In this way, we again ensure that our LPT calculations produce approximated dark matter fields with the same phases as the \textsc{Abacus} simulations and are not affected by cosmic variance}. This defines the fixed-resolution regime targeted in our coarse-mesh benchmarks; in terms of mass resolution the resulting approximate realisation is $\sim 10^{-4}$ times lower than the full-resolution reference.

%For a fair fixed-resolution comparison relevant to survey pipelines, we analyse all fields on a common $360^3$ mesh, corresponding to a cell size $\mathrm{d}L\simeq 5.55\,h^{-1}\mathrm{Mpc}$.

The RLPT forward model is run with a sparse particle load (a small subset of the full simulation particles), thereby isolating the regime where force/mesh resolution---rather than particle number---is the primary limitation and where post-processing short-range completions are most valuable.
This setup matches the practical conditions of recent subgrid modelling approaches \citep[e.g.][]{Forero_2024} and of survey-scale mock-production pipelines such as \texttt{HOLI-mocks} for DESI \citep[e.g.][]{KitauraSinigaglia_DESI_2026}.

%Figure~\ref{fig:dmV2000} illustrates the importance of mass assignment at coarse resolution. Plain CIC produces stronger granular, grid-scale fluctuations in low-density regions than the \textsc{Abacus} reference, whereas tetrahedral deposition yields a smoother, more spatially coherent representation of sheets and filaments. The corresponding improvement in field-level agreement is quantified by the cross-correlation coefficients in Fig.~\ref{fig:ckABACUS}.

The first finding comes from a visual impression by plotting slices of the matter field as shown in Figure \ref{fig:dmV2000}. The usual CIC mass assignment scheme yields fields which clearly have too much noise in low density regions as compared to the \textsc{Abacus}  solution.  
On such coarse meshes, the density-assignment scheme strongly impacts field-level agreement.

{\color{black} At this coarse resolution, plain CIC produces visibly stronger granular/grid-scale fluctuations in low-density regions than the Abacus reference. Tetrahedral deposition yields a smoother, more spatially coherent representation of sheets and filaments. The corresponding improvement is quantified by the cross-correlation spectra in Fig.~\ref{fig:ckABACUS}.}

The power spectra tell us a more complex story (see Figure \ref{fig:pkABACUS}). While it is not so difficult to achieve percent level accuracy beyond $k\simeq0.5\,h\,\mathrm{Mpc}^{-1}$ at high redshift even by ridging the 2LPT solution (see the red solid line -r2LPT- and the black solid line -rALPT- in the lower panel), we need to apply smooth particle ridging to achieve similar results at low redshift (see cyan curves in the upper panel).

However, the tetrahedra-based deposition exhibits a noticeable suppression of power at high wavenumbers.
While the {plain CIC} comparison is ``fair'' in the sense that both the \textsc{Abacus} and the LPT-based fields are measured with the same mass-assignment window, it becomes difficult to interpret differences in the power spectra when the underlying fields have been constructed with different deposition schemes.
In particular, changing the mass-assignment procedure effectively changes the small-scale smoothing kernel (and thus the window function) applied to the density field, so discrepancies at high $k$ can reflect window/aliasing effects rather than genuine differences in gravitational evolution.
For this reason, whenever we compare spectra across methods we either (i) enforce the same deposition for all fields (e.g.\ plain CIC for both reference and model), or (ii) explicitly account for the corresponding assignment windows (and any additional smoothing intrinsic to the tessellation) before attributing residual differences to the dynamics. A comparison with tetrahedra tesselation with \textsc{Abacus}  is difficult, as they have used a different number of particles.

Interestingly, the cross power spectra exhibit markedly different behaviour beyond the spherical Nyquist frequency $k_{\rm Nyquist}^{\rm sphere}\equiv \frac{\pi}{\mathrm{d}L}\simeq \frac{\pi}{5.55}\simeq 0.57\,h\,\mathrm{Mpc}^{-1}$ (see Figure \ref{fig:ckABACUS}). 
The Nyquist frequency of the box (maximum Fourier-mode magnitude at the grid corners) is given by $k_{\rm Nyquist}^{\rm box}=k_{\rm Nyquist}^{\rm sphere}\times\sqrt{3}$.
Whereas the {plain CIC} fields typically show increasingly irregular features beyond $k_{\rm Nyquist}^{\rm sphere}$---a signature of anisotropic aliasing and the fact that modes near the Cartesian Nyquist surface are not isotropically represented in spherically averaged statistics---the tetrahedra-based densities display a much smoother continuation across wavenumbers.
This behaviour is consistent with tetrahedra tessellation acting as a more geometrically faithful mass reconstruction from the particle distribution: by distributing mass continuously within tetrahedra rather than concentrating it at point particles with a compact cell window, the resulting density estimate suppresses small-scale sampling noise and reduces grid-direction artefacts, yielding a more stable cross spectrum even when approaching and formally exceeding the spherical Nyquist scale.
In practice, we therefore find tetrahedra deposition advantageous for field-level comparisons in the fixed-resolution regime, especially when the goal is to track the {relative} information content of approximate solvers across a wide $k$-range.

While the cross-correlation at $z=0.2$ increases noticeably when applying ridging on top of the tetrahedra-based solution, all approximations become much closer to each other at $z=1.1$.
This behaviour is expected in the fixed-resolution regime: at higher redshift the density field is less evolved and shell-crossing is less dominant, so a single-step perturbative transport already captures a larger fraction of the relevant modes, and the residual short-range correction becomes less critical.
In other words, the gain from an explicit Eulerian completion is largest when nonlinear sharpening below a few cells is the dominant deficiency (low redshift and/or coarse force resolution), whereas at $z\simeq 1$ the same mesh resolution probes a more weakly nonlinear field and the various LPT-based approximations yield similar field-level agreement.

The importance of ridging even at high redshifts becomes apparent once we move beyond the goal of reproducing the matter field on the mesh itself and instead use the evolved field as an ingredient of a {subgrid} tracer model.
In particular, when assigning tracer positions {within} cells (or relative to coarse cells) to control small-scale clustering, the relevant information is precisely the short-range morphology and local anisotropy that is only partially encoded at a fixed mesh resolution.
In this context, ridging provides a simple and deterministic way to enhance (or suppress) the short-range response of the underlying matter field while preserving its large-scale phases, thereby enabling controlled tuning of tracer clustering beyond the nominal mesh scale.

% ---------------------------------------------------------

\subsection{Proof of concept: subgrid halo positioning and small-scale control}
\label{sec:subgrid}

We present a proof-of-concept application in which ridging is used as a deterministic {subgrid response} model to control small-scale clustering beyond the nominal mesh resolution.
The setup is motivated by survey-mock pipelines in which the available information is often constrained on a coarse grid (e.g.\ halo counts per cell), while the scientific use case requires tracer positions with controlled clustering on substantially smaller scales. We stress that small scale clustering is crucial to accurately model systematic effects. The flexibility of ridging is demonstrated in Figure \ref{fig:pkspralptruns}, where the power of the ALPT solution is enhanced towards high wavenumbers with the choice of increasingly larger short range growth factors. 

The \texttt{PATCHY} mock method, which precedes the \texttt{HOLI-mocks} approach, assigned ALPT particle positions to match the number counts of luminous red galaxies (LRGs) and drew additional objects at random positions within a mesh cell when an insufficient number of particles was available \citep{Kitaura_2016}.
While this situation was relatively rare for BOSS-like number densities, the resulting ``in-cell randomisation'' can imprint unphysical small-scale structure (or de-structure) on scales comparable to the cell size, which becomes particularly problematic for observational effects that are highly sensitive to sub-Mpc separations, such as fibre collisions.
 In practice, this can bias the close-pair statistics and complicate the forward-modelling of fibre-assignment systematics \citep[e.g.][]{2017MNRAS.467.1940H}.

In 
%Balaguera et al. 2023 
\citep{Balaguera_2023}, we explored a deterministic ``collapse'' of particles to enhance clustering on small scales.
In that implementation, the target particle (or ``attractor'') within a cell/region was chosen in a simple, largely arbitrary manner, which already proved sufficient to boost small-scale power but did not optimally track the anisotropic collapse pattern of the underlying cosmic web.
This was improved in 
%Forero-Sánchez et al. 2024
\citep{Forero_2024}, where attractors were selected using collapsing-region information inferred from the Hessian of the density contrast within the \texttt{WebOn} framework, yielding a physically better-motivated relocation that aligns tracer positions with the local web geometry.

In the present work, RLPT provides a natural field-level generalisation of these ideas: the ``attractor'' behaviour emerges from an explicit short-range displacement operator (and its optional particle-based realisation), so that small-scale relocation is controlled by a small set of calibratable parameters while preserving the large-scale transport.

In this proof-of-concept experiment we separate three ingredients.
\begin{itemize}
    \item {Given} are coarse cell-scale constraints from the reference simulation, in particular halo number counts on a mesh of $\mathrm{d}L=5.55\,h^{-1}\mathrm{Mpc}$ (here taken directly from \textsc{Abacus} to isolate the positioning problem).
\item {Modelled} is the subgrid realisation: halos are placed on particle-level locations within each coarse cell (with cloning and, when needed, informed seeding), followed by a deterministic relocation step based on ridging or smooth-particle ridging (SPR).
\item {Tuned} is the resulting small-scale tracer clustering and close-pair statistics, controlled by the short-range displacement parameters (and, for SPR, by the kernel/viscosity-like regularisation and an optional small scatter), while preserving the imposed coarse constraints and the large-scale phases of the underlying transport.
\end{itemize}

We start from a low-resolution description of the tracer field given by halo number counts measured on a coarse mesh with cell size $\mathrm{d}L=5.55\,h^{-1}\mathrm{Mpc}$ (the same coarse resolution used in the \textsc{Abacus} comparison of Section~\ref{sec:abacus}).
We then populate each coarse cell by assigning halos to underlying dark-matter particle positions provided by an ALPT (or RLPT) matter field.
Operationally, halos are placed on particle locations {within the same cell}, cloning particles whenever the required halo count exceeds the number of available particles in that cell.
This preserves the large-scale phases and the cell-scale constraints by construction, while leaving freedom in the resulting {subgrid} distribution and thus in the small-scale clustering.

{ In very low-resolution forward models the situation can become more challenging.
In our \textsc{Abacus}-based setup, $\mathcal{O}(10\%)$ of coarse cells that contain halos in the reference catalogue contain {no} particles in the approximate realisation.
This occurs because the CIC+tet mass-assignment yields a continuous density field: tetrahedra can overlap a cell and produce a non-vanishing density even if no particle vertex lies inside that cell. Also, the approximate gravity solver introduces some deviations from the reference dark  matter field. 
As a consequence, a purely ``assign-halos-to-existing-particles'' strategy can fail locally.

We therefore complement it with an informed {seeding} step that introduces additional candidate particle locations in dynamically plausible regions---guided by the reconstructed density and the spectral hierarchical cosmic-web geometry---before the subsequent relocation (ridging / SPR) regulates the subgrid clustering.
In practice, we implement this by exploiting the cosmic-web classification together with region-dependent halo number densities, as detailed in the companion study \citep{KitauraSinigaglia_2026}.
}

To visualise the subgrid structure, we consider a $250\,h^{-1}\mathrm{Mpc}$ subvolume extracted from the \textsc{Abacus}  simulation and represent it on a $360^3$ mesh with cell size $\mathrm{d}L=0.69\,h^{-1}\mathrm{Mpc}$.
{ A representative slice is shown in Figure~\ref{fig:halos}.
The {upper left} panel illustrates the coarse cell-centred representation of the low resolution \textsc{Abacus}--halo number counts distribution (at $\mathrm{d}L=5.55\,h^{-1}\mathrm{Mpc}$), while the {upper right} panel shows the ALPT dark matter particle positions.
The lower panels show the high resolved halo number count distributions correspodning to the reference \textsc{Abacus} simulation on the left and to the reconstruction on the right. 
}

{To alleviate the geometrical artifacts arising from the mesh structure, we explore  the application of SPR as a subgrid model.} Figure~\ref{fig:halos} shows the corresponding halo distributions.
The {lower left} panel represents the \textsc{Abacus}  halo field on a high resolution.
The {lower right} panel shows a reconstruction obtained by taking halo counts on the coarse mesh ($\mathrm{d}L=5.55\,h^{-1}\mathrm{Mpc}$), assigning halos to ALPT particle positions (upper right panel) within each coarse cell (cloning particles as needed), and then applying a short relocation step consisting of SPR (see Section~\ref{sec:spr}) plus a small positional scatter. To avoid potential inaccuracies stemming from halo painting methods such as bias models \cite[see e.g.][]{Coloma_2024}, we do not predict halo number counts using approximated methods, but rely directly on the halo counts from the original \textsc{Abacus} halo catalogue.
For comparison, the {upper} panel shows the trivial alternative in which haloes are placed at the centres of their corresponding coarse cells, which erases subgrid information and strongly suppresses small-scale clustering.

The impact on clustering is quantified in Figure~\ref{fig:pkhalosprlpt}.
The solid curve shows the \textsc{Abacus}  halo power spectrum measured on the high-resolution mesh ($\mathrm{d}L=0.69\,h^{-1}\mathrm{Mpc}$).
The dash--dotted curve corresponds to the ALPT dark matter particle distribution from the coarse mesh gridded onto the high-resolution mesh, illustrating the baseline level of small-scale structure in the underlying fast-gravity field.
The subsequent DM--ALPT curves correspond to the same dark matter field {after} constraining it to match \textsc{Abacus}  halo number densities in progressively finer cosmic-web hierarchies, showing that hierarchical environment conditioning provides a deterministic route to regulate small-scale power \citep[][]{KitauraSinigaglia_2026}.
Finally, the magenta dotted curve shows the halo field obtained by placing halos at coarse cell centres, while the dashed blue curve corresponds to assigning halos to ALPT particles (with cloning when necessary) followed by smooth particle ridging.
The latter retains the coarse constraints and restores a substantial fraction of the small-scale clustering by relocating tracers coherently along local density ridges rather than by injecting purely random scatter.

This highlights an important practical point: even if the underlying dark-matter field is not percent-accurate to very high $k$, it can still be sufficiently faithful for a field-level bias model on the scales that anchor tracer formation; ridging then acts as an explicit post-processing response layer that deterministically adjusts the {tracer} clustering below a transition scale without requiring an equally accurate dark-matter solution on all resolved modes.

The magenta curve illustrates the {idealised} output of a field-level bias model at the cell level: it uses the true \textsc{Abacus}  halo number counts on the coarse mesh (i.e.\ perfect agreement in counts by construction).
In Fourier space this case exhibits the characteristic imprint of the underlying mesh, visible as a comb-like pattern of spikes at harmonics of the fundamental grid frequency.
Moreover, because all objects are effectively confined to cell centres, the field rapidly loses small-scale power, with a pronounced suppression already emerging around $k\simeq 0.1$--$0.2\,h\,\mathrm{Mpc}^{-1}$.

By contrast, the dashed blue curve highlights the potential of a deterministic post-processing relocation.
Assigning halos to particle positions within each coarse cell (including cloning when required) and then applying smooth-particle ridging (plus a small scatter) largely removes the grid imprint and yields a nearly unbiased power spectrum over a much wider range of wavenumbers, remaining close to the reference up to $k\sim 8\,h\,\mathrm{Mpc}^{-1}$ in this proof-of-concept configuration.

% =========================================================
\section{Discussion and outlook}
\label{sec:discussion}

We introduced {Ridged Lagrangian Perturbation Theory} (RLPT), a minimal two-step completion of LPT-based transport tailored to the fixed-resolution regime of modern survey pipelines.
RLPT is built around a simple operator split: (i) a fast long-range transport (e.g.\ $n$LPT or A-($n$)LPT/ALPT) that preserves the correct large-scale phases, followed by (ii) a single, explicitly Eulerian short-range completion applied as a post-processing step on the realised density field.
In practice, this completion is implemented by {\color{black} applying a spherical-collapse-inspired nonlinear remapping to the realised Eulerian density, isolating the short-range component of the resulting effective source, and reconstructing a short-range potential displacement via Poisson inversion}.
The resulting update costs essentially one additional solution of the Poisson equation (one forward and one inverse FFT on the mesh) and can therefore be deployed cheaply in production settings, including as a correction on top of single-step ALPT outputs at arbitrary redshifts.

The numerical studies demonstrate three robust outcomes.
First, across two high-resolution PM benchmarks, a {single} ridging step produces a large and systematic gain in field-level agreement beyond the improvement achieved by going from 2LPT to ALPT, while simultaneously recovering missing nonlinear power.
In the $L=100\,h^{-1}\mathrm{Mpc}$ case, ridging boosts the cross-correlation with the PM reference at $z=0$ from $54\%$ to $94\%$ at $k=0.5\,h\,\mathrm{Mpc}^{-1}$ and from $5\%$ to $\sim65\%$ at $k=1\,h\,\mathrm{Mpc}^{-1}$, with percent-level agreement in $P(k)$ up to $k\simeq0.3$--$0.4\,h\,\mathrm{Mpc}^{-1}$; similar trends hold at $z=1$ with reduced separation.
Second, the \textsc{Abacus}  comparison highlights that on coarse meshes the density assignment procedure is itself a leading-order part of the forward model: tetrahedral tessellation yields markedly smoother and more stable cross spectra (in particular beyond the spherical Nyquist scale) than plain CIC, and ridging provides an additional improvement on top of this more faithful coarse-grained reconstruction.
Third, and most importantly for mock production, ridging generalises naturally into a deterministic {subgrid response model}: once tracer constraints are imposed on a coarse grid (e.g.\ halo counts per cell), the same short-range operator (and its SPH-inspired particle realisation) can relocate tracers coherently within cells, thereby controlling small-scale clustering and close-pair statistics beyond the nominal mesh resolution in a way that is difficult to achieve with purely cell-based or random in-cell prescriptions.
A key practical lesson from the subgrid halo-positioning experiment is that placing tracers at coarse cell centres produces an artificial lattice imprint in Fourier space (spikes at grid harmonics) and a rapid loss of small-scale power already at $k\simeq 0.1$--$0.2\,h\,\mathrm{Mpc}^{-1}$, which is undesirable for applications sensitive to close pairs and fibre-collision systematics.
By contrast, assigning halos to particle locations within each cell (with cloning when needed) and applying a short smooth-particle ridging relocation step largely erases the grid imprint and restores small-scale clustering, yielding a near-unbiased power spectrum up to $k\sim 8\,h\,\mathrm{Mpc}^{-1}$ in this configuration.

Together, these results position RLPT as an explicit, calibratable short-range closure for field-level modelling: it preserves large-scale transport, injects controlled nonlinear sharpening below a tunable transition scale, and provides a practical ``knob'' to enhance or suppress short-scale clustering in a deterministic manner.
This makes RLPT well suited both for rapid generation of accurate dark matter fields (up to quasi-nonlinear scales relevant for cosmological analyses) and for tracer-level applications where subgrid positioning and small-scale systematics control are essential.
Even if the underlying dark matter field is not percent-accurate to very high $k$, it can still be sufficiently faithful on the scales that anchor a field-level bias model (large-scale phases and intermediate-scale morphology) \citep[][]{Coloma_2024}.
In that regime, ridging provides an efficient post-processing ``response layer'' that can be calibrated directly on the {tracer} statistics, enabling controlled enhancement or suppression of tracer clustering below a transition scale without requiring a commensurately accurate dark matter solution at all resolved wavenumbers.

Altering the small-scale clustering provides not only an additional degree of freedom to model nonlinearities in the standard Newtonian $\Lambda$CDM scenario, but also a way to mimic alternative theories leaving an imprint on the smaller scales, such as modified gravity and warm/hot dark matter \citep[see also the application of nonlocal bias to mimick $f(R)$ models:][]{Garcia-Farieta_2024}.

\subsection{Interpretation: RLPT as an explicit short-range closure}

A useful way to interpret RLPT is as an explicit short-range closure acting on the {field} rather than on correlators.
The method compresses unresolved nonlinear dynamics into a limited number of interpretable parameters---primarily a transition scale and a short-range growth amplitude---while preserving the transport phases laid down by the long-range $n$LPT/A-($n$)LPT backbone.
Because the ridging step is applied after mapping, it can be deployed flexibly: (i) as a post-processing improvement of an existing ALPT production run, (ii) as a controllable ``dial'' to increase or decrease clustering below a chosen scale (including deterministic sub-/super-Poisson behaviour in tracer models), and (iii) as a building block for tracer-level subgrid repositioning beyond the mesh.
This flexibility is particularly attractive for the regime where accelerated PM methods remain too costly to deploy at survey volume and sampling requirements, or where inference frameworks demand many forward evaluations.
RLPT is complementary to EFT: it is a field-level, scale-separated closure calibrated on reference statistics, whereas EFT remains a systematically improvable theory for correlators on mildly nonlinear scales.

\subsection{Limitations and practical guidance}

The numerical studies also clarify where RLPT is most effective and where it is less critical.
The gain from ridging is largest at low redshift and at fixed/coarse force resolution, where the baseline transport yields correct large-scale phases but insufficient nonlinear sharpening below a few cells.
At high redshift and very coarse grids, the mesh-level improvements can become modest, but ridging remains useful once one moves to tracer-level subgrid modelling, where the information needed to control small-scale clustering is precisely what the completion step is designed to provide.
Separately, the \textsc{Abacus}  comparison highlights that mass assignment should be treated as part of the model: on coarse meshes, CIC+tet can improve field-level behaviour as much as (or more than) refinements of the dynamical approximation itself, and it synergises well with ridging.

\subsection{Future work}

Several directions are immediate and are already motivated by the results shown here:

\begin{itemize}

\item \textbf{Systematic calibration across redshift and resolution:}
map $(r_s, D_{\rm SR})$ (and, where relevant, the spherical collapse remapping parameters) as functions of redshift and mesh scale, and provide ``default'' calibrations for production mock pipelines  (including automated optimisation and posterior sampling).

\item \textbf{Vorticity modelling:}
extend the proof-of-concept vortical correction to more general, physically motivated vorticity-source models, including non-local operators and higher-derivative terms, and calibrate on solenoidal displacement/velocity-vorticity spectra and field-level metrics across multiple boxes and redshifts.

\item \textbf{Iterated completion and stability:}
quantify the accuracy ladder obtained by a small number of repeated ridging steps, and establish robust stability prescriptions (caps and/or adaptive step sizes) that preserve large-scale transport while avoiding over-sharpening.

\item \textbf{SPRLPT development:}
formalise the SPH-inspired smooth-particle ridging as an operator family (kernel choices, neighbour strategies, pressure/viscosity prescriptions), benchmark it against the mesh-based ridging in cost and accuracy, and evaluate its performance in the subgrid tracer-placement regime where it is expected to be most advantageous; in particular, explore an iterated variant where the viscosity term damps updates relative to the previous iterate to improve stability and convergence.

\item \textbf{Spectral-hierarchy attractors for subgrid modelling:}
explore spectral hierarchical cosmic-web regions as physically motivated ``attractors'' for deterministic subgrid relocation, and quantify the impact on tracer positioning, anisotropic collapse alignment, and small-scale clustering beyond the mesh \citep[see][for the first level case]{horowitzDifferentiableCosmologicalHydrodynamics2025}.

\item \textbf{DESI-scale production and systematics:}
apply RLPT-based subgrid relocation to full survey-like mock catalogues, quantify improvements in small-scale clustering and close-pair statistics, and test impact on observational systematics modelling (including fibre collisions) in forthcoming DESI applications.

\item \textbf{Super- and sub-Poisson clustering modelling:}
investigate RLPT as a deterministic framework to increase (through positive short-range displacements) or decrease (through negative short-range displacements) tracer clustering, as an alternative to stochastic approaches such as negative-binomial models commonly used to enhance power at high wavenumbers \citep[][]{Kitaura_2014}.

\item \textbf{Tracer-property assignment:}
explore RLPT as a framework to enhance or suppress the clustering of tracer properties \citep[see related work in][]{Zhao_2015,Balaguera_2023}, particularly in regimes where assigning masses to the most massive objects is challenging because of exclusion effects and the resulting anti-correlations.

\end{itemize}

\subsection{Final remark}

Overall, the results support the main premise of RLPT: in the fixed-resolution regime relevant to survey pipelines, a single explicit Eulerian completion applied as a post-processing step can recover a substantial fraction of the missing short-range response of one-step LPT/ALPT baselines, yielding both improved small-scale clustering and increased field-level agreement at minimal computational overhead.
Together with accurate mass assignment (notably tetrahedral tessellation on coarse meshes) and with the possibility of calibrated solenoidal extensions and particle-local (SPH-inspired) variants, RLPT provides a practical and modular route to produce accurate dark matter fields and controllable tracer realisations for modern large-scale structure analyses.

% =========================================================

\section*{Acknowledgements}

The authors thank the \texttt{HOLI-mocks} team for useful discussions and testing the \textsc{WebON}-code, special mention to Ana Almeida, José María Coloma-Nadal, Ginevra Favole, Daniel Forero-Sánchez, Jorge García Farieta, Fernando Frías García Lago, Pere Rosselló Truyols, Yunyi Tang, Natalia Villa Nova Rodrigues, Cheng Zhao. 
FSK acknowledges the Instituto de Astrof\'isica de Canarias (IAC) for continuous support to the \textit{Cosmology with LSS probes} activities
and the Spanish Ministry of Innovation and Science / Agencia Estatal de Investigaci\'on for support of the project
\textit{Big Data of the Cosmic Web} (PID2020-120612GB-I00/AEI/10.13039/501100011033), under which this work has been conceived and carried out and  the \textit{FIRE (Field-level bayesian Inference to Reconstruct the univErse)} project, which has been recently funded by a \textit{Proyecto de Generaci\'on de Conocimiento 2024}, PID2024-160504NB-I00.
FS acknowledges support from the \textit{Institute for Fundamental Physics of the Universe} postdoctoral fellowship scheme. FS is also grateful to IAC for hospitality during part of the realization of this project.

\section*{Disclaimer}
The \texttt{WebOn} code used in this work has been mainly written by FSK in C++ without AI.

\texttt{WebOn} is a modular framework whose methodological components are presented across three companion papers: the present work on ridged Lagrangian perturbation theory; a dedicated paper on the spectral hierarchical cosmic-web  \citep{KitauraSinigaglia_2026}; a further paper describing the lightcone evolution mode within the \texttt{HOLI-mocks} pipeline {to produce the DESI covariance mocks used for the DR2 Key Projects} \citep{KitauraSinigaglia_DESI_2026} (see: \url{www.cosmic-signal.org}).
\texttt{WebOn} has already been used in three previous publications: one presenting the \emph{Hierarchical cosmic web and assembly bias} \citep{Coloma_2024} approach for effective field-level bias modelling exploiting the hierarchical cosmic web $k^{-2}$ and $k^0$ levels; one presenting \emph{Cosmomia: cosmic-web based redshift-space halo distribution} \citep{Forero_2024} for subgrid modelling exploiting the knots $k^0$  level as attractors; and \emph{Fast and accurate Gaia-unWISE quasar mock catalogs from LPT and Eulerian bias} \citep{2025arXiv250915890S}, using the  two-hundred 10 $h^{-1}\,\mathrm{Gpc}$ side  cubical volume simulations  performed with the lightcone evolution \textsc{WebOn} mode on the Leonardo@CINECA pre-exascale supercomputing facilities in September 2023 through the INAF\_C9A09 HPC proposal (637k core hrs). It has also served as a reference for its \texttt{JaX} differentiable version in \emph{Differentiable Fuzzy Cosmic-Web for Field Level Inference} \citep[][]{2025arXiv250603969R}, which is {under active development}.

\bibliographystyle{JHEP}
\bibliography{references}

\end{document}